\documentclass[14pt,candidate]{disser}

\usepackage[a4paper, includehead, includefoot, mag=1000,
            left=2.5cm, right=2cm, top=2cm, bottom=2cm]{geometry}
\usepackage[T2A]{fontenc}
\usepackage[utf8]{inputenc}
\usepackage[english,russian]{babel}

\usepackage{epstopdf}
\usepackage[pdftex,draft=true]{hyperref}
\usepackage{graphicx}
\usepackage[backend=biber, style=gost-numeric, sorting=ynt]{biblatex}
\usepackage{csquotes}

\usepackage{tipa}
\usepackage{bm}
\usepackage{times}
\usepackage{upgreek}
\usepackage{subfigure}
\usepackage{url}
\usepackage{array}
\usepackage{graphicx}
\usepackage{amsmath,amssymb,amsfonts}
\usepackage{float}
\usepackage[boxed]{algorithm2e}

\graphicspath{{fig/}}

\def\x{{\mathbf x}}
\def\f{{\mathbf f}}
\def\k{{\mathbf k}}
\def\b{{\mathbf b}}

\def\p{{\mathbf p}}
\def\s{{\mathbf s}}
\def\u{{\mathbf u}}
\def\R{{\mathbf R}}
\def\w{{\mathbf w}}
\def\I{{\mathbf I}}
\def\X{{\mathbf X}}

\def\a{{\mathbf a}}
\def\b{{\mathbf b}}
\def\P{{\mathbf P}}
\def\M{{\mathbf M}}
\def\Q{{\mathbf Q}}
\def\C{{\mathbf C}}
\def\P{{\mathbf P}}
\def\L{{\mathbf L}}
\def\A{{\mathbf A}}
\def\B{{\mathbf B}}
\def\C{{\mathbf C}}
\def\D{{\mathbf D}}
\def\U{{\mathbf U}}
\def\v{{\mathbf v}}

\DeclareMathOperator{\tr}{tr}

\def\0{{\bm 0}}

\newcommand{\stack}[3]{\mathrel{\mathop{\kern0pt #1}\limits_{#3}^{#2}}}

\renewcommand{\thesubfigure}{\roman{subfigure}}
\makeatletter
\renewcommand{\@thesubfigure}{(\thesubfigure)\space}
\renewcommand{\p@subfigure}{\thefigure}
\makeatother

\renewcommand\theappendix{\Asbuk{chapter}}

\begin{document}


\institution{
Министерство образования и науки Российской Федерации\\
Московский Авиационный Институт\\ (Государственный технический университет)
}
\libcatnum{621.396.96}
\topic{Разработка и исследование адаптивных фильтров селекции движущихся целей}
\title{ДИССЕРТАЦИЯ\\
на соискание ученой степени\\
кандидата технических наук}
\author{Орешкин Б.Н.}
\specnum{05.12.14} \spec{ Радиолокация и радионавигация}
%
\sa{Бакулев П.~А.}
\sastatus{д.~т.~н., проф.}

\city{Москва}
\date{2008}

\maketitle
\tableofcontents
\intro

\paragraph{Актуальность темы}

состоит в необходимости повышения качественных показателей
алгоритмов системы АСДЦ. В частности, многие существующие адаптивные
алгоритмы сходятся к оптимальному решению при неограниченном росте
объема обучающей выборки. Однако на практике объем обучающей выборки
сильно ограничен неоднородностью помехи. Кроме этого, влияние таких
факторов, как присутствие в обучающей выборке сигнала от цели
приводит к снижению скорости сходимости адаптивных алгоритмов к
оптимальному решению или разрушению их асимптотической
оптимальности, что приводит к значительным потерям при применении
адаптивной обработки в реальной ситуации. Таким образом, основными
направлениями совершенствования существующих адаптивных алгоритмов с
целью уменьшения указанных потерь являются увеличение скорости их
сходимости (уменьшение количества требуемых обучающих выборок) и
разработка адаптивных алгоритмов с пониженной чувствительностью к
присутствию в обучающей выборке полезного сигнала. Вопрос увеличения
скорости сходимости адаптивных алгоритмов достаточно хорошо
проработан в литературе. Однако часто увеличение скорости сходимости
адаптивных алгоритмов основано на определенных предположениях
относительно регулярности помех, либо применении искусственной
регуляризации оценок определенных параметров алгоритма. В результате
в первом случае ускоренный алгоритм может не быть устойчивым к
влиянию шумов, а во втором --- возникает необходимость определения
величины параметра регуляризации, который как правило трудно
определим аналитически. Следовательно поиск алгоритмов с повышенной
скоростью сходимости не полагающихся на регулярность помех, а также
методов отыскания оптимальных значений параметра регуляризации
являются актуальными направлениями исследования. С другой стороны,
исследованию влияния присутствия полезного сигнала в обучающей
выборке и методов борьбы со связанным с этим явлением подавления
адаптивным алгоритмом сигнала от цели посвящено сравнительно мало
работ. Существующие работы как правило полагаются на модель
проникновения полезного сигнала в обучающую выборку, основанную на
наличии ошибки при калибровке антенны, которая может быть оценена
аналитически до калибровки или экспериментально во время калибровки.
Однако данная модель не учитывает таких особенностей, как
возможность проникновения в обучающую выборку сигнала от быстро
маневрирующей или распределенной цели даже при идеально настроенной
антенне. Таким образом, поиск непараметрических подходов ослабления
эффекта подавления сигнала от цели не полагающихся на модель
проникновения сигнала от цели в обучающую выборку также
представляется актуальным направлением исследования.

\paragraph{Цель диссертационной работы}

заключается в повышении качественных показателей алгоритмов системы
АСДЦ при конечной длительности обучающей выборки и загрязнении
обучающей выборки сигналом от цели. Для достижения цели
диссертационной работы необходимо решить следующие задачи:
\begin{enumerate}
\item Проанализировать методы синтеза и анализа адаптивных
алгоритмов.
\item Проанализировать существующие алгоритмы режекции помех в системе
АСДЦ.
\item Разработать непараметрический алгоритм итеративной оптимизации коэффициента регуляризации
алгоритма РОКМ, основанный на максимизации эмпирического отношения
Рэлея.
\item Разработать и проанализировать алгоритм LMS с квадратичным ограничением на основе
упрощенной схемы стохастической аппроксимации множителя Лагранжа.
\item Провести моделирование разработанных алгоритмов при режекции
помех в составе системы АСДЦ и проанализировать полученные
экспериментальные результаты.
\end{enumerate}

\paragraph{Методы исследований,}

использованные в диссертации, основаны на статистической теории
обнаружения и оценивания, матричной алгебре, выпуклой оптимизации,
математическом и компьютерном моделировании. Главные результаты
получены на основе методов прикладной статистики, выпуклой векторной
оптимизации, аналитической и стохастической аппроксимации и
численных методов вычислительной математики.

\paragraph{Научная новизна}

диссертации заключается в следующем. По критерию максимума ОСПШ на
выходе разработаны адаптивные алгоритмы системы АСДЦ позволяющие
увеличить скорость адаптации и уменьшить влияние загрязнения
обучающей выборки полезным сигналом от цели. Разработан
непараметрический алгоритм оптимизации коэффициента алгоритма РОКМ,
основанный на линеаризации выражения для весов соответствующего
адаптивного фильтра и максимизации эмпирического отношения Рэлея.
Предложена соответствующая итеративная реализация этого алгоритма.
Разработан алгоритм LMS с квадратичным ограничением, использующий
упрощенную схему стохастической аппроксимации множителя Лагранжа.
Выяснена асимптотическая эквивалентность алгоритмов LMS c линейным и
квадратичным ограничением алгоритму LMS без ограничения в случае
отсутствия корреляции сигнала и помехи.

Моделирование показало, что применение предлагаемой оптимизации
коэффициента регуляризации алгоритма РОКМ приводит к выигрышу в 0--8
дБ в зависимости от входного ОСПШ по сравнению со стандартным РОКМ
использующим фиксированный коэффициент регуляризации в случае
загрязнения обучающей выборки полезным сигналом. В случае отсутствия
такого загрязнения потери незначительны и составляют 0--0,5 дБ.
Исследование разработанного алгоритма LMS с квадратичным
ограничением показало, что его скорость сходимости превышает
скорость сходимости алгоритма с линейным ограничением и позволяет
сократить размер обучающей выборки на 20--40\%. Кроме этого,
обнаружено, что применение квадратичного ограничения позволяет
снизить проникновение составляющих помехи на выход адаптивной
системы режекции на 30--35 дБ по сравнению с алгоритмом использующим
линейное ограничение при загрязнении обучающей выборки сигналом от
цели.

\paragraph{На защиту выносятся следующие положения:}
\begin{enumerate}
\item Алгоритм итеративной оптимизации коэффициента регуляризации алгоритма
РОКМ, основанный на максимизации эмпирического отношения Рэлея,
позволяющий получить выигрыш в выходном ОСПШ в 0--8 дБ  по сравнению
с РОКМ, использующим фиксированный коэффициент регуляризации, в
случае загрязнения обучающей выборки полезным сигналом.
\item Алгоритм LMS с квадратичным условием, использующий упрощенную схему стохастической
аппроксимации множителя Лагранжа и позволяющий сократить размер
обучающей выборки на 20--40\% и в то же время снизить просачивание
составляющих помехи на выход адаптивной системы режекции на 30--35
дБ по сравнению с алгоритмом использующим линейное ограничение при
загрязнении обучающей выборки сигналом от цели.
\end{enumerate}


\paragraph{Научное и практическое значение полученных результатов}

состоит в усовершенствовании алгоритмов системы АСДЦ при конечном
размере обучающей выборки и ее загрязнении полезным сигналом от
цели, что достигается соответствующей оптимизацией параметра
регуляризации оценки ковариационной матрицы помехи и модификацией
вида ограничения в алгоритме адаптивной подстройки весов фильтра
режекции помехи. Предложен непараметрический алгоритм оптимизации
коэффициента регуляризации алгоритма РОКМ, основанный на
линеаризации выражения весов адаптивного фильтра и максимизации
эмпирического отношения Рэлея и соответствующая итеративная
процедура вычисления оптимизированного коэффициента регуляризации.
Предложен алгоритм LMS с квадратичным условием и упрощенной
процедурой стохастической аппроксимации множителя Лагранжа, выяснены
асимптотические свойства алгоритма при отсутствии корреляции между
сигналом и помехой и проведены экспериментальные исследования,
показывающие выгоды применения алгоритма LMS с квадратичным
ограничением по сравнению с алгоритмом LMS с линейным ограничением.


\paragraph{Апробация работы}

произведена в форме научных докладов и дискуссий по основным
результатам работы на 4 всероссийских и международных конференциях.

\paragraph{По теме диссертации опубликовано}

8 печатных работ, в том числе одна публикация в центральной
печати~\cite{Ore07a} и 7 материалов
докладов~\cite{Ore06a,Ore06b,Ore06c,Ore07d,Ore07e,Ore08c,Ore08d} на
международных конференциях. Одна работа принята к публикации в
центральной печати \cite{Ore07b}, две работы представлены в редакцию
для публикации в центральной печати~\cite{Ore07c,Ore08f}, два
материала докладов представлено для публикации в оргкомитет
международной конференции~\cite{Ore08a,Ore08b}.


\paragraph{Структура и объем диссертации}

Диссертация состоит из введения, четырех глав, заключения,
библиографического списка из 184 наименований и приложения.
Диссертация содержит 149 страниц, в том числе 18 рисунков. Во
введении обоснована актуальность темы диссертации и представлены
основные положения, выносимые на защиту. Определены цели и задачи
диссертационного исследования, а также обоснована структура и объем
диссертации. В главе~\ref{chap:Ch1} представлен обзор методов
первичной обработки информации РЛС, осуществляемой с целью
подавления помех и обоснована необходимость применения адаптивной
обработки радиолокационной информации. В главе~\ref{chap:Ch2}
произведен обзор существующей по теме диссертации литературы,
приведена классификация существующих адаптивных алгоритмов, которые
могут быть применены в системе АСДЦ с целью подавления помех,
описаны существующие критерии качества, применяемые при синтезе и
анализе алгоритмов АСДЦ. В главе~\ref{chap:Ch3} описаны результаты
синтеза и теоретического анализа предлагаемых адаптивных алгоритмов
системы АСДЦ, а в главе~\ref{chap:Ch4} представлены результаты
моделирования предлагаемых алгоритмов и проделан сравнительный
анализ с некоторыми существующими решениями. В заключении подведены
итоги работы и сформулированы ее основные научные и практические
результаты. В приложении приведены список аббревиатур и тексты
компьютерных программ, использованных при моделировании.

\chapter{Методы обработки сигналов в РЛС с СДЦ} \label{chap:Ch1}

Обнаружение движущихся целей на фоне помех является одной из
важнейших задач радиолокации. Помехи на входе системы первичной
обработки сигналов РЛС состоят из аддитивных компонент собственного
белого шума приемника, отражений от местных предметов и атмосферных
образований, отражений от различного рода пассивных помех, например
облаков дипольных отражателей, сигналов активных помех, хаотической
импульсной помехи. Теория обнаружения квазидетерминированных
сигналов (отраженных сигналов), на фоне белого шума разработана в
достаточной мере и базируется на статистической теории принятия
решений в случае независимых выборок
\cite{Vai60,Lev66,Lev68,Lev76,Mid61,Sos92,Tih66,Tih82,Tih04,Shir81,Sko90,Tre01a,Tre01b}.
В большинстве случаев теоретические допущения относительно вида
распределения шума и сигнала на входе обнаружителя имеют место и на
практике. Кроме того, физическая природа белого шума в тракте РЛС,
который может быть, например, тепловым или дробовым, допускает
сравнительно простую параметризацию распределений. Что в свою
очередь облегчает осуществление математического моделирования и
синтеза оптимальных, робастных и адаптивных обнаружителей
\cite{Lev76}. В то же время и  Ван Трис и Левин (\cite{Tre01a} и
\cite{Lev76}) указывают в своих работах на отсутствие подобной
универсальности и завершенности в теории обнаружения
квазидетерминированных сигналов на фоне коррелированных помех (коими
являются отражения от местных предметов и атмосферных образований, а
также от различного рода искусственных образований типа облака
дипольных отражателей). В этой области существует несколько
фундаментальных проблем:
\begin{enumerate}
\item Статистически устойчивые оптимальные методы формирования тестовой
статистики разработаны только в предположении о бесконечности
интервала наблюдения;
\item Результаты синтеза оптимальных алгоритмов не могут быть напрямую
применены на практике, если нет полной уверенности, что модель
помехи, используемая при синтезе метода, полностью отражает
реальность;
\item В силу физической природы коррелированной помехи, вид
ее распределения и спектра подвержены сильной вариабельности во
времени и пространстве.
\end{enumerate}

Именно поэтому за последние 60 лет исследованию проблемы селекции
полезных сигналов (далее просто сигналов) на фоне коррелированных
помех (далее просто помех) посвящено много работ. Отличительной
особенностью проблемы СДЦ является тот факт, что борьба с помехами в
РЛС невозможна путем увеличения мощности передатчика \cite{Bak86}.
Поэтому основу принципа работы системы СДЦ составляет наличие
различий спектральных характеристик сигнала и помехи. Различие
спектров сигнала и помехи обусловлено эффектом Доплера, приводящего
к изменению частоты электромагнитных волн, отражаемых движущимся
объектом \cite{Bak86,Sko90,Sti98,Bar05}:
\begin{equation} \label{eqn:doppler}
f_\text{д} = \frac{2v_r}{c}f_0,
\end{equation}
где $f_\text{д}$~--- величина доплеровской поправки, $v_r$~---
относительная радиальная скорость цели, $c$~--- скорость света, а
$f_0$~--- несущая частота РЛС. Таким образом, в идеальном случае
сигналы, отраженные от неподвижных целей (местных предметов) будут
иметь нулевую доплеровскую поправку, а сигналы, отраженные от
движущихся целей будут иметь доплеровскую поправку отличную от нуля.
Очевидно, что сигналы могут быть отделены от помех посредством
пропускания их аддитивной смеси через высокочастотный фильтр
\cite{Kle98}. Однако в реальной ситуации кроме отражений от местных
предметов, сигнал может содержать отражения от атмосферных
образований, отражения от различного рода преднамеренных образований
типа облака дипольных отражателей, а также сигналы постановщиков
активных помех. Доплеровская поправка в спектре этих помех может
быть не равна нулю \cite{Nar94}, что может приводить к проникновению
большой части мощности спектральных компонент помех этого типа на
выход оптимального обнаружителя сигнала на фоне белого шума, что
недопустимо \cite{Bar05}. Кроме того, все перечисленные помехи
представляют собой случайные процессы с конечной (не нулевой)
шириной спектра, которая может превышать полосу подавления фильтра
помехи, что также приводит к значительному ухудшению характеристик
обнаружителя \cite{Gre01,Lom01}. Необходимо также отметить, что
мощность помех случайна и нестационарна по пространству и времени
(распределение помехи меняется случайным образом в элементах
разрешения РЛС как в течение одного обзора, так и между обзорами)
\cite{Gin99,Har01}. Таким образом, сложность проблемы заключается в
необходимости синтеза эффективных алгоритмов СДЦ в условиях слабой
априорной определенности относительно статистических свойств помех в
реальной помеховой обстановке \cite{Hay85a}. Проследим развитие
архитектуры систем СДЦ. Развитие архитектуры и алгоритмов систем СДЦ
было обусловлено в основном следующими факторами
\cite{Gal93,Ham01,Sko90,Sti98,Nat91}:
\begin{enumerate}
\item Совершенствование аппаратуры передатчика РЛС (переход от
некогерентных генераторов к когерентным усилителям СВЧ);
\item Совершенствование аппаратуры приемника (расширение
динамического диапазона приемного устройства и линеаризация тракта
вход приемника~--- вход системы СДЦ);
\item Переход с аналоговой на цифровую обработку сигналов
с последующим увеличением вычислительной мощности цифровых
процессоров;
\item Переход от антенн с механическим сканированием луча
к антеннам с электрическим сканированием луча (фазированным антенным
решеткам).
\end{enumerate}

\begin{figure*}[h]
\centering
\includegraphics[width=\columnwidth]{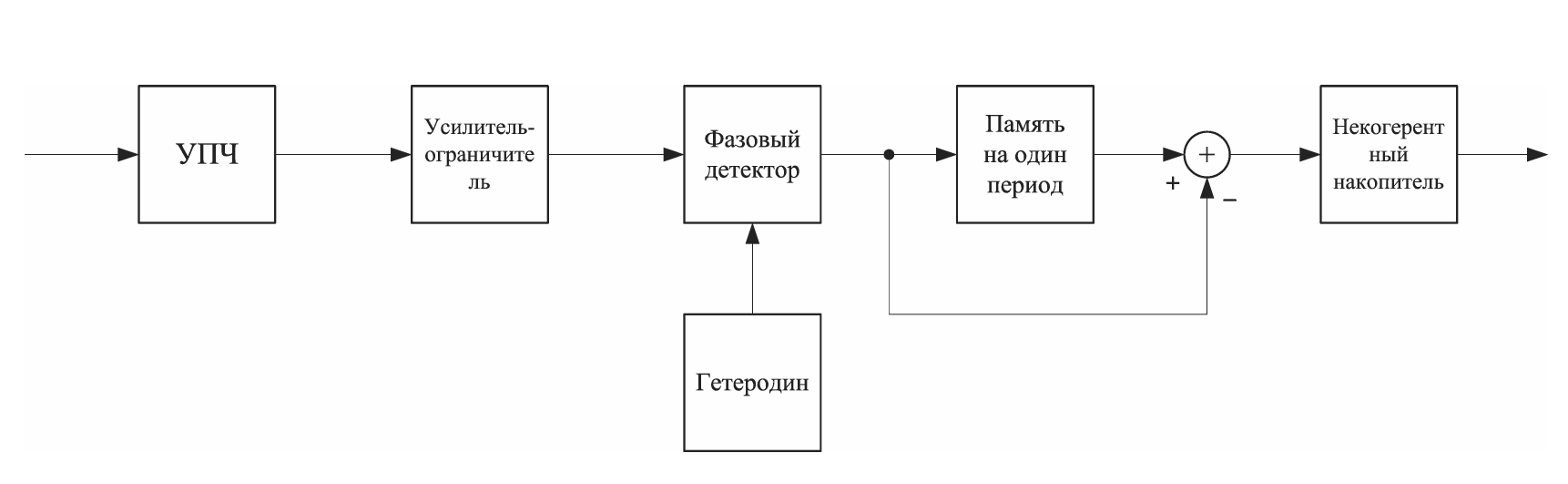} \vspace{-1cm}
\caption{Структурная схема системы раннего варианта СДЦ.}
\label{fig:MTI_old}
\end{figure*}

Структурная схема системы СДЦ, применявшейся на ранних этапах
развития радиолокации показана на Рис. \ref{fig:MTI_old}
\cite{Lar06}. Из--за применения некогерентных генераторов в такой
схеме используется некогерентная обработка и ограничитель после УПЧ
для обеспечения постоянного уровня ложных тревог.

\begin{figure*}[h]
\centering
\includegraphics[width=\columnwidth]{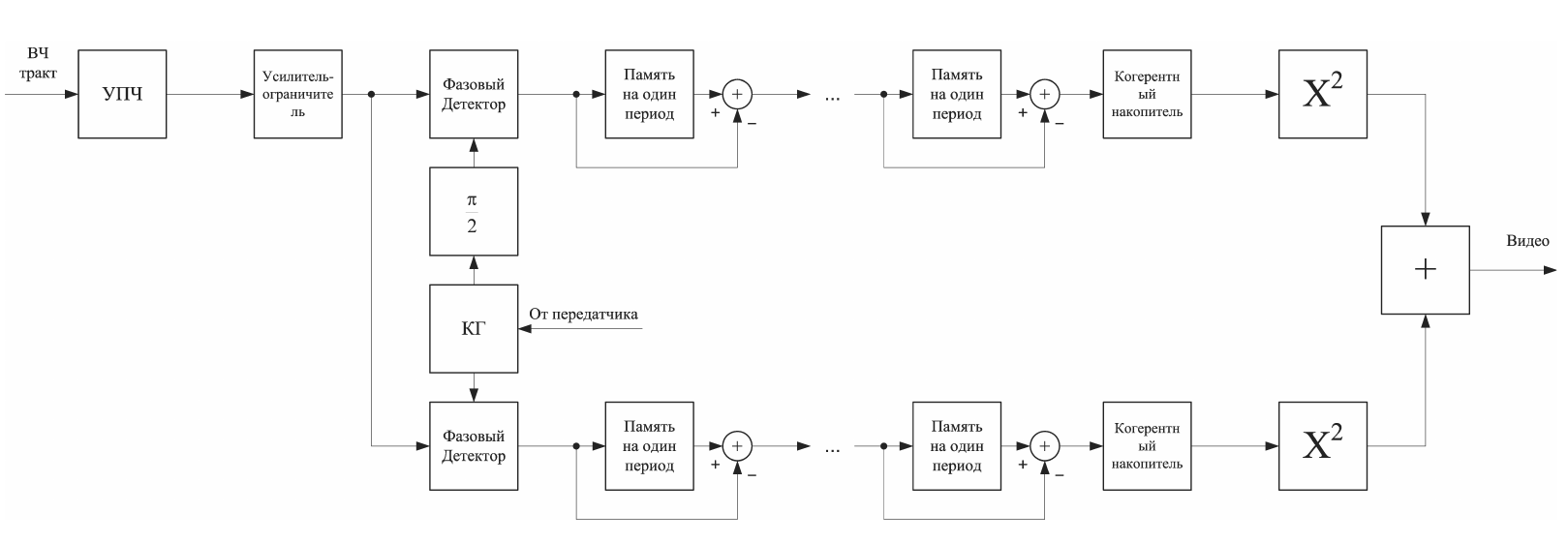} \vspace{-1cm}
\caption{Структурная схема когерентной аналоговой системы СДЦ.
КГ~--- когерентный гетеродин} \label{fig:MTI_2}
\end{figure*}

Внедрение СВЧ усилителей в практику радиолокации позволило перейти к
когерентной обработке в системе СДЦ. Появление более совершенных
аналоговых линий задержки и уменьшение погрешностей аппаратуры РЛС
позволило применять несколько схем ЧПВ, включенных последовательно
для увеличения степени подавления помех. Этот этап развития
архитектуры системы СДЦ может быть охарактеризован структурной
схемой на Рис. \ref{fig:MTI_2}.

Переход на цифровую обработку сигналов позволил реализовать
когерентный накопитель в виде гребенки фильтров БПФ. Развитие
технологии производства цифровой памяти и расширение динамического
диапазона РЛС позволило отказаться от усилителя-ограничителя  в
качестве устройства, обеспечивающего ПУЛТ. Было выяснено, что из-за
расширения спектра помехи в результате нелинейного преобразования,
ожидаемого улучшения работы системы СДЦ с ограничением сигнала не
происходит~\cite{Bak04}.

\begin{figure*}[h]
\centering
\includegraphics[width=\columnwidth]{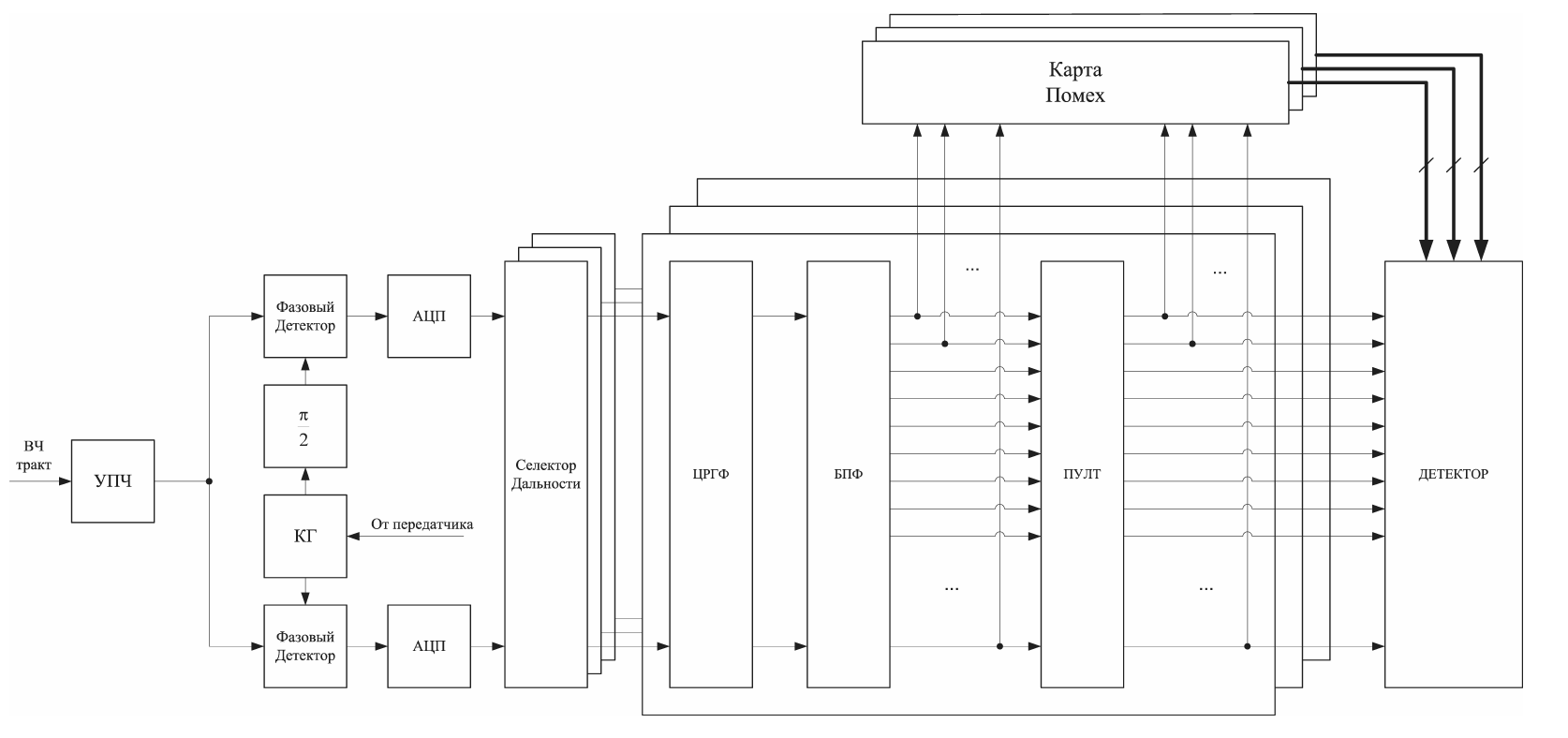} \vspace{-1cm}
\caption{Структурная схема цифровой системы СДЦ. ЦРГФ~--- цифровой
режекторный гребенчатый фильтр, АЦП~--- аналого--цифровой
преобразователь} \label{fig:MTI_3}
\end{figure*}

Поэтому для обеспечения ПУЛТ теперь применяется адаптивный порог в
обнаружителе, следующем за БПФ. Вычисление адаптивного порога
происходит по усредненному по соседним элементам разрешения значению
мощности помехи (именно этот тип обработки теперь принято называть
«ПУЛТ», а подобные устройства «ПУЛТ-процессорами» \cite{Bak86}).
Однако само по себе устройство ПУЛТ может работать только при
стационарном распределении мощности помехи в элементах разрешения.
Для обеспечения адекватной работы обнаружителя в условиях
нестационарной в пространстве помехи (например, на границах
помеховой зоны), в процессе установки адаптивных порогов
обнаружителя используется также усредненное во времени значение
мощности помехи в каждом из элементов разрешения по  частоте и
пространству. Такой принцип обработки принято называть обработкой с
использованием карт помехи. Приведенные рассуждения
проиллюстрированы на Рис. \ref{fig:MTI_3} \cite{Sko90}. Результаты
эксплуатации и испытаний РЛС с системой СДЦ указанного вида выявили
необходимость в применении более сложных, адаптивных, систем
обработки сигналов в системе СДЦ. Так как даже применение даже,
казалось бы, такой совершенной системы на практике далеко не всегда
позволяет решить проблему обнаружения слабого сигнала цели на фоне
мощных нестационарных помех \cite{War96}. Дальнейшее
усовершенствование аппаратуры ЦОС и воплощение принципов адаптивной
обработки сигналов привело к появлению нового поколения РЛС ---
адаптивных РЛС.

\begin{figure*}[tbp]
\centering
\includegraphics[width=\columnwidth]{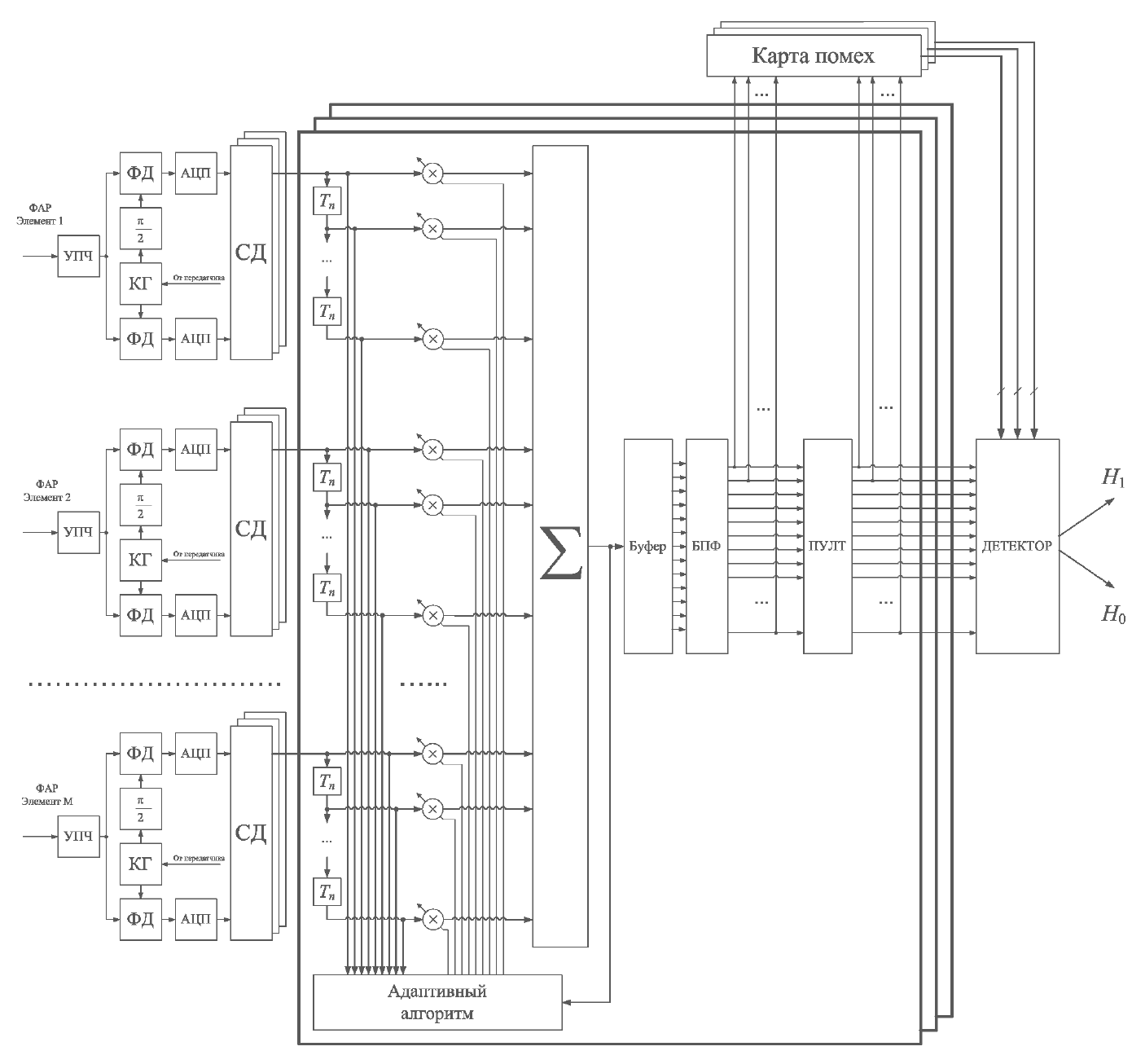} \vspace{-1cm}
\caption{Структурная схема цифровой системы СДЦ с STAP.  ФД~---
фазовый детектор, СД~--- селектор дальности, $T_\text{п}$~---
задержка на период повторения} \label{fig:MTI_stap}
\end{figure*}

В настоящее время большая часть исследований, направленных на
усовершенствование алгоритмов первичной ЦОС в РЛС ведется именно в
области адаптации РЛС к изменяющимся и априори неизвестным условиям
внешней среды. Это связано с упомянутыми выше особенностями,
характеризующими поведение помех в пространстве и времени. Следует
отметить, что применение ФАР в РЛС является неотъемлемым условием
развития в направлении адаптации РЛС к помеховой обстановке.
Применение ФАР в РЛС делает возможным осуществление так называемой
пространственно-временной адаптивной обработки, которая в зарубежной
литературе имеет акроним STAP, а у нас иногда еще называется ПДО или
ПВАО. Применение данного вида обработки в РЛС позволяет производить
одновременную адаптацию по пространственному и частотному
распределению помехи в пределах одного адаптивного алгоритма
\cite{Kle99}. Еще более общим и многообещающим подходом к адаптивной
обработке сигналов РЛС является использование в пределах одного
адаптивного алгоритма также и информации о поляризации \cite{Par06}.
Такая обработка называется
поляризационно--пространственно--временной \cite{Par06}.

Применение совместной ПВАО приводит к значительному улучшению
характеристик системы подавления помех по сравнению со случаем
раздельной пространственной и доплеровской обработки \cite{Sav00}.
Структурная схема РЛС с применением ФАР и ПВАО показана на Рис.
\ref{fig:MTI_stap} \cite{Lac01}. Заметим, что структурная схема,
показанная на Рис. \ref{fig:MTI_stap} является довольно общей. В
зависимости от конечной цели, преследуемой при разработке системы
подавления помех и конфигурации РЛС, в которой предполагается
использование системы СДЦ, ее архитектура может изменяться.
Например, совместная  система ПВАО может быть расчленена на
раздельную пространственную и доплеровскую обработку \cite{Su92}.
Вместо ФАР может быть использована система из нескольких антенн с
механическим (электронным) сканированием: «сигнальной» и
«помеховых», тогда подстройка весов адаптивного алгоритма
осуществляется с тем, чтобы скомпенсировать сигнал помехи в
сигнальной антенне сигналом в помеховых антеннах
\cite{Ger94b,Ger95}. Однако общим в современной концепции СДЦ при
любой ее архитектуре является адаптивный подход к подавлению помех.
То есть, оценки помеховой обстановки, такие как количество помех,
средние частоты мод спектра помехи, мощности каждой из помеховых
составляющих и ширина их спектра, полученные в реальном времени
используются для расчета коэффициентов адаптивного фильтра
подавления помехи. Расчет коэффициентов адаптивного фильтра
производится исходя из некоторого критерия качества работы системы
СДЦ, например минимизации мощности помехи или максимизации отношения
мощности сигнала к мощности помехи на выходе системы СДЦ.

\section{Заключение} \label{chap:sec:Ch1:concl} 

В данной главе рассмотрены основные принципы построения систем
обработки радиолокационной информации на фоне помех. Прослежено
историческое развитие таких систем и приведены типовые структурные
схемы систем обработки. Приведены аргументы, показывающие
необходимость применения адаптивных систем обработки в реальной
ситуации, когда спектральные свойства помехи могут быть неизвестными
или неопределенными.

\chapter{Адаптивные алгоритмы СДЦ} \label{chap:Ch2}

\section{Критерии эффективности систем СДЦ} \label{sec:MTI_effect}

Существуют различные параметры, с помощью которых осуществляется
оценка качества работы систем СДЦ \cite{Bak86,Bak04}. К ним прежде
всего относятся такие параметры, как коэффициент подавления помехи
($k_\text{п}$); коэффициент подпомеховой видимости ($k_\text{пв}$),
коэффициент улучшения ($k_\text{у}$); и наблюдаемость сигналов
движущихся целей ($q_\text{н}$). Коэффициент подавления помех
является простейшим критерием, определяющим качество подавления
помех:
\begin{equation} \label{eqn:Kp}
k_\text{п} = \frac{P_\text{п\ вх}}{P_\text{п\ вых}}
\end{equation}
Этот параметр не является объективным критерием оценки качества
работы системы СДЦ т.к. не учитывает прохождение полезного сигнала
через фильтр режекции помехи и влияние собственных шумов приемника.

{\it Коэффициент улучшения} отношения сигнала и помехи учитывает
ослабление или усиление полезного сигнала в устройстве подавления:
\begin{equation} \label{eqn:Ku}
k_\text{у} = \frac{\left(P_\text{c}/P_\text{п}
\right)_\text{вых}}{\left(P_\text{c}/P_\text{п} \right)_\text{вх}} =
\left[\frac{P_\text{c\ вых}}{P_\text{c\
вх}}\right]\left[\frac{P_\text{п\ вх}}{P_\text{п\ вых}}\right] =
q_\text{с}k_\text{п},
\end{equation}
где $q_\text{с} = P_\text{c\ вых}/P_\text{c\ вх}$ --- отношение
мощностей сигналов на выходе и входе.

Коэффициент подпомеховой видимости показывает во сколько раз при
заданных вероятностях правильного обнаружения $D$ и ложной тревоги
$F$ средняя мощность сигнала цели $\overline{P}_\text{c\ вх}$ может
быть меньше средней мощности пассивной помехи $\overline{P}_\text{п\
вх}$:
\begin{equation} \label{eqn:Kpv}
k_\text{пв} = \frac{P_\text{п\ вых}}{\overline{P}_\text{п\ вх}}
\left[\frac{P_\text{п}}{\overline{P}_\text{c}}\right]_\text{вых}
\left[\frac{\overline{P}_\text{c}}{P_\text{п}}\right]_\text{вых}
\left[\frac{\overline{P}_\text{c}}{P_\text{п}}\right]_\text{вх}^{-1}
= \frac{k_\text{у}}{q_\text{пор}},
\end{equation}
где $q_\text{пор}$ --- пороговое отношение сигнала к помехе по
мощности на входе порогового устройства, обеспечивающее обнаружение
с заданными вероятностями $D$ и $F$.

Наблюдаемость сигналов движущихся целей характеризуется отношением
мощности сигнала движущейся цели $P_\text{дц}$ к сумме мощностей
помех $P_\text{п}$ на выходе устройства СДЦ:
\begin{equation} \label{eqn:Qn}
q_\text{н} = \frac{P_\text{дц}}{P_\text{п}} =
\frac{P_\text{дц}}{\sum_{i=1}^N P_{\text{п}i}},
\end{equation}
где $P_{\text{п}i}$ - мощность $i$-й составляющей помехи.

\section{Классификация алгоритмов адаптивной фильтрации} \label{sec:MTI_class}

\begin{figure*}[tbp]
\centering
\includegraphics[width=\columnwidth]{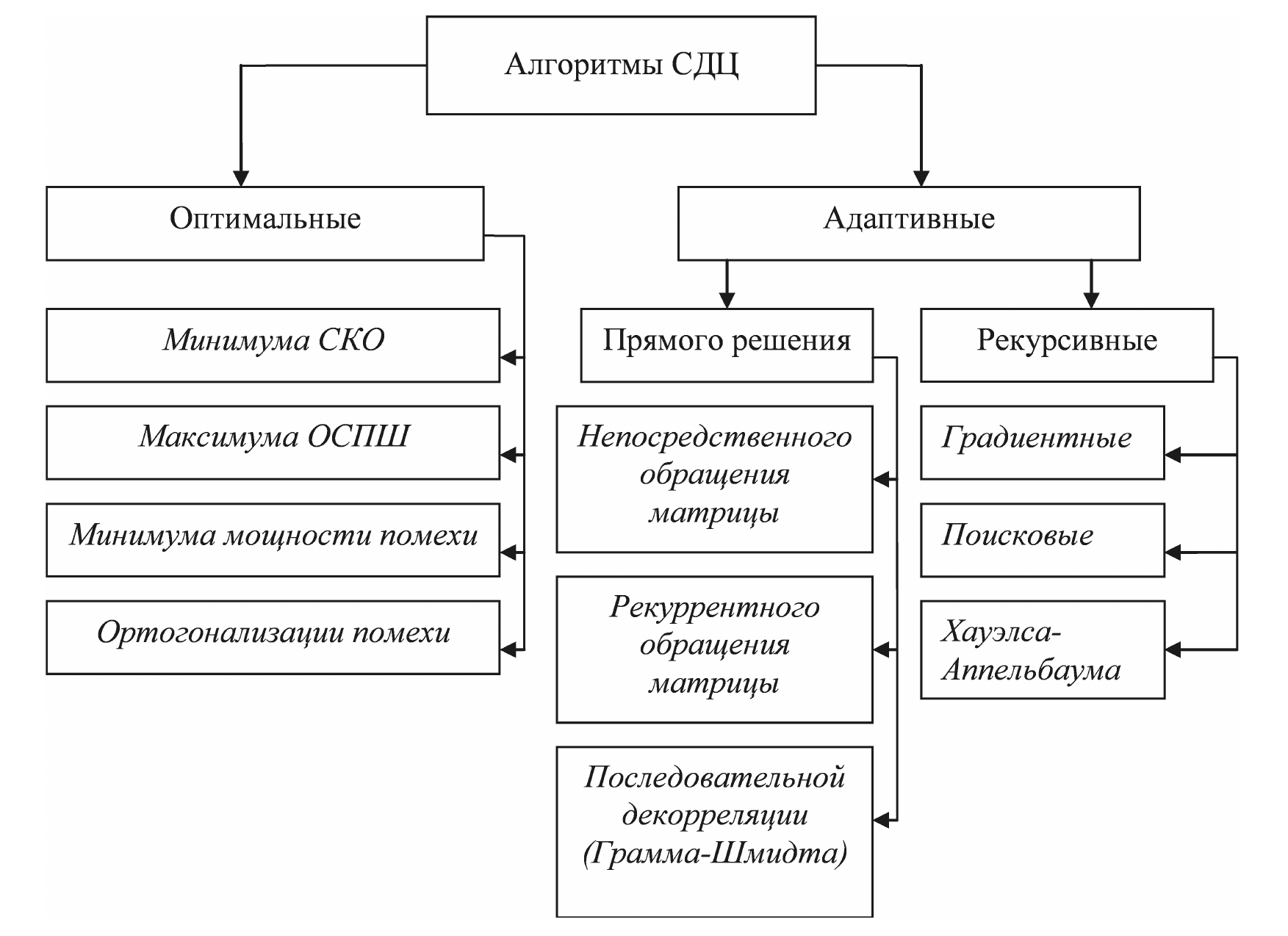} \vspace{-1cm}
\caption{Классификация алгоритмов адаптивной СДЦ}
\label{fig:Adapt_class}
\end{figure*}

Под алгоритмом адаптивной СДЦ понимается, прежде всего,
соответствующая процедура вычисления коэффициентов фильтра режекции
помехи. Адаптивные алгоритмы принято классифицировать по
применяемому при поиске оптимального вектора весовых коэффициентов
фильтра режекции помехи критерию качества или по структуре и
принципу организации вычислений в алгоритме. Соответствующая
классификация известных алгоритмов, применяемых при построении
адаптивных систем СДЦ, приведена на Рис. \ref{fig:Adapt_class}.
Выделяют следующие основные критерии качества работы адаптивных
алгоритмов:

\begin{enumerate}
\item МСКО между некоторым
эталонным сигналом  и выходным сигналом адаптивного алгоритма,
используемым для компенсации помехи;
\item Максимум ОСПШ на выходе
фильтра АСДЦ;
\item Минимума мощности помехи на выходе устройства
АСДЦ.
\end{enumerate}

Следует отметить, что критерии эффективности системы СДЦ, указанные
в разделе \ref{sec:MTI_effect} обычно используются для оценки
качества работы системы СДЦ.
В то же самое время критерии 1--3 могут быть непосредственно
использованны для синтеза алгоритмов СДЦ. При этом критерий 1
подразумевает наличие доступа к некоторому пилот--сигналу, который
может быть использован для настройки коэффициентов адаптивного
фильтра. Поэтому данный критерий чаще всего используется в связи,
где такой пилот--сигнал часто бывает доступен. Критерий 3 напрямую
соотносится с максимизацией коэффициента подавления помех
(\ref{eqn:Kp}). Как указывалась выше, данный коэффициент не
учитывает влияния фильтра СДЦ на мощность сигнала, поэтому этот
критерий в настоящее время редко используется для синтеза алгоритмов
АСДЦ. В свою очередь, как видно из (\ref{eqn:Ku}) и (\ref{eqn:Kpv})
критерий 2 является составной частью коэффициентов улучшения и
подпомеховой видимости. Данные коэффициенты наиболее объективно
оценивают степень улучшения характеристик РЛС, получаемую при
использовании того или иного алгоритма АСДЦ. Поэтому критерий
максимизации ОСПШ наиболее часто применяется при синтезе алгоритмов
АСДЦ.

\section{Обзор существующих алгоритмов адаптивной фильтрации} \label{sec:MTI_review}

Рассмотрим методы синтеза алгоритмов современной адаптивной системы
СДЦ (АСДЦ). Будем полагать, что при гипотезе $H_1$ о наличии
полезного сигнала, сигнал на входе АСДЦ представляет собой
аддитивную смесь сигнала, помехи, и белого шума приемника:
\begin{equation} \label{eqn:x_t_1}
H_1 : x(t) = s(t) + \upzeta(t) + \upxi(t),
\end{equation}
где $s(t)$~--- сигнал, $\upzeta(t)$~--- помеха, $\upxi(t)$~--- белый
Гауссовский шум с нулевым математическим ожиданием и дисперсией
$\upsigma^2_{\text{ш}}$ (далее просто шум). Когда же справедлива
гипотеза $H_0$ об отсутствии полезного сигнала, принимаемый сигнал
состоит из суммы шума и помехи, независимых друг от друга:
\begin{equation} \label{eqn:x_t_0}
H_0 : x(t) = \upzeta(t) + \upxi(t),
\end{equation}

\begin{figure*}[tbp]
\centering
\includegraphics[width=\columnwidth]{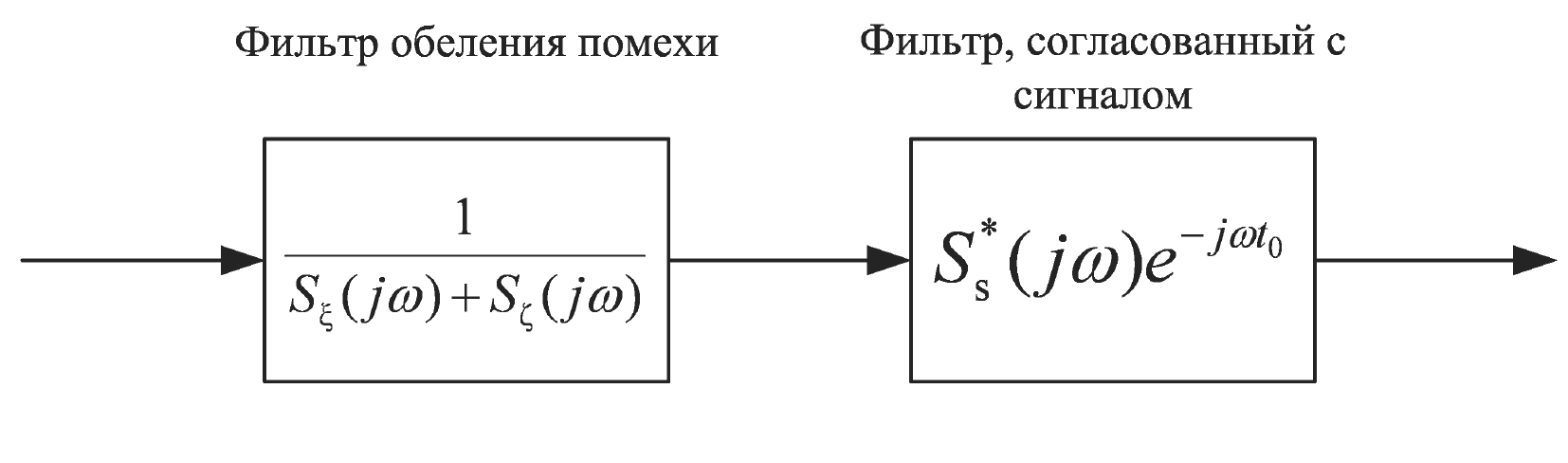} \vspace{-1cm}
\caption{Структурная схема оптимальной обработки сигнала в системе
СДЦ} \label{fig:Opt_MTI}
\end{figure*}

Из теории обнаружения и оценивания сигналов известно
\cite{Lev66,Lev68,Tre01a,Tre01b,Bak04}, что при заданных спектрах
$S_{s}(j\omega)$, $S_{\upzeta}(j\omega)$ и $S_{\upxi}(j\omega)$
сигнала, помехи и шума оптимальная обработка сигнала $x(t)$ состоит
в пропускании его через фильтр с частотной характеристикой вида:
\begin{equation} \label{eqn:K_go}
K(j\omega) = \frac{S_s^{*}(j\omega) e^{-j\omega
t_0}}{S_{\upzeta}(j\omega) + S_{\upxi}(j\omega)},
\end{equation}
где $S_s^{*}(j\omega)$ --- спектр комплексно сопряженный со спектром
сигнала, $t_0$ --- время прихода сигнала. Известная интерпретация
этого выражения заключается в представлении системы СДЦ каскадным
соединением обеляющего фильтра помехи и оптимального фильтра,
согласованного с сигналом \cite{Bak04,Sos92,Tih82}
Рис.~\ref{fig:Opt_MTI}.

С точки зрения корреляционной теории, обнаружение
известного сигнала $s(t)$ на фоне нормального коррелированного
случайного процесса с известной корреляционной функцией
$R_{\upxi+\upzeta}(y-t)$, удовлетворяющей интегральному уравнению:
\begin{equation} \label{eqn:int_eq}
\upvarphi_k(t) = \uplambda_k \int\limits_{-T}^T
R_{\upxi+\upzeta}(y-t) \upvarphi_k(y) dy, \ \ \ \ |t| \leq T,
\end{equation}
где $\uplambda_k$ и $\upvarphi_k(t)$ --- собственные числа и
собственные функции этого уравнения, состоит (в дискретном случае) в
формировании отношения правдоподобия \cite{Lev66}:
\begin{equation} \label{eqn:lik_rat}
\Lambda(\x) = \frac{p_{\x|H}(\x | H_1)}{p_{\x|H}(\x | H_0)}
\end{equation}
и соответствующей решающей статистики:
\begin{equation} \label{eqn:suf_stat}
\ln l(\x) = \s^H \R^{-1} \x.
\end{equation}
Здесь $\x = [x_1, x_2, \ldots, x_N]^T$~--- вектор выборочных
значений входного сигнала, $\s = [s_1, s_2, \ldots, s_N]^T$~---
вектор выборочных значений опорного сигнала, $\R^{-1}$~--- матрица,
обратная корреляционной матрице помехи, $N$~--- количество отсчетов
в анализируемой выборке. В случае применения совместной ПВАО,
размерность входного вектора сигнала системы СДЦ равна $N = K \times
L$, произвелению размерности антенной решетки $K$ и количества
импульсов в пачке $L$.

Выражение, называемое иногда решением интегрального уравнения
Винера--Хопфа для дискретного случая \cite{Hay96}:
\begin{equation} \label{eqn:win_hopf}
\w_\text{опт} = \R^{-1} \s,
\end{equation}
в практике адаптивной обработки сигналов принято ассоциировать с
вектором весов оптимального фильтра, максимизирующего ОСПШ. В этом
свете правило выбора решения, вытекающее из (\ref{eqn:suf_stat})
формулируется следующим образом \cite{Lev66}:
\begin{equation} \label{eqn:dec_rule}
\w_\text{опт}^H \s \stack{\gtrless}{H_1}{H_0} \upeta,
\end{equation}
где $\upeta$ --- порог, определяемый в соответствии с одним из
критериев качества обнаружения \cite{Poo94}. Стоит отметить, что в
независимости от вида ПРВ смеси помехи и шума, фильтр,
максимизирующий ОСПШ на его выходе однозначно определяется
ковариационной матрицей смеси помехи и шума $\R^{-1}$. Однако
максимизация ОСПШ еще не является гарантией хороших характеристик
обнаружения \cite{Far87}.

Статистика (\ref{eqn:suf_stat}) лежит в основе как многих
оптимальных, так и адаптивных алгоритмов подавления помех в РЛС с
СДЦ. Каждый из элементов этой процедуры можно найти на всех
рассмотренных выше структурных схемах систем СДЦ. Принципиальное
различие между рассмотренными оптимальными (субоптимальными) и
адаптивными структурными схемами систем СДЦ состоит в принципе,
руководящем расчетом значений вектора весовых коэффициентов $\w$
обнаружителя. В случае оптимальной (субоптимальной) обработки
коэффициенты $\w$ расчитываются заранее исходя из некоторой модели
сигнала и помех, а также критерия оптимальности системы. Однако на
практике, из-за недостатка априорной информации о статистиках
помехи, вместо оптимального или субоптимального фиксированного
фильтра с весами (\ref{eqn:win_hopf}) часто используются адаптивные
фильтры с весами, расчитываемыми непосредственно по входным данным.
При этом применительно к задаче АСДЦ считается, что несмотря на
нестационарность пространственно--временного распределения помехи,
имеется окружающий тестируемый элемент разрешения объем, в котором
помеха локально однородна \cite{Arm95}, а полезный сигнал от цели
отсутствует (справедлива гипотеза $H_0$). Это означает, что выборки
отражений из этого квази--однородного объема могут быть организованы
в матрицу обучающего сигнала $\X = [\x_1, \x_2, \ldots, \x_M]$ и
использованы для обучения адаптивного алгоритма обнаружения сигнала
на фоне шума с неизвестной мощностью и спектральными свойствами.
Существует несколько способов использования такой обучающей выборки
для преодоления подобной априорной неопределенности в алгоритмах
обнаружения \cite{Lev76,Poo94,Tre01b}. Одним из самых простых и
эффективных таких способов является формирование обобщенного
отношения правдоподобия \cite{Poo94,Gin02a,Gin02b,Gin95}, которое в
случае неизвестной ковариационной матрицы помехи принимает следующий
вид \cite{Mai05}:
\begin{equation} \label{eqn:glr}
\Lambda^{\prime}(\x) = \frac{\max\limits_{\widehat \R} p_{\x|H}(\x |
\R, H_1)}{\max\limits_{\widehat \R} p_{\x|H}(\x | \R, H_0)}.
\end{equation}
Откуда следует, что $\widehat \R$ является ОМП ковариационной матрицы смеси шума и помехи $\R$.
Учитывая допущения, сделанные выше о свойствах сигнала и шума
применительно к задаче АСДЦ, ОМП ковариационной матрицы помехи имеет
вид \cite{Kel85,Kel87}:
\begin{equation} \label{eqn:ML_R}
\widehat \R = \frac{1}{M} \X \X^H,
\end{equation}
Обсуждение альтернативных методов оценивания $\R$, основанных на
нормализации векторов обучающей выборки, использовании априорной
информации о характере неоднородности помехи, селекции обучающих
выборок по мощности сигнала помехи, или обработке в частотной
области может быть найдено например в
\cite{Kra07,Mai07b,McD07,Wan06,Blu06,Gur06,Ber06,Con06,Mel06,Far05,Gin95,Lin99,Rab99,Wan90}.
Cоответствующая обобщенному отношению правдоподобия (\ref{eqn:glr})
решающая статистика принимает следующую форму:
\begin{equation} \label{eqn:suf_stat_glr}
\ln l^{\prime}(\x) = \s^H \widehat \R^{-1} \x.
\end{equation}
Адаптивный детектор, основанный на (\ref{eqn:glr}) называется
обобщенным тестом отношения правдоподобия \cite{Con02}, а
соответствующий адаптивный алгоритм вычисления весов обнаружителя
\cite{Far84a,Far83}:
\begin{equation} \label{eqn:weights_okm}
\w_\text{окм} = \widehat \R^{-1} \s,
\end{equation}
носит название алгоритма непосредственного обращения корреляционной
матрицы помехи. В \cite{Kel85,Kel87} показано, что такой алгоритм
является асимптотически оптимальным, т. е. оптимальным при $M
\rightarrow \infty$. Потери, возникающие при конечном $M$ в среднем
не превышают 3 дБ, если полезный сигнал от цели отсутствует в
обучающей выборке, а $M \geq 2N$ \cite{Ree74,Kha87}. Если же
полезный сигнал присутствует в обучающей выборке, сходимость
алгоритма значительно замедляется \cite{Sto05,Car88}, а его
асимптотическая оптимальность разрушается. Кроме того, если полезный
сигнал присутствует в обучающей выборке, а мощность помехи в ней
мала, происходит подавление полезного сигнала в тестируемой выборке.
В этом случае адаптивная обработка приводит к значительным потерям в
ОСПШ \cite{Cai94}. Другой вид потерь связан с неоднородностью помехи
в элементах разрешения, используемых при оценке ковариационной
матрицы. Потери связанные с неоднородностью помехи могут
варьироваться в пределах от долей дБ до 16 дБ \cite{Mel00}. Более
того, в ситуации, когда помеха сильно неоднородна, алгоритм ОКМ
теряет свою асимптотическую оптимальность \cite{Nit90}. Одним из
эффективных способов борьбы с указанными проблемами при конечном
размере обучающей выборки $M$ является регуляризация оценки
ковариационной матрицы помехи \cite{Car88}. В простейшем случае
регуляризация оценки корреляционной матрицы состоит в добавлении
фиксированного смещения $\upalpha$ к элементам ее главной диагонали:
\begin{equation} \label{eqn:reg_R_hat}
\widehat \R^{\prime} = \widehat \R + \upalpha \I,
\end{equation}
Веса обнаружителя регуляризованного алгоритма обращения
корреляционной матрицы помехи в этом случае выражаются
следующим образом:
\begin{equation} \label{eqn:weights_rokm}
\w_\text{рокм} = \left( \widehat \R + \upalpha\I \right)^{-1} \s,
\end{equation}

Заметим, что процедура вычисления весов (\ref{eqn:weights_okm}) или
(\ref{eqn:weights_rokm}) является одной из самых вычислительно емких
операций обработки сигналов в адаптивной РЛС. В связи с этим можно
сформулировать основные задачи, решение которых необходимо при
синтезе алгоритмов АСДЦ. Во--первых, получение наиболее близкой к
оптимальной и статистически значимой оценки весов адаптивного
фильтра режекции помехи при конечном $M$. Во--вторых, минимизация
количества арифметических операции, необходимых для вычисления
значений весов адаптивного фильтра режекции помехи.

Существует множество алгоритмов, с помощью которых может быть
реализована операция вычисления весов адаптивного фильтра
(\ref{eqn:weights_okm}). В общем случае применения ПВАО с вобуляцией
периода повторения, матрица $\widehat \R$ вида:
\begin{equation} \label{eqn:cor_mtx}
\widehat \R = \left( \begin{array}{cccc}
 \hat r_{1,1} & \hat r_{1,2} & \cdots & \hat r_{1,N} \\
 \hat r_{2,1} & \hat r_{2,2} & \cdots & \hat r_{2,N} \\
 \cdots & \cdots & \cdots & \cdots \\
 \hat r_{N,1} & \hat r_{N,2} & \cdots & \hat r_{N,N} \end{array} \right).
\end{equation}
может и не быть Теплицевой~\cite{Sav00}. В этом случае процедура
вычисления весов (\ref{eqn:weights_okm}), например, методом Гаусса
занимает $\mathcal{O}(N^3)$ арифметических операций.

Для решения задачи обращения матрицы, в случае ее Теплицевости
используется быстрая процедура обращения матрицы Левинсона--Дербина
\cite{Mar90}. Эта процедура состоит в решении двух систем линейных
уравнений относительно первого:
\begin{equation} \label{eqn:lev_derb_1}
 \left[ \begin{array}{cccc}
 \hat r[0] & \hat r[-1] & \cdots & \hat r[-N+1] \\
 \hat r[1] & \hat r[0] & \cdots & \hat r[-N+2] \\
 \cdots & \cdots & \cdots & \cdots \\
 \hat r[N-1] & \hat r[N-2] & \cdots & \hat r[0] \end{array} \right]
 \left[ \begin{array}{c}
 \hat r^{-1}[1,1] \\
 \hat r^{-1}[2,1] \\
 \cdots \\
 \hat r^{-1}[N,1] \end{array} \right] =
 \left[ \begin{array}{c}
 1 \\
 0 \\
 \cdots \\
 0 \end{array} \right].
\end{equation}
и последнего столбцов обратной матрицы $\widehat \R^{-1}$:
\begin{equation} \label{eqn:lev_derb_N}
 \left[ \begin{array}{cccc}
 \hat r[0] & \hat r[-1] & \cdots & \hat r[-N+1] \\
 \hat r[1] & \hat r[0] & \cdots & \hat r[-N+2] \\
 \cdots & \cdots & \cdots & \cdots \\
 \hat r[N-1] & \hat r[N-2] & \cdots & \hat r[0] \end{array} \right]
 \left[ \begin{array}{c}
 \hat r^{-1}[1,N] \\
 \hat r^{-1}[2,N] \\
 \cdots \\
 \hat r^{-1}[N,N] \end{array} \right] =
 \left[ \begin{array}{c}
 1 \\
 0 \\
 \cdots \\
 0 \end{array} \right].
\end{equation}
где $\hat r[\cdot] = \hat r[i-j]$~--- элементы оценки корреляционной
матрицы помехи, а $\hat r^{-1}[i,j]$~--- элементы матрицы, обратной
корреляционной. Вводя подстановки:
\begin{eqnarray} \label{eqn:lev_derb_sub}
\a_N[k] &=& \frac{\hat r^{-1}[k,0]}{\hat r^{-1}[0,0]}, \\
\b_N[k] &=& \frac{\hat r^{-1}[N-k,N]}{\hat r^{-1}[N,N]}, \\
\uprho_N[k] &=& \frac{1}{\hat r^{-1}[0,0]} = \frac{1}{\hat
r^{-1}[N,N]},
\end{eqnarray}
где $k = 1 \ldots N$, можно получить следующую рекурсию ($l = 1
\ldots N$~--- номер шага рекурсии):
\begin{eqnarray} \label{eqn:lev_derb_rec}
\triangle_l &=& \sum\limits_{k=1}^l \hat r^{-1}[k] \a_{l-1}[l-k],\ \ \a_{l} = \left[\a_{l}[1], \ldots, \a_{l}[l]\right]^T,\ \ \a_{l}[0] = 1; \ \ \  \\
\bigtriangledown_l &=& \sum\limits_{k=1}^l \hat r^{-1}[-k] \b_{l-1}[l-k],\ \b_{l} = \left[\b_{l}[1], \ldots, \b_{l}[l]\right]^T, \ \b_{l}[0] = 1; \ \ \  \\
\a_{l}[l] &=& -\frac{\triangle_l}{\uprho_{l-1}}, \ \b_{l}[l] =
-\frac{\bigtriangledown_l}{\uprho_{l-1}},\ \  \uprho_{l} = \uprho_{l-1} - \frac{\triangle_l \bigtriangledown_l}{\uprho_{l-1}};\\
\a_{l}[k] &=& \a_{l-1}[k] + \a_{l}[l] \b_{l-1}[l-k], k = 1 \ldots l-1;\\
\b_{l}[k] &=& \b_{l-1}[k] + \b_{l}[l] \a_{l-1}[l-k], k = 1 \ldots
l-1.
\end{eqnarray}
Результатом выполнения рекурсии являются первый $\a$ и последний
$\b$ столбцы корреляционной матрицы помехи. Для вычисления остальных
элементов этой матрицы используется свойство персимметричности
матрицы, обратной Теплицевой. Выражение для этих элементов имеет
вид:
\begin{equation} \label{eqn:lev_derb_rest_elem}
\hat r_m^{-1}[j+1, k+1] = \hat r_m^{-1}[j, k] +
\frac{1}{\uprho_m}\left( \a_m[j+1] \b_m[k+1] - \a_m[N-j] \b_m[N-k]
\right),
\end{equation}
где $j \geq 0,\ \ k \leq N-1$. Данная процедура позволяет сократить
требуемое при обращении матрицы количество арифметических операций
до уровня $\mathcal{O}(N^2)$. Однако ее применение возможно только в
ситуации, когда корреляционная матрица помехи обладает свойством
Теплицевости. Существует ряд методов
\cite{Hun83,Zat97,Kim98,Ayo99,Hai91,Hai97,Kir94,Gol97a,Gol97b,Gol97c,Cha94,Jen93,Par96,Tuf82,Sto05},
позволяющих уменьшить количество требуемых вычислений при обращении
матриц общего вида. Основная идея, на которой основаны эти методы
состоит в том, что согласно \cite{Bre73}, оценка ковариационной
матрицы помехи имеет неполный ранг, который либо не превышает
количества разрешаемых источников помех (при условии, что эти
источники достаточно узкополосны), либо количества обучающих
выборок $M$ (при условии $M \leq N$).

В первой группе таких методов, размерность обращаемой матрицы
уменьшается путем применения алгоритма Ханга--Тернера
\cite{Hun83,Zat97,Kim98}, который в свою очередь основан на
непосредственном применении леммы об обратной матрицы
\begin{equation} \label{eqn:hung_tur_1}
\left[ \P^{-1} + \M \Q^{-1}\M^H \right]^{-1} = \P - \P\M \left[
\M^H\P\M + \Q \right]^{-1} \M^H\P
\end{equation}
к матрице вида (\ref{eqn:reg_R_hat}). При введении обозначений
\begin{equation} \label{eqn:hung_tur_2}
\P \triangleq \frac{1}{\upalpha}\I, \ \ \Q \triangleq M\I,\ \ \M
\triangleq \X,
\end{equation}
алгоритм вычисления весов адаптивного обнаружителя Ханга--Тернера
приобретает следующую форму:
\begin{equation} \label{eqn:hung_tur_3}
\w_\text{хт} = \s - \X \left( \X^H\X + M\upalpha\I
\right)^{-1}\X^H\s
\end{equation}
Из (\ref{eqn:hung_tur_3}) видно, что применение алгоритма
Ханга--Тернера позволяет снизить вычислительные затраты на обращение
ковариационной матрицы помехи с $\mathcal{O}(N^3)$ до
$\mathcal{O}(M^3)$. Во многих случаях это может приводить к
значительной экономии вычислительных ресурсов, однако выигрыш может
быть получен только если выполняется условие $M \leq N$.
Альтернативой подходу Ханга--Тернера является применение метода
проецирования обучающей выборки в подпространство меньшей
размерности путем применения шаблонной матрицы преобразования,
обеспечивающей минимальную потерю информации, заключенной в
обучающей выборке \cite{Li06}. Подавление помехи затем
осуществляется в подпространстве меньшей размерности, что приводит к
экономии вычислительных ресурсов процессора. Однако, если алгоритм
Ханга--Тернера является точным и не ведет к потерям подавления, то
применение метода \cite{Li06} связано с одновременным снижением
вычислительных затрат и ухудшением подавления.

Другая группа методов обращения матрицы в пространстве меньшей
размерности
\cite{Ayo99,Hai91,Hai97,Kir94,Gol97a,Gol97b,Gol97c,Cha94,Jen93,Par96,Tuf82,Sto05}
использует тот факт, что только часть собственных векторов оценки
ковариационной матрицы соответствуют помеховому пространству. Другая
часть этих векторов соответствует шуму и просачивающемуся в оценку
ковариационной матрицы сигналу. Оценка ковариационной матрицы (в
случае ее Теплицевости) может быть всегда представлена в
ортонормированном базисе, составленном из ее собственных векторов
$\u_i$ полученных путем разложения по собственным числам
$\lambda_i$:
\begin{equation} \label{eqn:eig_canc_1}
\widehat\R = \sum\limits_{i=1}^{N} \lambda_i\u_i\u_i^H
\end{equation}
Полученные в результате разложения собственные векторы и собственные
числа затем используются при вычислении вектора весов лежащего вне
пространства помехи. При применении подобного метода подавления
помехи делается предположение о том, что размерность помехового
пространства $M_\text{п}$ известна и равна количеству источников
помех. Очевидно, что размерность помехового пространства заранее
известна быть не может, поэтому использование оценки этой
размерности неизбежно приводит к потерям. Кроме этого, реальные
сигналы представляют собой сравнительно широкополосные случайные
процессы, поэтому предположение о равенстве размерности помехового
пространства количеству источников помех также часто нарушается на
практике \cite{Goo07}. Однако в \cite{Hai96} было показано, что
применение подобного метода приводит не только к уменьшению
вычислительных затрат до уровня $\mathcal{O}(M_\text{п}^3)$, но и к
увеличению скорости сходимости алгоритма основанного на разложении
по собственным числам по сравнению с алгоритмом ОКМ. Большинство
статей на эту тему также подтверждают этот результат путем
моделирования. Сравнение характеристик и вычислительных затрат на
реализацию алгоритма основанного на разложении по собственным числам
с другими субоптимальными методами обращения ковариационной матрицы
помехи можно найти в \cite{Zat97,Gie02}. Вопросы оптимизации
вычислений и выбора оптимального помехового подпространства
рассмотрены в \cite{Yan07}.

Принципиально иным методом формирования решающей статистики
(\ref{eqn:suf_stat}) при обнаружении сигнала на фоне помех является
алгоритм на основе процедуры ортогонализации Грама--Шмидта
\cite{Lin86,Ger91,Ger90a,Ger90b,Hua01,Jen93,Far84b}. Структура
алгоритма следует из следующего наблюдения. Матрица $\R^{-1}$ в
(\ref{eqn:suf_stat}) может быть представлена в виде произведения
нижней и верхней треугольных матриц $\L$ и $\U$ в результате
операции известной как LU-факторизация \cite{Far83}:
\begin{equation} \label{eqn:GS_1}
\R^{-1} = \L \U
\end{equation}
Причем в случае, когда ковариационная матрица $\R$ является
квадратной и Эрмитовой нижняя и верхняя треугольные матрицы связаны
операцией Эрмитового сопряжения
\begin{equation} \label{eqn:GS_ERM}
\U = \L^H
\end{equation}
Тогда (\ref{eqn:GS_1}) принимает форму разложения Холецкого:
\begin{equation} \label{eqn:GS_2}
\R^{-1} = \L \L^H
\end{equation}
В этом случае формирование (\ref{eqn:suf_stat}) происходит следующим
образом:
\begin{equation} \label{eqn:GS_3}
\ln l(\x) = \s^H \L \L^H \x.
\end{equation}

\begin{figure*}[tbp]
\centering
\includegraphics[width=16cm]{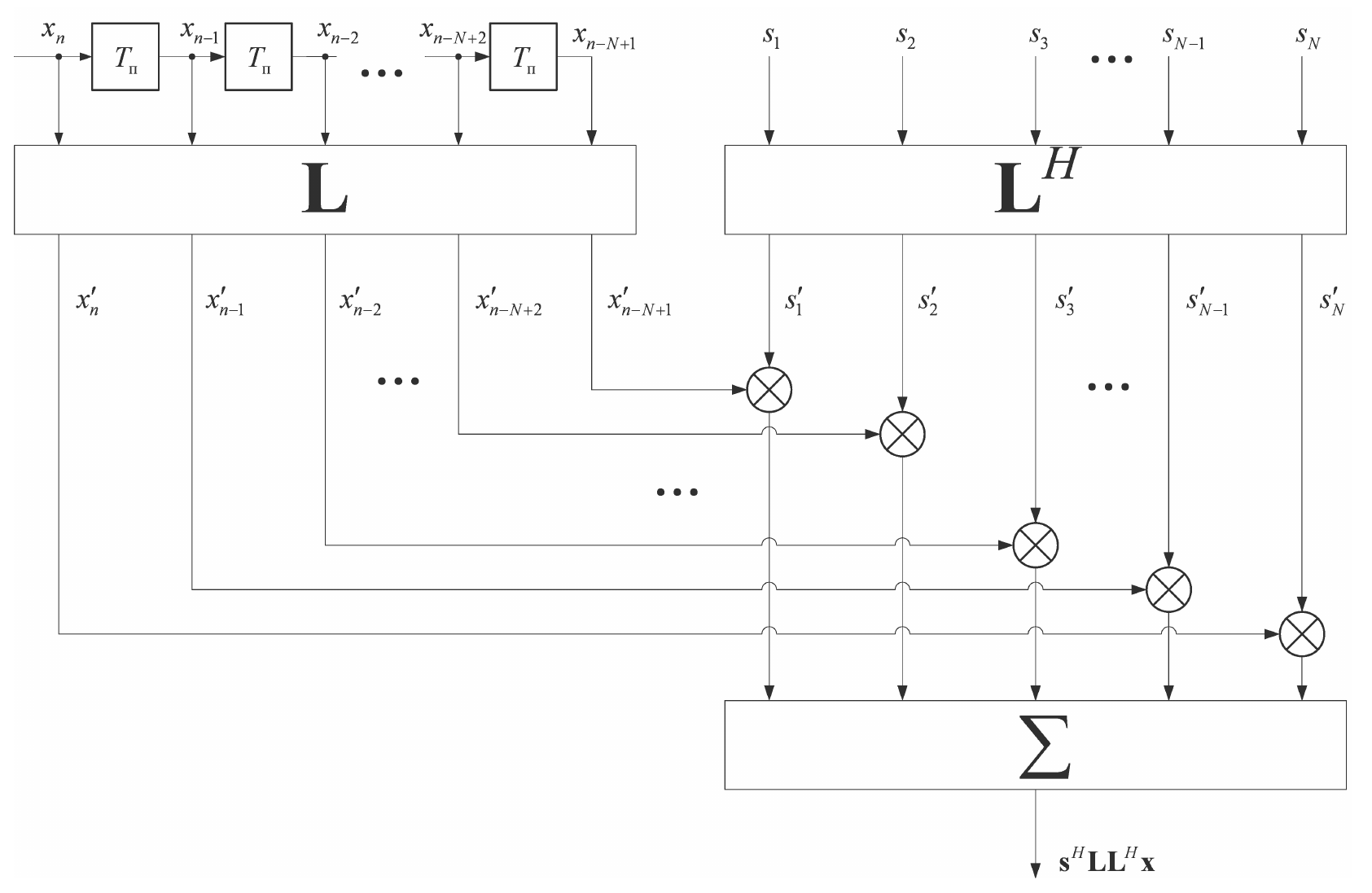}
\caption{Факторизация оптимального процессора} \label{fig:GS_farina}
\end{figure*}

Таким образом, операция вычисления обратной матрицы для формирования
отношения правдоподобия может быть заменена эквивалентной операцией
вычисления нижней треугольной матрицы $\L$ (Рис.
\ref{fig:GS_farina}). В свою очередь матрица  может быть вычислена
при помощи эффективной процедуры ортогонализации ГШ. Процедура
ортогонализации ГШ используется в теории функционального анализа для
получения ортогонального векторного базиса из изначально произвольно
заданных векторов.

Применение алгоритма ГШ к случайным процессам дает следующие
уравнения для вычисления весов (элементов матрицы $\L$) и выходных
значений обеляющего фильтра:
\begin{equation} \label{eqn:GS_4}
x^\prime_1 = \frac{x_1}{ \sqrt{\mathbb{E}\{x_1 x^{*}_1 \}}}
\end{equation}
\begin{equation} \label{eqn:GS_5}
x^\prime_i = \frac{x_i - \sum_{m=1}^{i-1}\mathbb{E}\{ x_i
x_m^{\prime*} \}}{ \sqrt{\mathbb{E}\left\{ \left|x_i -
\sum_{m=1}^{i-1}\mathbb{E}\{ x_i x_m^{\prime*}\}x_m^{\prime}
\right|^2 \right\}}}, \ \ i = 1,\ldots,N
\end{equation}

Смысл операций (\ref{eqn:GS_4}), (\ref{eqn:GS_5}) заключается в
нормализации мощности первого выходного отсчета $x_1^\prime$ и
последовательной ортогонализации каждой последующей компоненты
$x_i^\prime$ выходного вектора процедуры ГШ $\x^\prime$ относительно
всех предыдущих компонент $x_{i-1}^\prime, i = 1\ldots N$.
Вследствие этого осуществление процедуры ГШ эквивалентно операции
$\x^\prime = \L^H \x$. Где матрица $\L^H$ образует импульсную
характеристику обеляющего матричного фильтра.

\begin{figure*}[tbp]
\centering
\includegraphics[width=14cm]{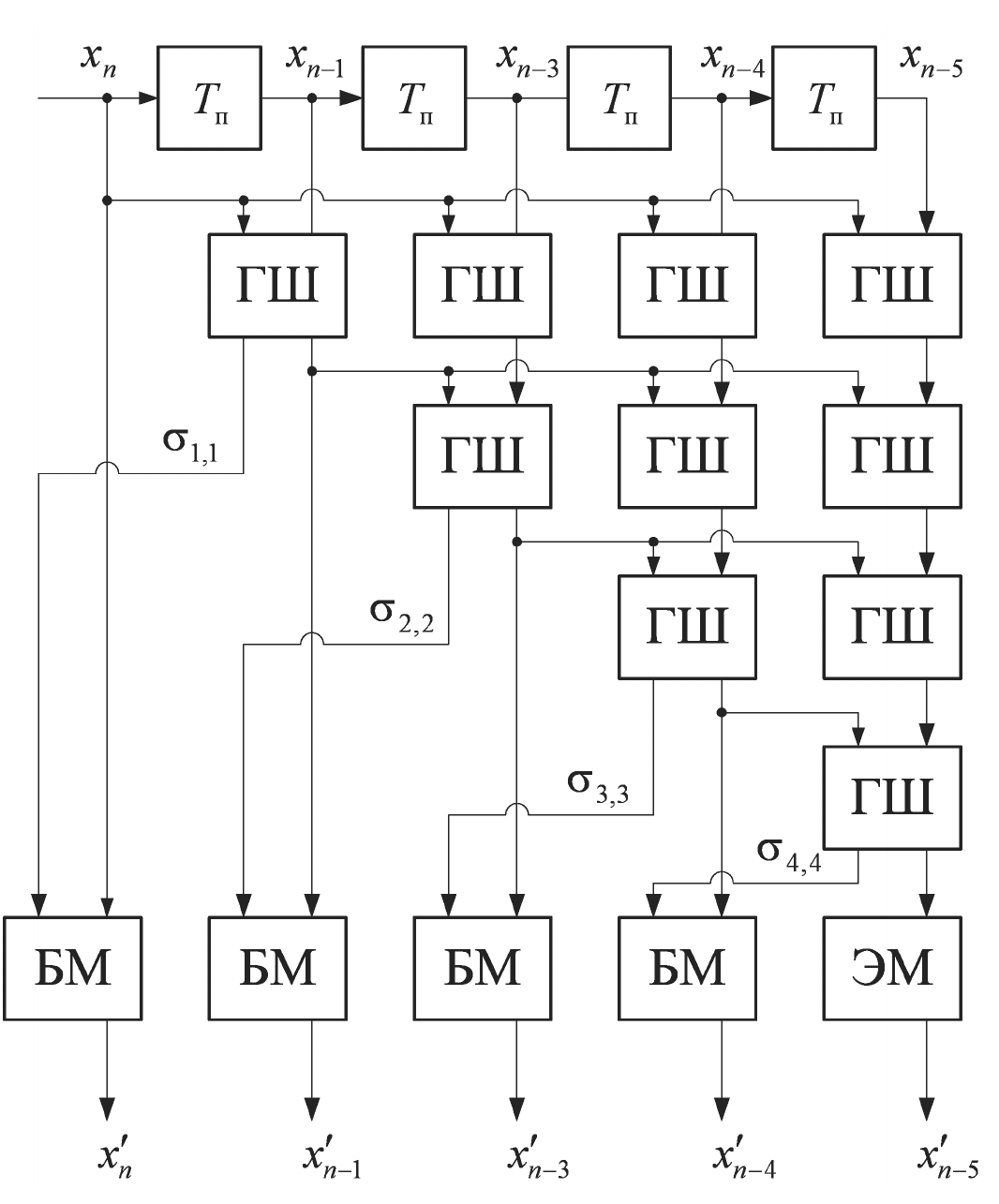}
\caption{Структурная схема реализации операции $\x^\prime = \L^H \x$
на базе процедуры ГШ. ЭМ --- эквалайзер мощности, БМ --- блок
масштабирования} \label{fig:GS_alg1}
\end{figure*}

\begin{figure*}[tbp]
\centering
\includegraphics[width=14cm]{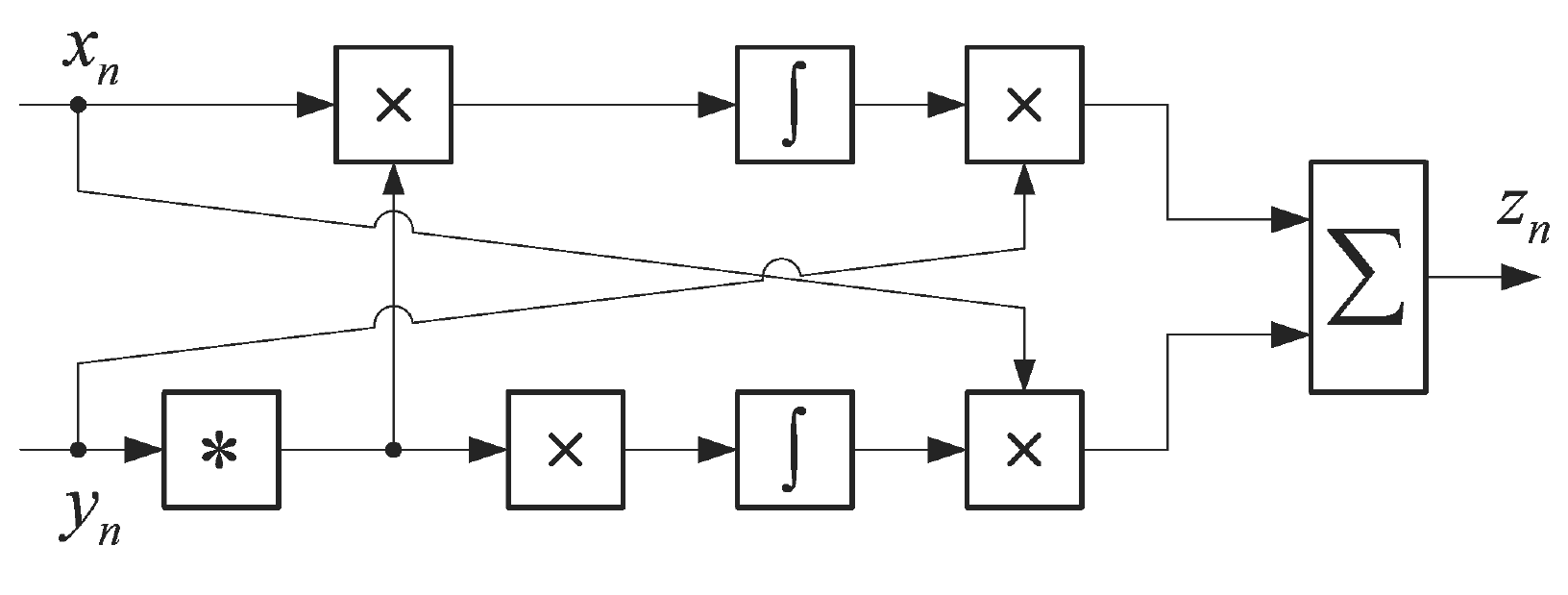}
\caption{Структурная схема блока ГШ, осуществляющего декорреляцию
сигнала $x_n$ относительно сигнала $y_n$. Блок $*$ осуществляет
комплексное сопряжение сигнала} \label{fig:GS_alg2}
\end{figure*}

Более подробно структурная схема процессора, осуществляющего
операцию   показана на Рис. \ref{fig:GS_alg1} и Рис.
\ref{fig:GS_alg2}. Структурная схема модуля ГШ, выполняющего
декорреляцию соседних отсчетов показана на Рис. \ref{fig:GS_alg2}.

Основными отличительными чертами алгоритма на основе ГШ являются
\cite{Far84b}:
\begin{enumerate}
\item алгоритм непосредственно следует из теории оптимальной обработки
на фоне помех;
\item высокая скорость адаптации;
\item модульная структура (хорошо подходит для практической реализации);
\item вычислительная сложность $\mathcal{O}(M^2)$.
\end{enumerate}

\begin{figure*}[tbp]
\centering
\includegraphics[width=\columnwidth]{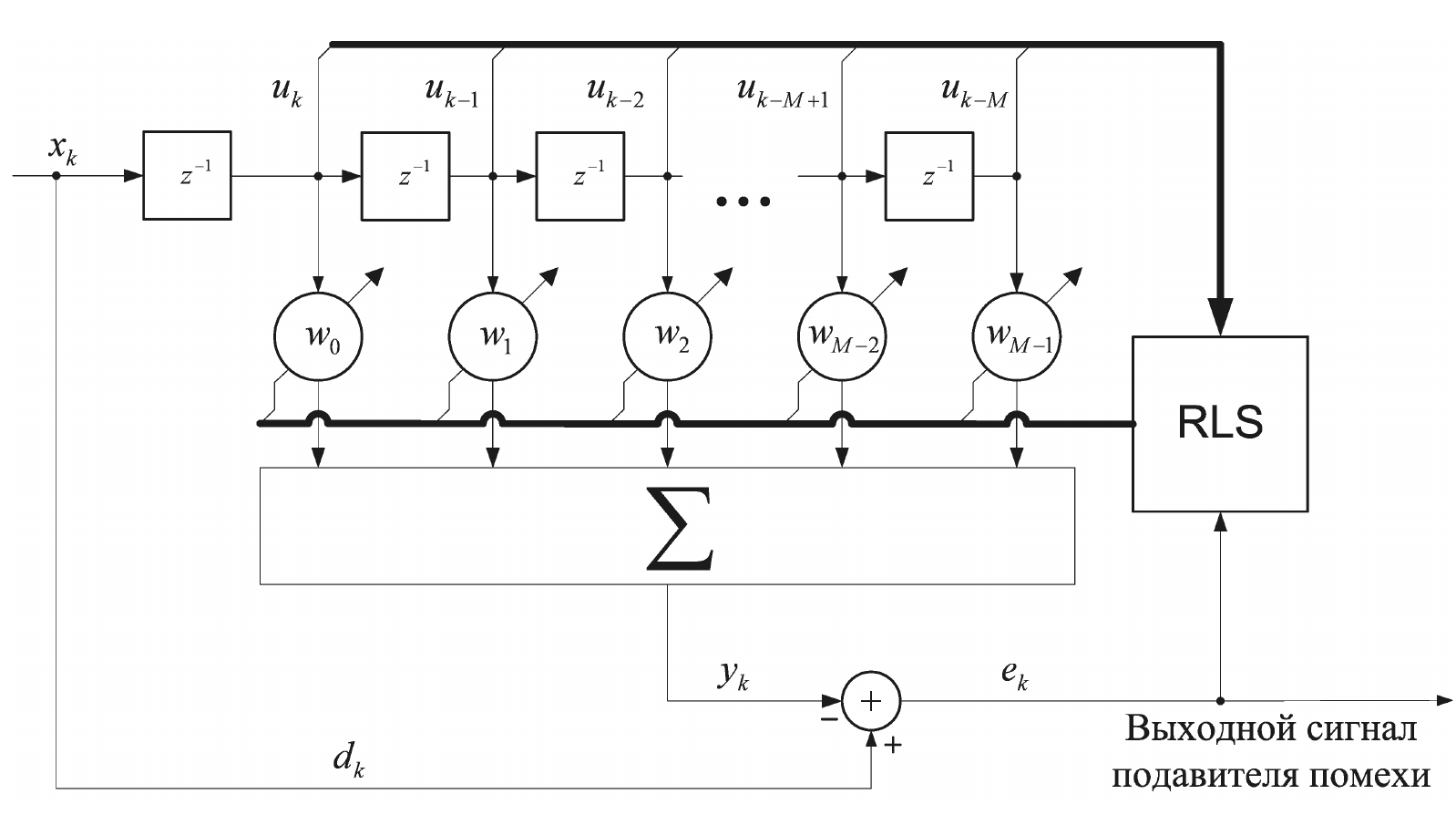}
\caption{Обеляющий фильтр на базе алгоритма RLS} \label{fig:RLS}
\end{figure*}

Рекурсивной реализацией (\ref{eqn:suf_stat}) является адаптивный
алгоритм RLS \cite{Hay96,Hay02,Leu05}. Известно также \cite{Hay97},
что этот алгоритм является частным случаем фильтра Калмана. Принцип
действия алгоритма RLS основан на применении леммы об обращении
матрицы \cite{Hay96,Hay02}. Если $\A$ и $\B$ являются положительно
определенными квадратными матрицами размерности $M \times M$,
связанными следующим соотношением:
\begin{equation} \label{eqn:RLS_1}
\A = \B^{-1} + \C\D\C^H
\end{equation}
где $\D$ - матрица размерности $N \times N$ , а $\C$ - матрица
размерности $M \times N$, то лемма об обращении матрицы может быть
записана следующим образом \cite{Hay96,Hay02}:
\begin{equation} \label{eqn:RLS_2}
\A^{-1} = \B - \B\C\left( \D + \C^H\B\C \right)^{-1}\C^H\B
\end{equation}
Применение (\ref{eqn:RLS_2}) позволяет избежать операции
непосредственного обращения корреляционной матрицы помехи, путем
замены ее на соответствующую рекурсию. Чтобы сформировать эту
рекурсию в рамках последовательного алгоритма RLS необходимо связать
переменные в (\ref{eqn:RLS_2}) с переменными в (\ref{eqn:ML_R})
следующим образом:
\begin{eqnarray} \label{eqn:RLS_3}
\A &=& \widehat\R_k,\\
\B^{-1} &=& \lambda \widehat\R_k,\\
\C &=& \x_k,\\
\D &=& 1
\end{eqnarray}
где $k$ --- номер шага рекурсии $\x_k$ --- $k$--ый столбец матрицы
обучающего сигнала $\X$, $\lambda$ --- коэффициент, определяющий
скорость реакции фильтра на новые данные (фактор памяти), а
$\widehat\R_k$ пропорциональна ОМП матрицы $\R$, использующей $k$
первых столбцов матрицы $\X$ (выборок) обучающего сигнала и имеющая
вид:
\begin{equation} \label{eqn:RLS_4}
\widehat\R_k = \sum\limits_{i=1}^k \x_i \x_i^H
\end{equation}
Подставляя (\ref{eqn:RLS_3}) в (\ref{eqn:RLS_2}) получаем следующее
выражение:
\begin{equation} \label{eqn:RLS_5}
\widehat\R_k^{-1} = \lambda^{-1}\widehat\R_{k-1}^{-1} -
\frac{\lambda^{-2}\widehat\R_{k-1}^{-1}\x_k\x^H_k\widehat\R_{k-1}^{-1}}{1
+ \lambda^{-1}\x^H_k\widehat\R_{k-1}^{-1}\x_k}
\end{equation}
Если положить для удобства:
\begin{equation} \label{eqn:RLS_6}
\P_k = \R^{-1}_k, \text{\hspace{1cm} и \hspace{1cm}} \k_k =
\frac{\lambda^{-1}\widehat\R_{k-1}^{-1}\x_k}{1 +
\lambda^{-1}\x^H_k\widehat\R_{k-1}^{-1}\x_k}
\end{equation}
где $\k_k$ известен как коэффициент усиления Калмана, то выражение
для рекурсивного вычисления обратной корреляционной матрицы помехи
можно записать следующим образом:
\begin{equation} \label{eqn:RLS_7}
\P_k = \lambda^{-1}\P_{k-1} - \lambda^{-1}\k_k
\x^H_k\widehat\R_{k-1}^{-1}
\end{equation}
Для того, чтобы получить уравнения для обновления оценки вектора
весов адаптивного алгоритма RLS необходимо использовать выражение
для оценки вектора взаимной корреляции входного и опорного сигналов:
\begin{equation} \label{eqn:RLS_8}
\widehat\p_k = \lambda\widehat\p_{k-1} + \x_k d^*_k
\end{equation}
где $\widehat \p_k$ --- оценка вектора взаимной корреляции входного
и опорного сигналов, $d_k$ --- отсчет некоторого опорного сигнала.
Известно \cite{Hay96}, что:
\begin{equation} \label{eqn:RLS_9}
\w_k = \widehat\R^{-1}_k \p_k = \lambda \P_k\p_{k-1} + \P_k\x_k
d^*_k
\end{equation}
Тогда рекурсия для весов $\w_k$ может быть записана с использованием
(\ref{eqn:RLS_9}) следующим образом:
\begin{eqnarray} \label{eqn:RLS_10}
\w_k &=& \w_{k-1} - \k_k\x^H_k\w_{k-1} + \P_k\x_kd^*_k \\
&=& \w_{k-1} + \k_k\left(d^*_k - \x^H_k\w_{k-1}\right)\\
&=& \w_{k-1} + \k_k e^*_k
\end{eqnarray}
где $e_k$ --- априорная ошибка предсказания:
\begin{equation} \label{eqn:RLS_11}
e_k = d_k - \w^H_{k-1}\x_k
\end{equation}
Структурная схема обеляющего фильтра на базе алгоритма RLS показана
на Рис. \ref{fig:RLS}.

Заметим, что параметр $\lambda \neq 1$ имеет смысл только в условиях
неоднородности сигнала помехи~\cite{Mai07a}, либо при использовании
коэффициентов фильтра одной ячейки разрешения для инициализации
алгоритма в соседней ячейке разрешения или при инициализации
алгоритма в текущем обзоре его параметрами, полученными в предыдущем
обзоре. Кроме того, было показано \cite{Hay97}, что в случае
нестационарных сигналов необходимо использовать не RLS c
экспоненциальным взвешиванием (описанный выше), а фильтр Калмана.
Однако для использования фильтра Калмана в реальной ситуации
необходимо знание переходной матрицы процесса, управляющего
параметрами помехи в пределах некоторой временной (или
пространственной) зоны \cite{Hay97}.

Основными отличительными чертами алгоритма на базе алгоритма RLS
являются \cite{Hay96,Hay97}:
\begin{enumerate}
\item быстрое время адаптации не зависящее от разброса собственных чисел оценки корреляционной
матрицы помехи;
\item полное использование информации о корреляционных свойствах
помехи;
\item вычислительная сложность $\mathcal{O}(M^2)$.
\end{enumerate}

Несмотря на то, что базовая модификация алгоритма RLS имеет
вычислительную сложность $\mathcal{O}(M^2)$, имеются ссылки на
быстрые алгоритмы RLS, имеющие сложность $\mathcal{O}(11M)$,
$\mathcal{O}(7M)$, и даже $\mathcal{O}(5M)$ \cite{Cio84}. Кроме
этого, алгоритм RLS может быть неустойчив при использовании
арифметики с фиксированной точкой. Поэтому представляет интерес
обсуждение более устойчивых к ошибкам квантования модификаций RLS,
основанных на QR--факторизации \cite{Lin91}.

\begin{figure*}[tbp]
\centering
\includegraphics[width=\columnwidth]{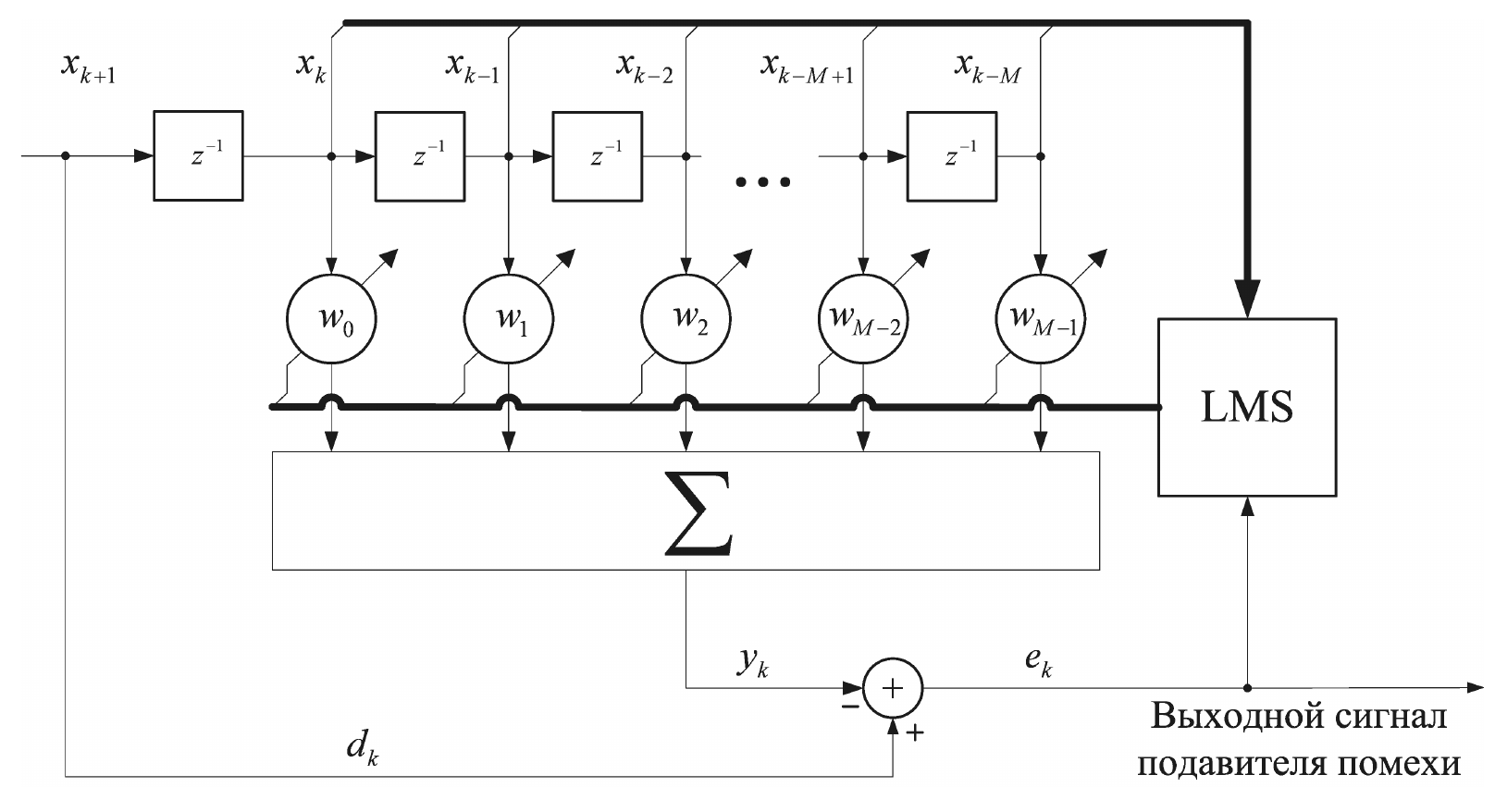}
\caption{Обеляющий фильтр на базе алгоритма LMS} \label{fig:LMS}
\end{figure*}

Другим рекурсивным алгоритмом вычисления весов адаптивного фильтра с
вычислительной сложностью $\mathcal{O}(M)$ является алгоритм LMS,
его обобщение --- алгоритм NLMS \cite{Hay96,Wer04,Jo05}, а также
алгоритмы типа наименьших средних квадратов, работающие со
статистиками более высокого порядка \cite{Hub07,Kiv06}. Имеется
большое количество публикаций, посвященных исследованию
статистических свойств этого класса алгоритмов
\cite{Wid85,Hay96,Hay02,Hub07,Kiv06,Tob06,Vic06,Are06,Zha06,God05,Rao05,Shi04,Tob04}
и вопросов их реализации \cite{Zho06,Ree81,God05}. Ряд публикаций
посвящен вопросам его применения в задаче адаптивного подавления
помех \cite{Wid85,Kle04,Kle02,Mon04,Mar87,Vic06,God05} Структурная
схема обеляющего фильтра на базе алгоритма LMS показана на  Рис.
\ref{fig:LMS}.

Принцип работы алгоритма LMS основан на минимизации СКО между
некоторым опорным сигналом $d_k$ и выходным сигналом адаптивного
фильтра, который как и ранее равен $y_k = \w^H_k\x_k$. В этом случае
функционал качества работы алгоритма может быть выражен через
корреляционные свойства входных воздействий:
\begin{eqnarray} \label{eqn:LMS_1}
J_k &=& \mathbb{E}\{ e_k e^*_k \} = \mathbb{E}\left\{ \left(d_k - y_k\right) \left(d_k - y_k\right)^* \right\}\\
&=& \mathbb{E}\left\{ \left(d_k - \w^H_k\x_k\right) \left(d_k - \w^H_k\x_k\right)^* \right\}\\
&=& \mathbb{E}\left\{ d_k d^*_k - d_k \w^H_k \x_k - d^*_k \x^H_k \w_k + \w^H_k \x_k \x^H_k \w_k \right\} \\
&=& \upsigma^2_d - \p^H\w_k - \w_k^H\p + \w_k^H \R_k \w_k
\end{eqnarray}
где $\upsigma^2_d$ --- мощность опорного сигнала, $\p$ --- вектор
взаимной корреляции опорного и входного сигналов.

Минимизация указанного функционала качества градиентным методом
\cite{Wid85,Hay96} приводит к уравнению обновления весов вида:
\begin{equation} \label{eqn:LMS_2}
\w_k = \w_{k-1} + \Delta\w_k
\end{equation}
где $\Delta\w_k$ --- поправка на шаге $k$. Для решения задачи
нахождения поправки $\Delta\w_k$ в реальном времени при априори
неизвестных свойствах помехи в алгоритме LMS используется метод
наискорейшего спуска. Суть метода состоит в том, что величина
поправки выбирается пропорциональной градиенту функционала качества:
\begin{eqnarray} \label{eqn:LMS_3}
\Delta\w_k &=& -\nabla J_k = \left[ \begin{array}{cccc}
\frac{\partial J_k}{\partial \w_k(0)} & \frac{\partial J_k}{\partial \w_k(1)} & \ldots & \frac{\partial J_k}{\partial \w_k(M-1)} \end{array} \right]^T\\
&=& -\nabla \left( \upsigma^2_d - \p^H\w_k - \w_k^H\p + \w_k^H \R_k \w_k \right) \\
&=& 2\p - 2\R\w_k
\end{eqnarray}
Таким образом, рекурсивная процедура расчета коэффициентов
адаптивного фильтра по алгоритму наискорейшего спуска с учетом
(\ref{eqn:LMS_2}) и (\ref{eqn:LMS_3}) может быть записана в виде:
\begin{equation} \label{eqn:LMS_4}
\w_k = \w_{k-1} + \frac{\upmu}{2} \left( 2 \p - 2 \R\w_k \right)
\end{equation}
где $\upmu$ --- шаг адаптации.

Заметим, что при отсутствии шумовой компоненты в сигнале значение
градиента функционала качества в пределе стремится к нулю:
\begin{equation} \label{eqn:LMS_5}
\lim\limits_{k \rightarrow \infty} \nabla J_k = 0
\end{equation}
Что соответствует минимально достижимому значению СКО. При этом
коэффициенты адаптивного фильтра LMS сходятся в пределе к
винеровскому решению задачи оптимальной фильтрации сигнала на фоне
коррелированной помехи, известному как уравнение Винера--Хопфа
\cite{Hay96}:
\begin{equation} \label{eqn:LMS_6}
\lim\limits_{k \rightarrow \infty} \w_k = \R^{-1}\p
\end{equation}
Основной проблемой при применении данного результата в обработке
сигналов в реальном времени, когда данные на вход алгоритма LMS
поступают последовательно, является то, что статистические свойства
помехи, т. е. корреляционная матрица $\R$ и вектор взаимной
корреляции опорного сигнала и помехи $\p$ априори неизвестны. Однако
данное затруднение можно преодолеть, аппроксимировав точные значения
этих параметров в (\ref{eqn:LMS_4}) их оценками. Тогда
(\ref{eqn:LMS_4}) можно записать в виде:
\begin{equation} \label{eqn:LMS_7}
\widehat\w_k = \widehat\w_{k-1} + \frac{\upmu}{2} \left( 2
\widehat\p - 2 \widehat\R_k\widehat\w_k \right)
\end{equation}
Принимая во внимание тот факт, что:
\begin{equation} \label{eqn:LMS_8}
\widehat\R_k = \x_k \x^H_k \text{\ и\ } \widehat\p_k = d_k \x_k
\end{equation}
уравнение обновления весов адаптивного фильтра в алгоритме LMS может
быть окончательно записано в виде:
\begin{eqnarray} \label{eqn:LMS_9}
\widehat\w_k &=& \widehat\w_{k-1} +  \upmu \left( -d_k \x_k + \x_k
\x^H_k \widehat\w_k \right)\\
&=& \widehat\w_{k-1} - \upmu \x_k \left( d_k - \x^H_k \widehat\w_k
\right)
\end{eqnarray}

Основными отличительными чертами алгоритма на базе алгоритма LMS
являются \cite{Hay96}:
\begin{enumerate}
\item медленное время адаптации, зависящее от разброса собственных
чисел оценки корреляционной матрицы помехи;
\item вычислительная сложность $\mathcal{O}(M)$;
\item сходимость алгоритма гарантирована только если $\upmu < \frac{1}{\lambda_\text{max}}$ .
\end{enumerate}

Для обеспечения сходимости алгоритма в условиях, когда максимальное
собственное число $\lambda_\text{max}$ корреляционной матрицы помехи
заранее неизвестно, применяется алгоритм NLMS \cite{Yam02}. Алгоритм
NLMS отличается от алгоритма LMS наличием переменного шага адаптации
$\upmu_k$. Для обеспечения сходимости и одновременно увеличения
скорости адаптации до максимально возможной применяется следующая
аппроксимация значения максимального собственного числа
корреляционной матрицы помехи. Известно, что:
\begin{equation} \label{eqn:LMS_10}
\lambda_\text{max} \leq \sum\limits_{i=1}^{N} \lambda_i = \tr\left(
\widehat\R_k \right)
\end{equation}
где $\lambda_i$ --- собственные числа, а $\tr(\cdot)$  --- след
матрицы. В свою очередь след матрицы равен мощности сигнала в линии
задержки адаптивного фильтра:
\begin{equation} \label{eqn:LMS_11}
\tr\left( \widehat\R_k \right) = \x^H_k \x_k
\end{equation}
Поэтому следующая процедура обновления шага адаптации:
\begin{equation} \label{eqn:LMS_12}
\upmu_k = \frac{\upmu_0}{\x^H_k \x_k}
\end{equation}
где $\upmu_0$ выбирается из условия требуемой точности подстройки
коэффициентов, гарантирует сходимость алгоритма NLMS при любых
значениях $\lambda_\text{max}$. Однако применение этой процедуры не
снимает проблемы зависимости скорости сходимости от разброса
собственных чисел корреляционной матрицы помехи.

Существуют и другие варианты регулирования значения шага адаптации и
увеличения скорости адаптации алгоритма LMS, например, так
называемый метод ускоренного градиента \cite{Mon04,Apo00}. Одним из
алгоритмов ускоренного градиента является алгоритм, предложенный
Пауэлом \cite{Pow62}. Каждый шаг алгоритма Пауэла заключается в
выборе такого направления движения от вектора $\w$ в направлении
отрицательного градиента $\v$, которое было бы ортогонально всем
направлениям движения, выбранным на более ранних шагах:
\begin{equation} \label{eqn:Powel_1}
\frac{\partial}{\partial \upmu} \left( J(\w + \upmu\v) \right) = 0
\end{equation}
где $J$ --- поверхность, по которой осуществляется движение
(функционал качества). СКО выходного сигнала такого алгоритма:
\begin{equation} \label{eqn:Powel_2}
\mathbb{E}\{ e e^{*} \} = \mathbb{E}\{ d^2 \} + \w^H \p + \p^H \w +
\w^H \R \w
\end{equation}
Градиент $\mathbb{E}\{ e e^{*}(\w + \upmu\v) \}$ по отношению к
$\upmu$ имеет следующее выражение:
\begin{equation} \label{eqn:Powel_3}
\nabla_{\upmu} \mathbb{E}\{ e e^{*} \} = \v^H \p + \p^H \v  + \v^H
\R \w + \w^H \R \v + 2\upmu \v^H \R \v
\end{equation}
Градиент $\mathbb{E}\{ e e^{*}(\w + \upmu\v) \}$ равен нулю, если
шаг адаптации равен:
\begin{equation} \label{eqn:Powel_4}
\upmu = -\frac{\v^H \p + \v^H \R \w}{\v^H \R \v}
\end{equation}
Поскольку ни $\p$, ни $\R$ априори неизвестны, необходимо
использовать в (\ref{eqn:Powel_2})--(\ref{eqn:Powel_4}) их оценки.

Это семейство алгоритмов не получило широкого распространения в
практике радиолокации по следующим причинам \cite{Mon04}:
\begin{enumerate}
\item Значительное увеличение вычислительных затрат по сравнению с
алгоритмом LMS;
\item Несмотря на значительное увеличение скорости адаптации,
алгоритм адаптируется все же не так быстро, как, скажем RLS;
\item Все алгоритмы, основанные на методе ускоренного
градиента очень чувствительны к присутствию шума в выборке.
\end{enumerate}

Градиентные алгоритмы, рассмотренные выше являются особенно
подходящими для ситуаций, когда критерием качества является
квадратичный (или по крайней мере унимодальный) функционал
\cite{Mon04}. Для некоторых задач теории управления  математическое
соотношение между неизвестными параметрами и критерием качества либо
неизвестно, либо слишком сложно, чтобы быть полезным при разработке
адаптивных алгоритмов. При решении подобных задач можно пользоваться
либо методами систематического поиска, либо методами случайного
поиска \cite{Mon04}. Оба класса алгоритмов способны производить
глобальный поиск и нечувствительны к наличию седловых точек на
поверхности критерия качества. Методы систематического поиска
требуют слишком большой вычислительной мощности адаптивного
процессора. Методы случайного поиска просты и обладают указанными
выше достоинствами. Существуют три основные модификации алгоритмов
случайного поиска \cite{Mon04}: линейный случайный поиск,
ускоренный случайный поиск, управляемый ускоренный случайный
поиск. Уравнение рекуррентного обновления весов в алгоритме
ЛСП можно записать следующим образом:
\begin{equation} \label{eqn:LSP_1}
\w_k = \w_{k-1} + \upmu \left( \widehat{J}(\w_k) -  \widehat{J}(\w_k
+ \Delta\w_k) \right) \Delta\w_k
\end{equation}
где $\widehat{J}(\cdot)$ --- оценка значения функционала качества.
Алгоритм называется линейным, потому что величина приращения
значений весов алгоритма прямо пропорциональна значению приращения
критерия качества. На каждой итерации алгоритма происходит генерация
вектора приращения $\Delta\w_k$, имеющего Гауссовское распределение
с нулевым средним и дисперсией $\upsigma^2$. После этого происходит
оценка приращения критерия качества, вызванного приращением
$\Delta\w_k$. В соответствии с измеренным приращением критерия
качества происходит подстройка весов адаптивного фильтра. Параметры
алгоритма $\upmu$ и $\upsigma^2$ выбираются исходя из условия
устойчивости и требуемой скорости адаптации фильтра:
\begin{eqnarray} \label{eqn:LSP_2}
0 < \upmu \upsigma^2 < \frac{1}{\lambda_\text{max}} \\
\uptau_\text{max} = \frac{1}{2 \upmu \upsigma^2 \lambda_\text{min}}
\end{eqnarray}
где $\uptau_\text{max}$ --- постоянная времени обучающей кривой.
Очевидно, что, как и LMS, алгоритм ЛСП чувствителен к разбросу
собственных значений корреляционной матрицы помехи.

Ускорение процесса подстройки весов адаптивного фильтра в алгоритме
УСП осуществляется за счет увеличения шага адаптации    при движении
в направлении благоприятного изменения критерия качества. Уравнение
рекуррентного обновления весов в алгоритме УСП описывается
выражением:
\begin{equation} \label{eqn:USP_1}
\w_k = \w_{k-1} + \upmu_k  \Delta\w_k
\end{equation}
Случайный вектор $\Delta\w_k$ задает направление поиска и
определяется как:
\begin{equation} \label{eqn:USP_2}
\Delta\w_k^i = \cos(\uptheta_i) + j \sin(\uptheta_i), \ \ i =
1,\ldots,M
\end{equation}
где $\uptheta_i$ --- случайный угол, равномерно распределенный на
интервале $[0, 2\uppi]$. В исходном состоянии вектор $\w_0$
устанавливается равным $\0$, а шаг адаптации $\upmu = \upmu_0$. На
каждом шаге алгоритма определяется $\widehat{J}(\w_{k})$,
генерируется $\Delta\w_k$ и оценивается $\widehat{J}(\w_{k+1})$.
Если критерий качества изменяется в благоприятном направлении, то
$\Delta\w_{k+1} = \Delta\w_k$ сохраняет свое значение на следующем
шаге, а шаг адаптации удваивается: $\upmu_{k+1} = 2\upmu_k$. В
противном случае, вектор весовых коэффициентов остается неизменным,
$\w_{k+1} = \w_k$, шаг адаптации сбрасывается $\upmu_{k+1} =
\upmu_0$ и происходит случайный выбор нового направления движения
$\Delta\w_{k+1}$. Возврат к старому значению $\w_k$ обеспечивает
устойчивость алгоритма при любых значениях $\upmu_0$.

В алгоритме УУСП не только направление, но и размер шага адаптации
являются случайными. Уравнение обновления весов имеет вид:
\begin{equation} \label{eqn:UUSP_1}
\w_k = \w_{k-1} + \Delta\w_k
\end{equation}
Случайное приращение $\Delta\w_k$ является Гауссовским вектором с
нулевым средним и дисперсией $\upsigma^2$, вычисляемой по формуле:
\begin{equation} \label{eqn:UUSP_2}
\upsigma^2 = K_1 + J_\text{min}K_2
\end{equation}
где $K1$ и $K2$ --- константы.

Управление поиском в параметрическом пространстве сводится к
чередованию случайной и детерминированной фазы. Случайным в этом
случае является не только направление, но и шаг поиска. В случайной
фазе фиксируется значение вектора весовых коэффициентов и
производится поиск направления улучшения эффективности. Когда
направление определено, происходит переход к детерминированной фазе
поиска: фиксируется $\Delta\w_k$ и на каждом шаге удваивается его
модуль. Если эффективность начинает ухудшаться, происходит возврат к
случайной фазе, в которой приращение $\Delta\w_k$ будет значительно
меньше по модулю. Следует особо подчеркнуть, что скорость сходимости
алгоритмов случайного поиска определяется не только временем
переходного процесса и количеством итераций, но и временем
вычисления критерия эффективности на каждом шаге. Чтобы получить
оценку критерия, необходимо статистически обрабатывать некоторое
количество выборок его значения.

Описанные выше непараметрические подходы к использованию информации,
заключенной в корреляционной матрице помехи имеют ограниченную
разрешающую способность \cite{Mar90}. АКП оценивается по конечной (в
практике радиолокационной обработки сигналов очень короткой)
выборке. Поэтому из--за ограниченной разрешающей способности
непараметрических методов, оценки параметров спектра помехи могут
значительно отличаться от параметров реального спектра. Особенно
сильно этот эффект проявляется при узкополосной помехе. Для
увеличения разрешающей способности Берг \cite{Bur75} предложил
использование в спектральном оценивании метода максимальной
энтропии, в западной литературе --- MEM. Увеличение разрешающей
способности метода оценки весов адаптивного фильтра может приводить
к значительному увеличению точности компенсации помехи \cite{Dad83}.
Идея метода заключается в исключении предположения о периодичности
или равенстве нулю данных за пределами интервала наблюдения
\cite{Mar90}.

Проблема спектрального оценивания по методу ММЭ формулируется
следующим образом \cite{Pap81}. Необходимо вычислить оценку
СПМ $S_{\bm{\upzeta}}(f)$
некоторого случайного процесса (в данном случае помехи
$\bm{\upzeta}$), которая максимизировала бы выражение для энтропии
процесса с гауссовской ПРВ
\cite{Pap81}:
\begin{equation} \label{eqn:MEM_entr}
H_{\bm{\upzeta}} = \ln\sqrt{2\uppi e} +
\frac{1}{4f_N}\int\limits_{-4f_N}^{4f_N} \ln S_{\bm{\upzeta}}(f) df
\end{equation}
где $f_N$ --- частота повторения импульсов РЛС.

\begin{figure*}[tbp]
\centering
\includegraphics[width=14cm]{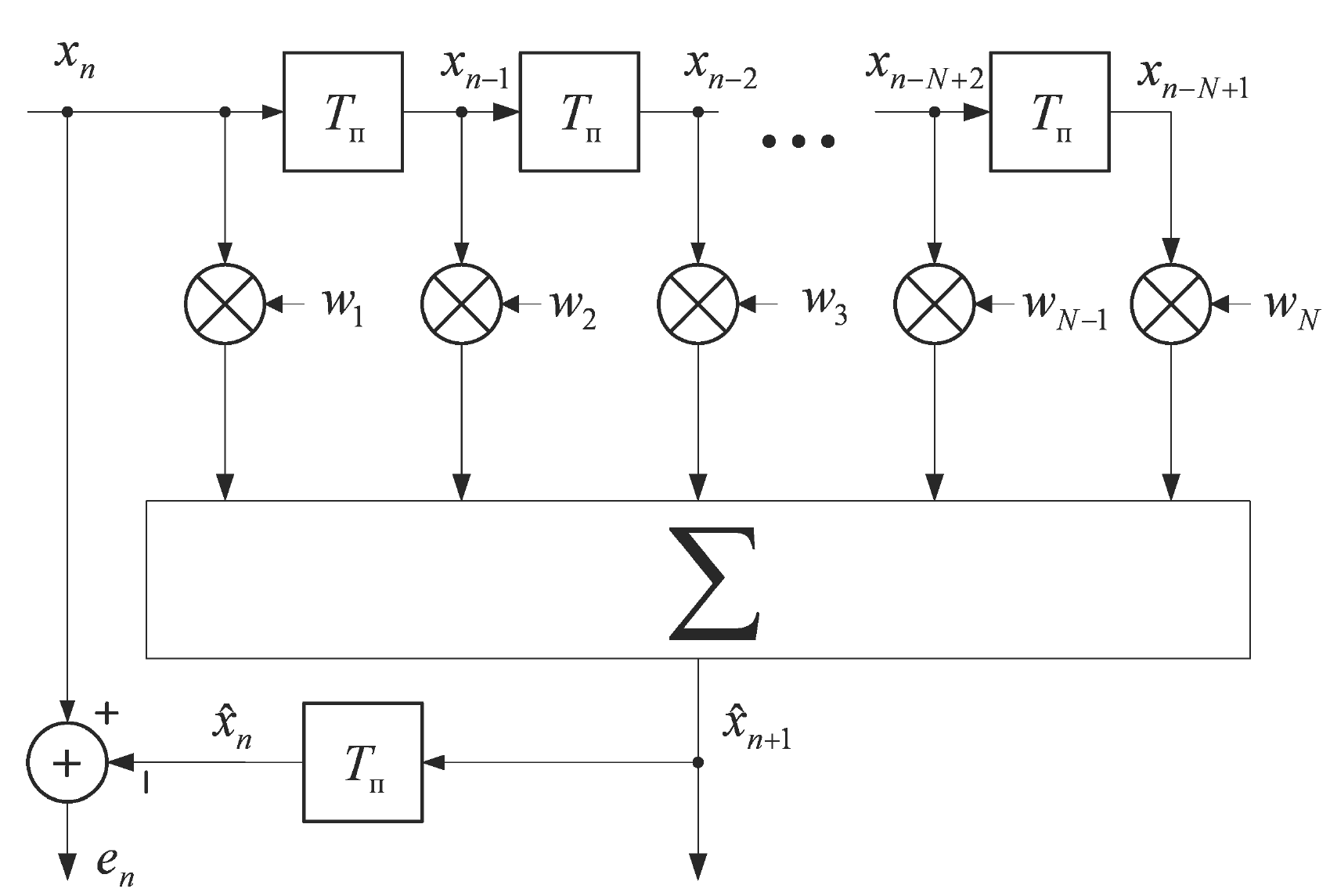} \vspace{-1cm}
\caption{Структурная схема обеляющего фильтра ММЭ} \label{fig:MEM}
\end{figure*}

Применение метода множителей Лагранжа дает следующее
параметризованное выражение для оценки СПМ
$\widehat{S}_{\bm{\upzeta}}^{\text{ММЭ}}(f)$ методом ММЭ:
\begin{equation} \label{eqn:MEM_est}
\widehat{S}_{\bm{\upzeta}}^{\text{ММЭ}}(f) =
\frac{T_\text{п}P_\text{п}}{\left| 1 + \sum_{i=1}^{N-1} w_i
\exp(-j2\uppi i f T_\text{п}) \right|^2}
\end{equation}
где $P_\text{п}$ --- мощность помехи, $T_\text{п}$ --- период
повторения зондирующих импульсов, а параметры $w_i, i = 1\ldots N$
находятся решением следующей системы уравнений, известной еще под
названием уравнений Юла-Уолкера \cite{Mar90}:
\begin{equation} \label{eqn:MEM_Yule_Walk}
 \left[ \begin{array}{cccc}
 r[0] & r[-1] & \cdots & r[-N+1] \\
 r[1] & r[0] & \cdots & r[-N+2] \\
 \cdots & \cdots & \cdots & \cdots \\
 r[N-1] & r[N-2] & \cdots & r[0] \end{array} \right]
 \left[ \begin{array}{c}
 1 \\
 w_1 \\
 \cdots \\
 w_{N-1} \end{array} \right] =
 \left[ \begin{array}{c}
 P_\text{п} \\
 0 \\
 \cdots \\
 0 \end{array} \right].
\end{equation}

Для того, чтобы практически использовать ММЭ для борьбы с помехами,
можно включить его в схему предсказателя на один шаг вперед (Рис.
\ref{fig:MEM}) \cite{Dad83} и использовать один из быстрых
алгоритмов вычисления параметров $w_i$ \cite{Mar90}. В этом случае
для формирования весов подавителя используется первый столбец
инверсии ковариационной матрицы помехи. В статьях \cite{Uen83,Kre85}
был обсужден вопрос применения других столбцов этой матрицы. Однако
в \cite{Far88} показано, что это не приводит к выигрышу в ОСПШ на
выходе ЦРГФ. Cогласно Рис. \ref{fig:MEM} ошибка предсказания
$e_N(n)$ равна:
\begin{equation} \label{eqn:MEM_err}
e_N(n) = x(n) - \widehat x_N(n)
\end{equation}
где $\widehat x_N(n)$ --- оценка значения процесса на $n$ шагу с
использованием $N$ его предыдущих отсчетов:
\begin{equation} \label{eqn:MEM_est_N}
\widehat x_N(n+1) = \sum\limits_{i=1}^{N} w_i x_{n+1-i}.
\end{equation}
Показано \cite{Pap81}, что для процессов АР типа ошибка
(\ref{eqn:MEM_err}) предсказателя с весами (\ref{eqn:MEM_Yule_Walk})
имеет характер белого шума (происходит выбеливание помехи). Более
того, применение весов типа (\ref{eqn:MEM_Yule_Walk}) приводит к
минимизации мощности помехи с ковариационной матрицей $\R$. Поэтому
алгоритм ММЭ принадлежит к классу алгоритмов минимизирующих мощность
помехи на выходе устройства АСДЦ.

Алгоритм ММЭ с весами удовлетворяющими (\ref{eqn:MEM_Yule_Walk})
предполагает знание корреляционной матрицы помехи. Cуществует и
практически более значимый вариант, в котором оценки коэффициентов
вычисляются рекуррентно, непосредственно по отсчетам входной выборки
\cite{Dad83,Far87}. Данную процедуру вычисления коэффициентов $\C_M$
($M$ --- порядок фильтра):
\begin{equation} \label{eqn:MEM_est_N_1}
\widehat x(n) = \sum\limits_{k=1}^{N} \C_M(k) x(n-k).
\end{equation}
можно представить как два одновременно работающих фильтра
предсказания на один шаг: вперед с коэффициентами $[1, -C_M(1), …,
-C_M(M)]^T$ и назад с коэффициентами $[1, -C^{*}_M(1), …,
-C^{*}_M(M)]^T$ \cite{Yon00}. Вычисление этих коэффициентов
осуществляется по критерию минимизации суммарной ошибки фильтра
предсказания вперед $\f_M$ и фильтра предсказания назад $\b_M$:
\begin{equation} \label{eqn:MEM_SE}
\C_M = \arg\min\limits_{\C_M} \mathbb{E}\{ e^2_M \}
\end{equation}
Минимизация $\mathbb{E}\{ e^2_M \}$ в (\ref{eqn:MEM_SE}) приводит к
рекурсивному алгоритму вычисления коэффициентов $\C_M(k)$ следующего
вида:
\begin{eqnarray} \label{eqn:MEM_rec}
\C_M(k) &=& \C_{M-1}(k) + \C_{M}(M)\C_{M-1}^{*}(M-k)\\
\overline{e^2}(M) &=& \left(1 - \left| \C_{M}(M) \right|^2\right)\overline{e^2}(M-1)\\
\f_M(k) &=& \f_{M-1}(k) + \C_{M}(M)\b^{*}_M(k-1)\\
\b_M(k) &=& \b_{M-1}(k) + \C^{*}_{M}(M)\f_M(k-1)
\end{eqnarray}
Таким образом, самая сложная операция состоит в вычислении
коэффициента $\C_{M}(M)$:
\begin{equation} \label{eqn:MEM_C_MM}
\C_{M}(M) = \frac{-2 \sum_{k=1}^{N-M}
\b^{*}_{M-1}(k)\f_{M-1}(k)}{\sum_{k=1}^{N-M} \left[ |\b_{M-1}(k)|^2
+ |\f_{M-1}(k)|^2 \right]}
\end{equation}
Инициализация алгоритма осуществляется следующим образом:
\begin{eqnarray} \label{eqn:MEM_rec_init}
\overline{e^2}(0) &=& \frac{1}{N} \sum\limits_{k=1}^{N} |x(k)|^2\\
\f_0(k) &=& \b_0(k) = x(k)
\end{eqnarray}

Другим, более общим методом моделирования помехи является ее
представление моделью АРСС \cite{Kos94,Kos96}. В этом методе, в
отличие от ММЭ, рассматривающем помеху как случайный процесс
авторегрессии, помеха моделируется случайным процессом
авторегрессии--скользящего среднего. Теоретические проблемы  и
вопросы практического применения методов ММЭ и АРСС в алгоритмах
АСДЦ рассмотрены в таких работах как
\cite{And94,Kos94,Kos96,Vos03,Kos01,Hui84,Bur75,Dad83,Pap81,Mar90,Kre85,Hay82,Hay76,Kar87,Par03}.
Более простые, но субоптимальные методы подавления помех основанные
на оценивании таких параметров спектра помехи, как средняя частота и
ширина спектра рассмотрены в
\cite{Bak86,Kve96,Hua00,Har01,Yon99,Yon05,Nie06}.

Помимо таких несомненных достоинств как невысокая вычислительная
сложность $\mathcal{O}(M^2)$, высокая разрешающая способность (и
следовательно хорошая работоспособность на короткой выборке), ММЭ
обладает следующими недостатками, присущими многим другим
параметрическим методам спектрального оценивания. Сложность выбора
значения порядка модели $M$. Существующие процедуры, такие как
информационный критерий Акаике \cite{Mar90} часто не обеспечивают
адекватного оценивания порядка фильтра $M$, особенно при малом
размере обучающей выборки \cite{Fis05}. Альтернативные подходы могут
требовать значительных вычислительных затрат, как например метод
\cite{Cha84}, использующий процедуру ГШ, или метод \cite{Bet04},
использующий поиск в пространстве статистик обнаружения,
сконструированном с учетом вероятных значений порядка модели. Выбор
слишком большого значения $M$ часто приводит к появлению лишних
полюсов в оцениваемом спектре, что обуславливает увеличение мощности
ошибки и нарушению процесса выбеливания помехи. В свою очередь
уменьшение $M$ ниже оптимального значения влечет за собой потерю
разрешающей способности (основного преимущества ММЭ). Кроме этого,
представление помехового процесса $\bm{\upzeta}$ моделью АР или, в
более общем виде, моделью АРСС \cite{Mar90} приводит к сужению
класса случайных процессов к обработке которых применимы алгоритмы
класса ММЭ (АРСС). Как известно, АРСС модель адекватно описывает
только случайные процессы с полиномиальным спектром \cite{Bog75}.

Заметим также, что принцип подавления помехи Рис. \ref{fig:RLS},
Рис. \ref{fig:LMS} и Рис. \ref{fig:MEM} один и тот же. Разница между
алгоритмами RLS, LMS, ММЭ и подобными им заключается в способе
расчета коэффициентов адаптивного фильтра.

\section{Заключение} \label{chap:sec:Ch2:concl}

В данной главе представлен обзор существующих подходов к решению
задачи АСДЦ на фоне коррелированных помех с неизвестными
спектральными характеристиками. Сформулированы критерии
эффективности систем СДЦ и приведена классификация существующих
адаптивных алгоритмов. Проведена параллель между критериями
эффективности системы СДЦ, используемыми для оценки эффективности ее
работы и критериями качества, используемыми при синтезе адаптивных
алгоритмов. Приведено статистически оптимальное решение задачи
обнаружения полезного сигнала на фоне коррелированной помехи с
известными спектральными свойствами. Изложен субоптимальный подход к
решению задачи обнаружения полезного сигнала при априорной
неопределенности относительно спектральных свойств помехи,
основанный на обобщенном отношении правдоподобия. После этого
рассмотрен ряд распространенных алгоритмов ПВАО, в большой степени
основанных на этом подходе. В том числе, алгоритм непосредственного
обращения оценки ковариационной матрицы помехи и вопросы его
эффективной реализации на основе рекурсии Левинсона--Дербина;
алгоритм ортогонализации Грама--Шмидта и его реализация на основе
факторизации Холецкого; ряд градиентных и квазиградиентных
алгоритмов, основанных на методе наискорейшего спуска; рекурсивный
алгоритм наименьших квадратов с экспоненциальным взвешиванием, ряд
алгоритмов случайного поиска, гарантирующих сходимость к глобальному
минимуму, но имеющих низкую скорость сходимости; а также
параметрические алгоритмы, основанные на методе максимальной
энтропии, и полагающиеся на АР(АРСС) модель помехи.

\chapter{Результаты синтеза адаптивных алгоритмов} \label{chap:Ch3}

В данной главе мы представляем результаты синтеза предлагаемых
адаптивных алгоритмов предназначенных для подавления помех с
неизвестной ковариационной матрицей в системе АСДЦ.

\section{Оптимизация уровня регуляризации для ослабления подавления сигнала от цели в алгоритме РОКМ} \label{sec:SMI_reg}

\subsection{Постановка задачи}

Как указывалось ранее, регуляризация оценки ковариационной матрицы
помехи (\ref{eqn:reg_R_hat}) используется в алгоритме РОКМ
(\ref{eqn:weights_rokm}) для уменьшения потерь в выходном ОСПШ
возникающих при конечном размере обучающей выборки $M$. Карлсон
\cite{Car88} предложил фиксировать $\upalpha$ на уровне $10$ дБ выше
уровня шума. Такая регуляризация позволяет значительно сократить
потери в широком диапазоне изменения параметров помеховой
обстановки, но не является оптимальной во всем диапазоне изменения
параметров помеховой обстановки. В частности, в ситуации, когда
полезный сигнал присутствует в обучающей выборке, а мощность помехи
в ней мала, оптимальной с точки зрения максимизации отношения сигнал
помеха шум (ОСПШ) является значительно более высокая величина
$\upalpha$. Ряд статей \cite{Bec07,Kim98,Li03,Lor05,Vor03,Vin04}
посвящен оптимизации $\upalpha$ с учетом реальных параметров
помеховой обстановки. Однако в этих статьях авторы полагаются на то,
что сигнал в обучающей выборке появляется из-за ошибок калибровки
приемной антенны \cite{You01,Goe06}. Ошибка калибровки может быть
оценена или учтена заранее \cite{Vor03,Vin04}, во время калибровки
\cite{Goe06} или в процессе адаптации \cite{Lee05}. Поэтому в своих
статьях эти авторы используют этот параметр для получения алгоритмов
оптимизации $\upalpha$. Кроме этого, в некоторых из этих статей
делается предположение о том, что размерность помехового
пространства известна и также используют этот факт при разработке и
анализе алгоритмов. Ясно, что такой подход не универсален. Полезный
сигнал далеко не всегда проникает в обучающую выборку из--за ошибок
калибровки антенны. К примеру, проникновение полезного сигнала от
цели возможно в случае, если цель является быстро маневрирующей или
распределенной. В тоже время, размерность помехового пространства на
практике бывает сложно оценить. Поэтому в данном разделе мы получаем
непараметрический алгоритм оптимизации величины регуляризации
$\upalpha$, не используя при этом модели проникновения полезного
сигнала в обучающую выборку или знания размерности помехового
пространства. Данный алгоритм предложен нами в
\cite{Ore07c,Ore08b,Ore08c}.

\subsection{Оптимизации величины $\upalpha$}

Мы формулируем задачу оптимизации величины $\upalpha$ как задачу
максимизации ОСПШ заданного в виде отношения Релея:
\begin{equation} \label{eqn:lsmi_Ray}
\gamma_\text{вых} = \frac{|\w^H \s|^2}{\w^H \R \w}.
\end{equation}
Здесь $\w$, как и ранее
\begin{equation} \label{eqn:lsmi_w}
\w \triangleq \w_\text{рокм} = \left(\widehat \R + \upalpha\I
\right)^{-1}\s.
\end{equation}
Является вектором весов соответствующего адаптивного фильтра.
Поскольку непосредственная максимизация (\ref{eqn:lsmi_Ray})
затруднена, данную задачу обычно переформулируют в эквивалентную
задачу условной минимизации знаменателя в правой части
(\ref{eqn:lsmi_Ray}):
\begin{equation} \label{eqn:lsmi_const_opt}
\upalpha = \arg\min\limits_{\upalpha} \w^H \R \w, \text{\ \ при
условии\ \ } |\w^H \s|^2 = 1
\end{equation}

Для решения задачи (\ref{eqn:lsmi_const_opt}) мы применяем метод
линеаризации выражения для весов   сходный с методом, примененым
например в \cite{Tia01}. Заметим, что $\upalpha$ обычно мал\'{o} и
(\ref{eqn:lsmi_w}) может быть разложено в ряд Тейлора в окрестности
точки $\upalpha = 0$:
\begin{equation} \label{eqn:lsmi_w_ser}
\w = \left[\w\right]_{\upalpha = 0} + \upalpha \left[
\frac{\partial}{\partial \upalpha} \w \right]_{\upalpha = 0} +
\frac{\upalpha^2}{2!} \left[ \frac{\partial^2}{\partial^2 \upalpha}
\w \right]_{\upalpha = 0} + \frac{\upalpha^3}{3!} \left[
\frac{\partial^3}{\partial^3 \upalpha} \w \right]_{\upalpha = 0} +
\ldots
\end{equation}
Вычисляя первую производную (\ref{eqn:lsmi_w}) в точке $\upalpha =
0$:
\begin{equation} \label{eqn:lsmi_w_1st_dir}
\left[ \frac{\partial}{\partial \upalpha} \w \right]_{\upalpha = 0}
= \left[ -\left(\widehat \R + \upalpha\I
\right)^{-1}\I\left(\widehat \R + \upalpha\I \right)^{-1}\s
\right]_{\upalpha = 0} = -\widehat \R^{-2}\s,
\end{equation}
и ограничиваясь первыми двумя членами ряда (\ref{eqn:lsmi_w_ser}) мы
получаем следующую линейную аппроксимацию (\ref{eqn:lsmi_w}):
\begin{equation} \label{eqn:lsmi_w_approx}
\w \simeq \widehat \R^{-1}\s - \upalpha \widehat \R^{-2}\s,
\end{equation}
которая после введения обозначений
\begin{equation} \label{eqn:lsmi_not}
\widetilde \w \triangleq \widehat\R^{-1}\s \text{\ и \ }
\widetilde\v \triangleq \widehat\R^{-1}\widetilde\w
\end{equation}
принимает следующий вид:
\begin{equation} \label{eqn:lsmi_w_not_approx}
\w \simeq \widetilde\w - \upalpha\widetilde\v.
\end{equation}
Подстановка (\ref{eqn:lsmi_w_not_approx}) в
(\ref{eqn:lsmi_const_opt}), применение аппроксимации неизвестной
корреляционной матрицы помехи ее оценкой $\R \simeq \widehat\R$ и
применение метода множителей Лагранжа для решения поставленной
задачи приводит к следующему критерию качества:
\begin{equation} \label{eqn:lsmi_J}
J(\upalpha) = \widetilde\w^H\s - \upalpha \widetilde\w^H\widetilde\w
-\upalpha\widetilde\v^H\s + \upalpha^2\widetilde\v^H\widetilde\w +
\lambda(\widetilde\w^H\s - \upalpha\widetilde\v^H\s - 1)
\end{equation}
где $\lambda$ --- множитель Лагранжа. Градиент критерия качества
относительно $\upalpha$ выражается следующим образом:
\begin{equation} \label{eqn:lsmi_J_grad}
\frac{\partial}{\partial\upalpha}J(\upalpha) =
-\widetilde\w^H\widetilde\w - \widetilde\v^H\s +
2\upalpha\widetilde\v^H\widetilde\w - \lambda\widetilde\v^H\s.
\end{equation}
Приравнивая (\ref{eqn:lsmi_J_grad}) к нулю мы получаем следующее
выражение для $\upalpha$:
\begin{equation} \label{eqn:lsmi_alpha}
\upalpha = \frac{\widetilde\w^H\widetilde\w + \widetilde\v^H\s +
\lambda\widetilde\v^H\s}{2\widetilde\v^H\widetilde\w}.
\end{equation}
Выражение для множителя Лагранжа $\lambda$ может быть найдено путем
подстановки (\ref{eqn:lsmi_alpha}) в условие $|\w^H \s|^2 = 1$ из
(\ref{eqn:lsmi_const_opt}):
\begin{eqnarray} \label{eqn:lsmi_lam1}
\w^H\s & = &  (\widetilde\w^H - \upalpha\widetilde\v^H)\s \nonumber \\
 & = & \widetilde\w^H\s - \left( \frac{\widetilde\w^H\widetilde\w + \widetilde\v^H\s +
\lambda\widetilde\v^H\s}{2\widetilde\v^H\widetilde\w} \right)^{*}\widetilde\v^H\s  \\
 & = & \widetilde\w^H\s - \frac{\widetilde\w^H\widetilde\w\widetilde\v^H\s + \s^H\widetilde\v\widetilde\v^H\s +
\lambda\s^H\widetilde\v\widetilde\v^H\s}{2\widetilde\v^H\widetilde\w}
= 1\nonumber
\end{eqnarray}
Наконец, выражение для $\lambda$ приобретает следующий вид:
\begin{equation} \label{eqn:lsmi_lam2}
\lambda = -1 - \frac{2\widetilde\w^H\widetilde\v(1 -
\widetilde\w^H\s) +
\widetilde\w^H\widetilde\w\widetilde\v^H\s}{|\s^H\widetilde\v|^2}
\end{equation}

\begin{figure*}
\centering
\begin{tabular}{|l|}
\hline ~\\

\begin{minipage}[l]{16.5cm}

\begin{enumerate}

\item Оценивание корреляционной матрицы помехи \ \ $\widehat \R = \frac{1}{M} \X
\X^H$ \ \ \ (\ref{eqn:ML_R});

\item Инициализация итеративного алгоритма \ \ $\upalpha_0 = \upsigma^2_{\text{ш}}$;

\item Для всех $i = 1\ldots T$
    \begin{itemize}
    \item $\widetilde \w_i = \left(\widehat\R + \upalpha_i\I \right)^{-1}\s$   \ \ \
(\ref{eqn:lsmi_not});
    \item $\widetilde\v_i
= \left(\widehat\R + \upalpha_i\I \right)^{-1}\widetilde\w_i$ \ \ \
(\ref{eqn:lsmi_not});
    \item $\lambda_i = -1 - \left({2\widetilde\w_i^H\widetilde\v_i(1 -
\widetilde\w_i^H\s) +
\widetilde\w_i^H\widetilde\w_i\widetilde\v_i^H\s}\right)/{|\s^H\widetilde\v_i|^2}$
\ \ \ (\ref{eqn:lsmi_lam2});
    \item $\upalpha_{i+1} = ({\widetilde\w_i^H\widetilde\w_i + \widetilde\v_i^H\s +
\lambda_i\widetilde\v_i^H\s})/({2\widetilde\v_i^H\widetilde\w_i})$ \
\ \ (\ref{eqn:lsmi_alpha});
    \end{itemize}
\item Оценивание вектора весов $\w = \left(\widehat\R + \upalpha_{K+1}\I
\right)^{-1}\s$ \ \ \ (\ref{eqn:lsmi_w});
\end{enumerate}

\end{minipage}

~\\~\\\hline
\end{tabular}

\caption{Итеративный алгоритм определения оптимального уровня
регуляризации $\upalpha$ и оценивания вектора весов адаптивного
алгоритма РОКМ.} \label{fig:it_rokm}
\end{figure*}

Таким образом, алгоритм определения $\upalpha$, максимизирующего
ОСПШ (\ref{eqn:lsmi_Ray}) состоит в определении $\widetilde\w$ и
$\widetilde\v$ по формуле (\ref{eqn:lsmi_not}), $\lambda$ и
$\upalpha$ по формулам (\ref{eqn:lsmi_lam2}) и
(\ref{eqn:lsmi_alpha}) и, наконец, $\w$ по формуле
(\ref{eqn:lsmi_w}). Однако, во--первых, $\widehat\R^{-1}$ в
(\ref{eqn:lsmi_not}) может быть плохо обусловлена, поэтому в
(\ref{eqn:lsmi_not}) необходима регуляризация. Мы предлагаем
использовать в (\ref{eqn:lsmi_not}) регуляризацию на уровне шума,
т.е. в десять раз меньшую предложенной в \cite{Car88}. Во--вторых,
линейная аппроксимация в (\ref{eqn:lsmi_w_approx}) препятствует
точному вычислению оптимального $\upalpha$ за одну итерацию такого
алгоритма, если оптимальное $\upalpha$ велико. Поэтому на практике
может потребоваться несколько итераций. Применение нескольких
итераций алгоритма вычисления оптимального $\upalpha$ приводит к
тому, что на каждой итерации ряд (\ref{eqn:lsmi_w_ser}) получается
путем разложения выражения для весов в точке $\upalpha = \upalpha_i$
согласно обозначений на Рис. \ref{fig:it_rokm}. Следовательно, если
на каждой итерации значение $\upalpha$ приближается к оптимальному,
то точность аппроксимации вектора весов выражением
(\ref{eqn:lsmi_w_approx}) возрастает. Результаты моделирования
говорят о том, что значительное улучшение характеристик
предлагаемого алгоритма по сравнению с алгоритмом РОКМ использующем
фиксированное значение $\upalpha$ возможно уже при количестве
итераций $T$ равном 3. Итеративный алгоритм определения показан на
Рис. \ref{fig:it_rokm}. Видно, что алгоритм сравнительно прост в
реализации. Результаты моделирования предлагаемого алгоритма и
сравнение его характеристик с оптимальным алгоритмом и алгоритмом
РОКМ, использующим фиксированный коэффициент регуляризации приведены
в разделе \ref{sec:SMI_reg:mod}.

\section{Алгоритм LMS с квадратичным ограничением} \label{sec:LMS_sq}

\subsection{Постановка задачи}

Как уже указывалось ранее, адаптивные алгоритмы широко применяются в
радиолокации и связи для подстройки весов системы ПВАО отраженного
сигнала с целью подавления помех. Алгоритм РОКМ является одним из
таких алгоритмов. Вычислительная сложность этого алгоритма в
некоторых случаях препятствует его применению в реальных системах,
поэтому исследования в направлении создания итеративных алгоритмов
меньшей сложности ведутся с конца пятидесятых годов \cite{Mon04}.
Одним из первых предложенных итеративных алгоритмов является
градиентный алгоритм Уидроу \cite{Wid85} известный теперь под
аббревиатурой LMS, а также его модификации \cite{Mon04}, включающие
в том числе нормализованный алгоритм LMS --- NLMS. Применение
алгоритма LMS встретило трудности, выражающиеся в неконтролируемом
подавлении полезного сигнала в случае, когда помеха коррелирована с
сигналом и привело к появлению версии LMS с ограничением на усиление
полезного сигнала \cite{Fro72}. Дальнейшие работы привели к
появлению алгоритмов, подобных алгоритму LMS с ограничением
\cite{Fro72}, обобщенных на случай широкополосного приема
\cite{Sla07,Buc86,Gri82}, влиянию неточностей при калибровке антенны
\cite{Cox87} и неравномерности среды распространения \cite{Gan01}.
Далее исследования шли по пути увеличения скорости сходимости
градиентных алгоритмов. Обзор существующих методов приведен в
\cite{Mor04}. Отличительной чертой LMS с линейным ограничением,
наряду с простотой, является наличие слагаемого в уравнении
обновления весов, отвечающего за выполнение этого ограничения. В
ряде случаев величина этого слагаемого может оказаться слишком
большой, что может приводить к замедлению адаптации весов по
сравнению с алгоритмом без ограничения. В других случаях, величина
этого слагаемого может оказаться слишком малой, что приводит к
проникновению помеховых составляющих в вектор весов. Несмотря на
большой объем литературы по алгоритму LMS, этому вопросу уделено
мало внимания. Решение проблемы может быть найдено при использовании
квадратичного ограничения. В свою очередь оптимизация градиентного
алгоритма с квадратичным ограничением эквивалентна, как будет
показано далее, задаче разработки итеративного алгоритма оценивания
главного обобщенного собственного вектора \cite{Mor04}. Существует
большое количество алгоритмов решения этой задачи, к примеру
\cite{Yan06,Kar84,Yu85}. Однако эти методы сложны и требуют
$\mathcal{O}(N^2)$ операций на каждой итерации. В \cite{Kwo99}
описано решение задачи, схожей с поставленной нами. Однако
разработанный в \cite{Kwo99} алгоритм требует порядка $9N$ операций
и предназначен для работы в качестве эквалайзера системы CDMA.
Поэтому алгоритм предложенный в \cite{Kwo99} требует наличия
дополнительных сигналов с выходов корреляторов системы CDMA для
подстройки весов. Предлагаемый алгоритм \cite{Ore07a,Ore08a},
использующий более простой метод аппроксимации множителя Лагранжа,
требует порядка $6N$ операций и не полагается на наличие каких бы то
ни было дополнительных сигналов, кроме матрицы обучающего сигнала.
Данный алгоритм предназначен для подстройки весов ФАР. Алгоритм
предложенный в \cite{Che07} использует квадратичное условие для
устранения подавления полезного сигнала алгоритмом РОКМ при наличии
расстройки в направлении на цель. Однако в этом алгоритме
используется привязка к двум углам слева и справа от цели, а
ограничение задано в виде неравенства, поэтому точное выполнение
равенства мощности сигнала от цели определенной заданной величине
невозможно. Применение квадратичного ограничения в алгоритмах
отличных от РОКМ в этой статье не рассмотрено.

Обозначим веторный дискретный во времени комплексный сигнал с выхода
линейной антенной решетки, состоящей из $N$ элементов, $\x = [x_1,
x_2, \ldots, x_N]^T$. Пусть как и ранее этот сигнал представляет
собой аддитивную смесь полезного сигнала $\s = [s_1, s_2, \ldots,
s_N]^T$, помехи $\bm{\upzeta} = [\upzeta_1, \upzeta_2, \ldots,
\upzeta_N]^T$, и белого Гауссовского шума с нулевым средним
$\bm{\upxi} = [\upxi_1, \upxi_2, \ldots, \upxi_N]^T$:
\begin{equation} \label{eqn:x_d_1}
\x = \s + \bm{\upzeta} + \bm{\upxi},
\end{equation}

Сигнал на выходе формирователя луча с вектором коэффициентов $\w$,
представляет собой взвешенную сумму элементов вектора $\x$:
\begin{equation} \label{eqn:y_d_1}
y = \w^H\x
\end{equation}

ОСПШ на выходе формирователя луча как и ранее задается отношением
Релея вида (\ref{eqn:lsmi_Ray}), а задача состоит в разработке
итеративного алгоритма обновления весов $\w$, асимптотически
максимизирующих выходной ОСПШ. Решение такой задачи при условии
единичного усиления в направлении полезного сигнала $\s$, полученное
Фростом \cite{Fro72} при условии, что направление на полезный сигнал
соответствует нормали к антенне $\s = [1, 1, \ldots, 1]^T$, приводит
к следующему алгоритму LMS с ограничением:
\begin{equation} \label{eqn:LMS_f_1}
\w(m+1) = \w(m) - \upmu y^H(m)\x(m) -
\frac{1}{N}\sum\limits_{i=1}^N\w_i(m) + \upmu y^H(m)
\frac{1}{N}\sum\limits_{i=1}^N \x_i(m) + \frac{1}{N}
\end{equation}
здесь $k$~--- номер обучающей выборки (номер итерации), $\w(m) =$ $
[w_1(m),$ $ w_2(m),$ $ \ldots,$ $ w_N(m)]^T$~--- вектор весов на
$m$-ой итерации, $\x(m)$~--- $m$-ый столбец матрицы обучающего
сигнала $\X = [\x_1, \x_2, \ldots, \x_M]$, $\upmu$~--- шаг
адаптации, $y^H(m) = \w^H(m)\x(m)$~--- выходной сигнал формирователя
луча на $m$-ом шаге.

В \cite{Fro72} указано, что осуществление минимизации знаменателя
отношения Релея (\ref{eqn:lsmi_Ray}) при линейном ограничении
$\w^H(m)\s = 1$, эквивалентно максимизации ОСПШ. Это подразумевает,
что итеративные алгоритмы, полученные путем наложения линейного и
квадратичного ограничений должны быть асимптотически эквивалентны
\cite{Cox87} при $M \rightarrow \infty$ в смысле величины
получаемого на выходе ОСПШ. Как показывают результаты моделирования,
этот вывод действительно справедлив. Однако с другой стороны,
результаты моделирования говорят о том, что градиентный алгоритм,
полученный посредством применения квадратичного ограничения,
позволяет получить большую скорость сходимости весов к оптимальному
значению при конечном размере обучающей выборки $M$.

\subsection{Синтез алгоритма LMS с квадратичным ограничением}

Сформулируем задачу максимизации ОСПШ как задачу минимизации
знаменателя (\ref{eqn:lsmi_Ray}) аналогичную
(\ref{eqn:lsmi_const_opt}) при квадратичном ограничении следующим
образом:
\begin{equation} \label{eqn:lms_quad_const}
\w = \arg\min\limits_{\w} \w^H \R \w, \text{\ \ при условии\ \ }
\w^H \R_s \w = 1,
\end{equation}
где как и ранее $\R \triangleq \R_{\upxi+\upzeta}$ ---
ковариационная матрица суммы помехи и шума, а $\R_s = E\{ \s \s^H \}
= \s \s^H$ --- корреляционная матрица сигнала.

Данная задача может быть решена стандартным методом множителей
Лагранжа путем перехода к задаче безусловной минимизации следующего
функционала (критерия качества):
\begin{equation} \label{eqn:lms_J}
J(\w) = \w^H \R \w - \lambda\left(\w^H \R_s \w - 1\right)
\end{equation}
Здесь $\lambda$ --- множитель Лагранжа. Градиент критерия качества
(\ref{eqn:lms_J}), $\nabla_J~\triangleq \nabla_{\w}J(\w)$, равен:
\begin{equation} \label{eqn:lms_nabla_J}
\nabla_J = 2 \R \w - 2 \lambda \R_s \w
\end{equation}
Решение задачи минимизации (\ref{eqn:lms_J}) сводится к нахождению
решения уравнения
\begin{equation} \label{eqn:lms_nabla_J_0}
\nabla_J = 0
\end{equation}
и приводит к необходимости решения обобщенной проблемы нахождения
максимального собственного вектора и собственного числа матричного
уравнения вида
\begin{equation} \label{eqn:lms_eiq_eq}
\R\w = \lambda \R_s \w
\end{equation}

Как было указанно ранее, существующие неградиентные методы решения
этой задачи \cite{Kar84,Yu85} требуют $\mathcal{O}(N^2)$ операций на
каждой итерации. Аппроксимация же решения градиентным методом
приводит к итеративному алгоритму обновления весов, требующему
$\mathcal{O}(N)$ операций, и имеющему следующий вид:
\begin{equation} \label{eqn:lms_grad_approx}
\w(m+1) = \w(m) - \upmu^\prime \nabla_J
\end{equation}
который после подстановки (\ref{eqn:lms_nabla_J}) в
(\ref{eqn:lms_grad_approx}) принимает следующий вид:
\begin{equation} \label{eqn:lms_grad_alg}
\w(m+1) = \w(m) - \upmu (\R \w - \lambda \R_s \w)
\end{equation}
где $\upmu = 2\upmu^\prime$. Поскольку матрица $\R$ априори не
известна, в (\ref{eqn:lms_grad_alg}) можно использовать ее
стохастическую аппроксимацию $\widehat\R(m)$
\begin{equation} \label{eqn:lms_R_hat_m}
\widehat\R(m) = \x(m)\x^H(m)
\end{equation}
Тогда (\ref{eqn:lms_grad_alg}) принимает следующую форму:
\begin{equation} \label{eqn:lms_grad_alg_approx}
\w(m+1) = \w(m) - \upmu \left( \x(m)\x^H(m)\w(m) -
\lambda(m)\s\s^H\w(m) \right)
\end{equation}

Множитель Лагранжа $\lambda(m)$ в (\ref{eqn:lms_grad_alg_approx})
также неизвестен. Применение одношаговой аппроксимации
(\ref{eqn:lms_R_hat_m}) матрицы $\R$ в (\ref{eqn:lms_eiq_eq}), а
также умножение обеих частей (\ref{eqn:lms_eiq_eq}) на $\w^H(m)$
приводит к тому, что $\lambda(m)$ может быть аппроксимирован на
каждой итерации следующим образом:
\begin{equation} \label{eqn:lms_lam_approx}
\lambda(m) \simeq \frac{ \w^H(m) \widehat\R(m) \w(m) }{ \w^H(m)
\R_s(m) \w(m) } = \frac{ \left| \x^H(m) \w(m) \right|^2 }{\left|
\s^H(m) \w(m) \right|^2}
\end{equation}
Подстановка (\ref{eqn:lms_lam_approx}) в
(\ref{eqn:lms_grad_alg_approx}) дает окончательную форму алгоритма
LMS с квадратичным ограничением и произвольным направляющим вектором
$\s$:
\begin{equation} \label{eqn:lms_grad_alg_s}
\w(m+1) = \w(m) - \upmu \left[ \x(m)\x^H(m)\w(m) - \frac{ \left|
\x^H(m) \w(m) \right|^2 }{\left| \s^H(m) \w(m) \right|^2}\s\s^H\w(m)
\right]
\end{equation}

Приведем алгоритм (\ref{eqn:lms_grad_alg_s}) к форме алгоритма
Фроста (\ref{eqn:LMS_f_1}) с направляющим вектором совпадающим с
нормалью к антенне $\s = [1, 1, \ldots, 1]^T$:
\begin{equation} \label{eqn:lms_grad_alg_s_all_1}
\w(m+1) = \w(m) - \upmu y^H(m)\x(m) + \upmu \frac{\left| y(m)
\right|^2}{\sum_{i=1}^N \w_i(m)}
\end{equation}

\subsection{Анализ скорости сходимости}

Как видно из сравнения (\ref{eqn:lms_grad_alg_s_all_1}) и
(\ref{eqn:LMS_f_1}), эти два алгоритма отличаются только членами,
отвечающими за вид ограничений. В алгоритме Фроста
(\ref{eqn:LMS_f_1}) этот член зависит от выходного сигнала антенной
решетки линейно, а в алгоритме (\ref{eqn:lms_grad_alg_s_all_1}) ---
квадратично, что является естественным следствием типа ограничения.
В обоих случаях член, отвечающий за выполнение ограничения
представляет собой скаляр, добавляемый к каждому из весов
адаптивного фильтра на каждой итерации. Виртуально наличие такого
скаляра можно интерпретировать как присутствие в обучающей выборке
полезного сигнала, если заметить, что полезный сигнал в
рассматриваемом случае представляет собой $\s = [1, 1, \ldots,
1]^T$. Однако это справедливо и при более общей форме $\s$, как
следует к примеру из (\ref{eqn:lms_grad_alg_s}). Этот факт
становится еще более очевидным, если
(\ref{eqn:lms_grad_alg_s_all_1}) и (\ref{eqn:LMS_f_1}) преобразовать
к следующему виду:
\begin{eqnarray}
\w(m+1) &=& \w(m) - \upmu y^H(m) \left[ \x(m) -
\frac{1}{N}\sum\limits_{i=1}^N \x_i(m) \right] \nonumber \\
&-& \frac{1}{N}\sum\limits_{i=1}^N\w_i(m) + \frac{1}{N}, \label{eqn:LMS_f_2}\\
\w(m+1) &=& \w(m) - \upmu y^H(m) \left[\x(m) -
\frac{y(m)}{\sum_{i=1}^N \w_i(m)}\right]
\label{eqn:lms_grad_alg_s_all_2}
\end{eqnarray}
где (\ref{eqn:LMS_f_2}) соответствует (\ref{eqn:LMS_f_1}), а
(\ref{eqn:lms_grad_alg_s_all_2}) соответствует
(\ref{eqn:lms_grad_alg_s_all_1}). Нетрудно видеть, что в обновлении
весов адаптивного фильтра участвует не сам сигнал $\x(m)$, как это
имело бы место в случае алгоритма LMS без ограничений, а некоторый
сигнал $\x^\prime(m) = \x(m) + \upnu\s$, где $\upnu$~--- комплексный
скаляр. Известно, что присутствие полезного сигнала $\s$ в обучающей
выборке помехи приводит к замедлению настройки адаптивного фильтра
на подавление этой помехи. Поэтому естественным является вывод о
том, что чем меньше абсолютная величина скаляра $\upnu$, тем быстрее
будет проходить адаптация. Порядок величины $\upnu$ может быть
оценен следующим способом. Положим, что помеха $\bm{\upzeta}$
некоррелирована с сигналом $\s$. В этом случае при оптимальной
инициализации обоих алгоритмов $\w(0) = \s/N$ \cite{Fro72},
ограничение выполняется автоматически при любом $m$. Это приводит к
тому, что (\ref{eqn:LMS_f_2}) и (\ref{eqn:lms_grad_alg_s_all_2})
могут быть с хорошей точностью представлены в виде
(\ref{eqn:LMS_f_3}) и (\ref{eqn:lms_grad_alg_s_all_3})
соответственно:
\begin{eqnarray}
\w(m+1) &=& \w(m) - \upmu y^H(m) \left[ \x(m) -
\frac{1}{N}\sum\limits_{i=1}^N \x_i(m) \right], \label{eqn:LMS_f_3}\\
\w(m+1) &=& \w(m) - \upmu y^H(m) \left[\x(m) - y(m)\right]
\label{eqn:lms_grad_alg_s_all_3}
\end{eqnarray}
Предположим, что адаптация в (\ref{eqn:lms_grad_alg_s_all_3}) не
приводит к заметному подавлению помехи, что эквивалентно
утверждению, что $\w(m) \approx \s/N$ для любого $m$. В этом случае
\begin{equation} \label{eqn:lms_y_m_2}
y(m) = \w^H \x \simeq \frac{1}{N}\sum\limits_{i=1}^N \x_i(m)
\end{equation}
В предположении о том, что $N$ четно и достаточно велико, обучающие
выборки не содержат полезного сигнала, а помеха состоит из одной
составляющей, величина
\begin{equation} \label{eqn:lms_x_bar_1}
\overline{x}_N(m) = \frac{1}{N}\sum\limits_{i=1}^N \x_i(m)
\end{equation}
может быть аппроксимированна следующим выражением \cite{Mon04}:
\begin{equation} \label{eqn:lms_x_bar_2}
\overline{x}_N(m) \approx \frac{1}{N}\sum\limits_{i=1}^N
\bm{\upzeta}_i(m) = \frac{1}{N}\sum\limits_{i=0}^{N-1} \exp \left(
j\upvarphi \left( i - \frac{N-1}{2} \right) \right)
\end{equation}
где $\upvarphi$~--- угол прихода помехи. Использование формул Эйлера
приводит к представлению (\ref{eqn:lms_x_bar_2}) в виде комплексного
тригонометрического ряда:
\begin{eqnarray} \label{eqn:lms_x_bar_3}
\overline{x}_N(m) & \approx & \frac{1}{N} \left[
\cos\left(-\frac{N-1}{2}\upvarphi\right) +
j\sin\left(-\frac{N-1}{2}\upvarphi\right) + \ldots + \right. \nonumber \\
&+& \cos\left(-\frac{3}{2}\upvarphi\right) +
j\sin\left(-\frac{3}{2}\upvarphi\right) +
\cos\left(-\frac{1}{2}\upvarphi\right) +
j\sin\left(-\frac{1}{2}\upvarphi\right)\nonumber \\
&+& \cos\left(\frac{1}{2}\upvarphi\right) +
j\sin\left(\frac{1}{2}\upvarphi\right)
+\cos\left(\frac{3}{2}\upvarphi\right) +
j\sin\left(\frac{3}{2}\upvarphi\right) + \ldots + \nonumber \\
&+&\left. \cos\left(\frac{N-1}{2}\upvarphi\right) +
j\sin\left(\frac{N-1}{2}\upvarphi\right) \right]
\end{eqnarray}
Использование свойств четности и нечетности функций косинуса и
синуса приводит к следующему упрощению (\ref{eqn:lms_x_bar_3})
\begin{eqnarray} \label{eqn:lms_x_bar_4}
\overline{x}_N(m) & \approx & \frac{2}{N} \left[
\cos\left(\frac{1}{2}\upvarphi\right) +
\cos\left(\frac{3}{2}\upvarphi\right) + \ldots +
\cos\left(\frac{N-1}{2}\upvarphi\right) \right]
\end{eqnarray}
Применение формулы для суммы тригонометрических функций \cite{Pou99}
позволяет получить выражение для $\overline{x}_N(m)$ в замкнутой
форме:
\begin{equation} \label{eqn:lms_x_bar_5}
\overline{x}_N(m) \approx
\frac{1}{N}\frac{\sin\frac{N\upvarphi}{2}}{\sin\frac{\upvarphi}{2}}
\end{equation}
Поскольку следующие условия выполняются:
\begin{eqnarray} \label{eqn:lms_x_bar_6}
\left| \sin\frac{N\upvarphi}{2} \right| &\leq& 1, \ \ \forall N,
\upvarphi \\
\left| \frac{1}{\sin\frac{\upvarphi}{2}} \right| &<& \infty, \ \
\forall \upvarphi \neq \uppi k, k = 0,1,2,\ldots
\end{eqnarray}
то асимптотически при $N \rightarrow \infty$ мы имеем следующий
результат:
\begin{equation} \label{eqn:lms_x_bar_7}
\lim\limits_{N \rightarrow \infty} \left| \overline{x}_N(m) \right|
\leq \lim\limits_{N \rightarrow
\infty}\frac{1}{N}\left|\frac{1}{\sin\frac{\upvarphi}{2}}\right| = 0
\end{equation}

Откуда следует, что $\lim\limits_{N \rightarrow \infty}
\left|\overline{x}_N(m)\right| = 0$, т. е. при больших размерах
антенной решетки алгоритмы (\ref{eqn:lms_grad_alg_s_all_1}) и
(\ref{eqn:LMS_f_1}) эквивалентны по скорости сходимости алгоритму
LMS без ограничений. Однако на практике, обновление весов
адаптивного фильтра на каждом шаге в среднем приводит к увеличению
ОСПШ на выходе фильтра. Отсюда можно сделать следующий вывод:
\begin{equation} \label{eqn:lms_x_bar_8}
\left|E\{ y(m) \}\right| \leq \left|E\{ \overline{x}_N(m) \}\right|
\end{equation}

Это означает, что в среднем абсолютная величина упомянутого выше
коэффицента $\upnu$ , отражающего степень замедления алгоритма LMS с
условием, для алгоритма (\ref{eqn:lms_grad_alg_s_all_1}) на каждом
шаге адаптации меньше или равна значению этой величины для алгоритма
(\ref{eqn:LMS_f_1}). Т. е. скорость адаптации алгоритма с
квадратичным условием в среднем должна быть больше или равна
скорости адаптации алгоритма с линейным ограничением. Поэтому мы
считаем применение алгоритма с квадратичным условием более
целесообразным и в разделе \ref{sec:LMS_sq:mod} приводим результаты
моделирования, подтверждающие верность приведенных выше рассуждений.
Кроме этого, мы также демонстрируем результаты моделирования
ситуации, когда полезный сигнал присутствует в обучающей выборке. В
этой ситуации член отвечающий за выполнения условия неподавления
полезного сигнала в адаптивном алгоритме LMS с квадратичным
ограничением имеет существенно большую величину, чем в алгоритме LMS
с линейным ограничнием. Это приводит к значительному снижению
проникновения помехи в вектор весов и выгодно отличает алгоритм LMS
с квадратичным ограничением от алгоритмов без ограничения или с
линейным ограничением.

\section{Заключение} \label{chap:sec:Ch3:concl}

В данной главе рассмотрены результаты синтеза предлагаемых
адаптивных алгоритмов АСДЦ.

Во--первых, в разделе \ref{sec:SMI_reg} синтезирован итеративный
алгоритм оптимизации величины коэффициента регуляризации алгоритма
РОКМ, предложенный в \cite{Ore07c,Ore08b,Ore08c}. Целью такой
оптимизации является ослабление эффекта подавления полезного сигнала
в ситуации, когда этот сигнал проникает в обучающую выборку.
Сформулирована и решена задача условной оптимизации коэффициента
регуляризации, основанная на максимизации эмпирического отношения
Рэлея. Для решения задачи применен метод линеаризации выражения для
весов адаптивного фильтра. Ошибки, вызванные такой линеаризацией,
препятствуют ее решению в случае, когда оптимальное значение
коэффициента регуляризации относительно велико. Поэтому на основе
полученного решения предложен итеративный алгоритм вычисления
коэффициента регуляризации. Данный итеративный алгоритм компенсирует
ошибки использованной линеаризации.

Во--вторых, в разделе \ref{sec:LMS_sq} синтезирован алгоритм LMS с
квадратичным ограничением, предложенный в \cite{Ore07a,Ore08a}.
Сформулирована задача безусловной оптимизации весов оптимального
фильтра на фоне коррелированного шума с известной ковариационной
матрицей. Затем эта задача переформулирована в задачу условной
оптимизации с квадратичным условием. Найден градиент
соответствующего функционала качества, который в отличие от многих
стандартных алгоритмов содержит информацию именно о квадратичном, а
не о линейном условии. Данный градиент аппроксимирован с помощью
стандартных выражений для одношаговых оценок ковариационной матрицы.
Найдена аппроксимация множителя Лагранжа имеющая более простой вид,
чем в других адаптивных алгоритмах полагающихся на квадратическое
ограничение. На основе найденных аппроксимаций сконструирован
итеративный градиентный алгоритм. Показано, что алгоритмы с линейным
и квадратичным ограничением при определенных условиях асимптотически
эквивалентны алгоритму без ограничений, если помеха и полезный
сигнал не коррелированы. Кроме этого, даны некоторые интуитивные
соображения, подтверждающие тот факт, что при конечной размерности
обработки синтезированный алгоритм с квадратичным условием имеет
скорость сходимости большую или равную скорости сходимости алгоритма
с линейным ограничением.

\chapter{Результаты моделирования синтезированных алгоритмов} \label{chap:Ch4}

В данной главе мы представляем результаты моделирования предлагаемых
адаптивных алгоритмов предназначенных для подавления помех с
неизвестной ковариационной матрицей в системе АСДЦ.

\section{Моделирование алгоритма оптимизации уровня регуляризации в алгоритме РОКМ} \label{sec:SMI_reg:mod}

В данной части мы приводи результаты моделирования алгоритма
оптимизации уровня регуляризации в алгоритме РОКМ синтезированного в
разделе \ref{sec:SMI_reg}. Сравнение качества работы алгоритма РОКМ
с коэффициентом регуляризации вычисляемым в соответствии с
алгоритмом на Рис. \ref{fig:it_rokm}, фиксированном на 10 дБ выше
уровня шума и идеальным алгоритмом выбеливания помехи с полностью
известной корреляционной матрицей произведено методом компьютерного
моделирования. Исследованы характеристики указанных алгоритмов при
различных ОСПШ на входе, наличии или отсутствии сигнала от цели в
обучающей выборке, различном количестве отсчетов (импульсов в пачке)
$N$ в тестируемой выборке и различном количестве отсчетов $M$ в
обучающей выборке. Результаты моделирования говорят о том, что
предлагаемый алгоритм оценивания $\upalpha$ действительно позволяет
ослабить эффект подавления сигнала от цели со стандартным алгоритмом
РОКМ. Для моделирования были использованы следующие параметры
помеховой обстановки: частота повторения импульсов равна 20 кГц,
помеха состоит из двух компонент со средними частотами 0 и 1003 Гц и
шириной спектрального пика 500 Гц, спектральная огибающая спектра
помехи --- Гауссовская, ОСШ равно 10 дБ. Доплеровский сдвиг частоты
полезного сигнала равен 4 кГц. Распределение амплитуды сигнала от
цели в обучающей выборке принято Релеевским, а средняя мощность
принята равной мощности полезного сигнала в тестируемой выборке.
Количество итераций $T$ при определении оптимального $\upalpha$
равно 3.

\begin{figure*}[tbp]
\centering \subfigure[ОСПШ 20 дБ]{
\label{fig:reg_aut_1:20} 
\includegraphics[width = 5.2cm]{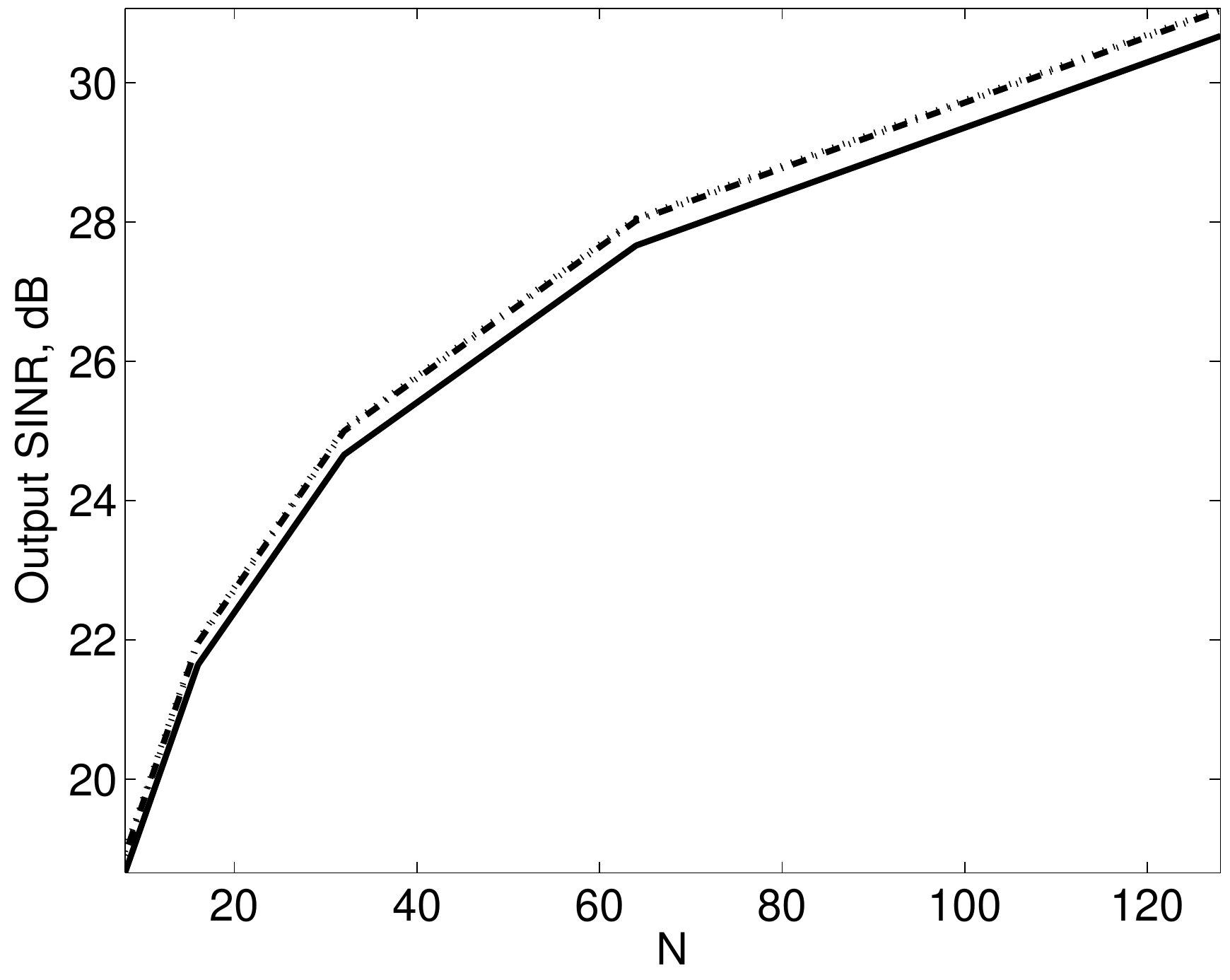}}
\hfill \subfigure[ОСПШ 10 дБ]{
\label{fig:reg_aut_1:10} 
\includegraphics[width = 5.2cm]{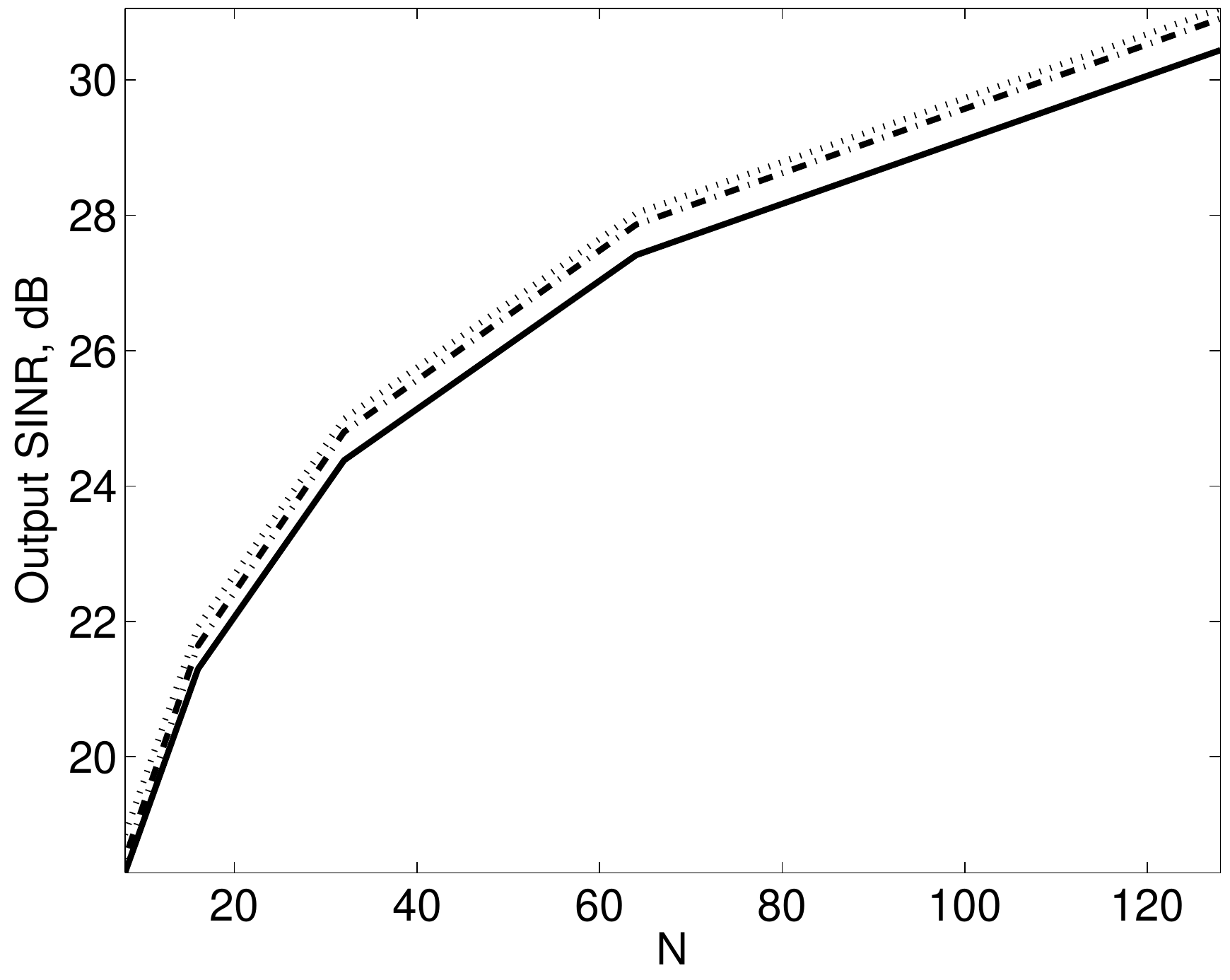}}
\hfill \subfigure[ОСПШ 0 дБ]{
\label{fig:reg_aut_1:0} 
\includegraphics[width = 5.2cm]{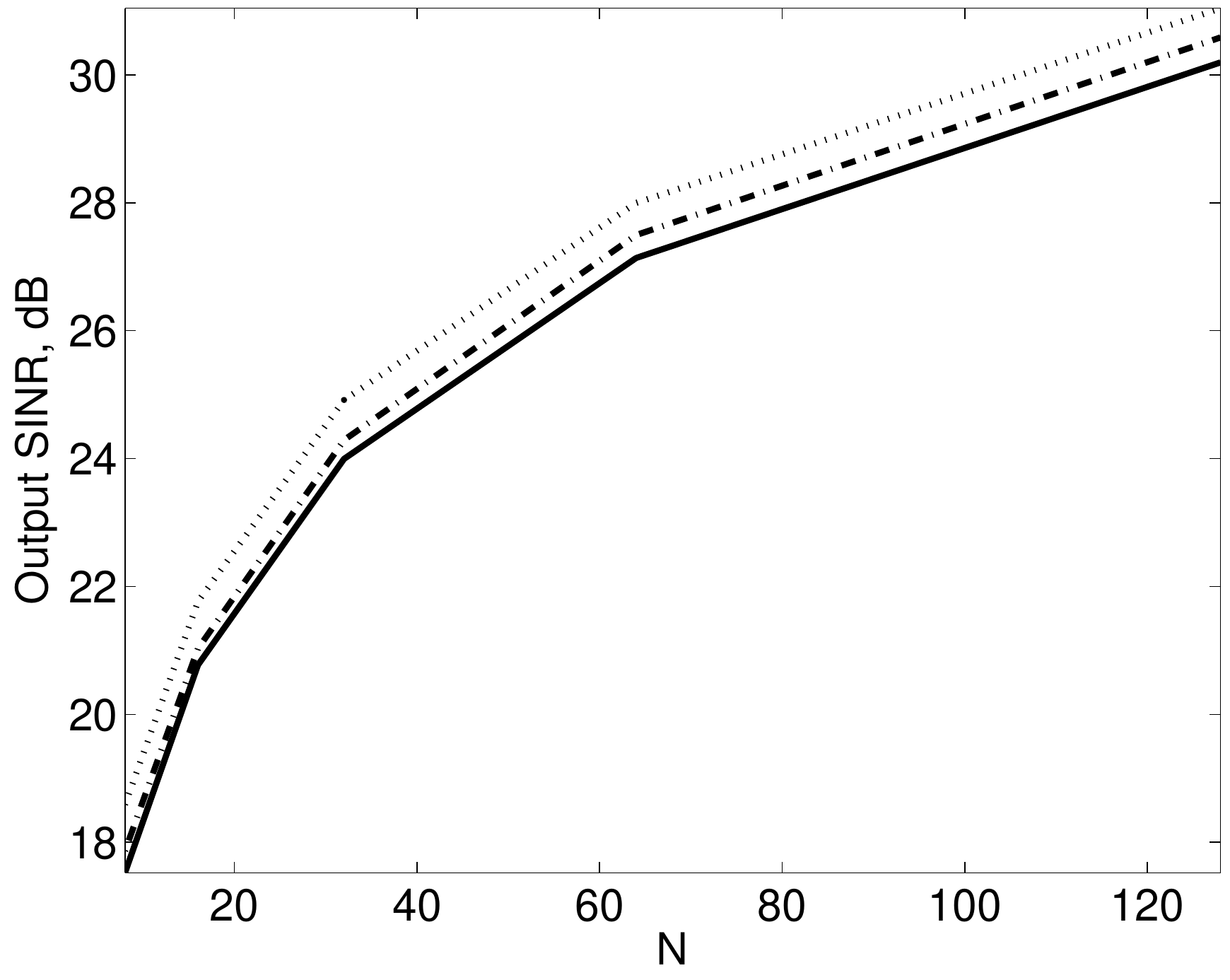}}
\hfill \subfigure[ОСПШ -10 дБ]{
\label{fig:reg_aut_1:-10} 
\includegraphics[width = 5.2cm]{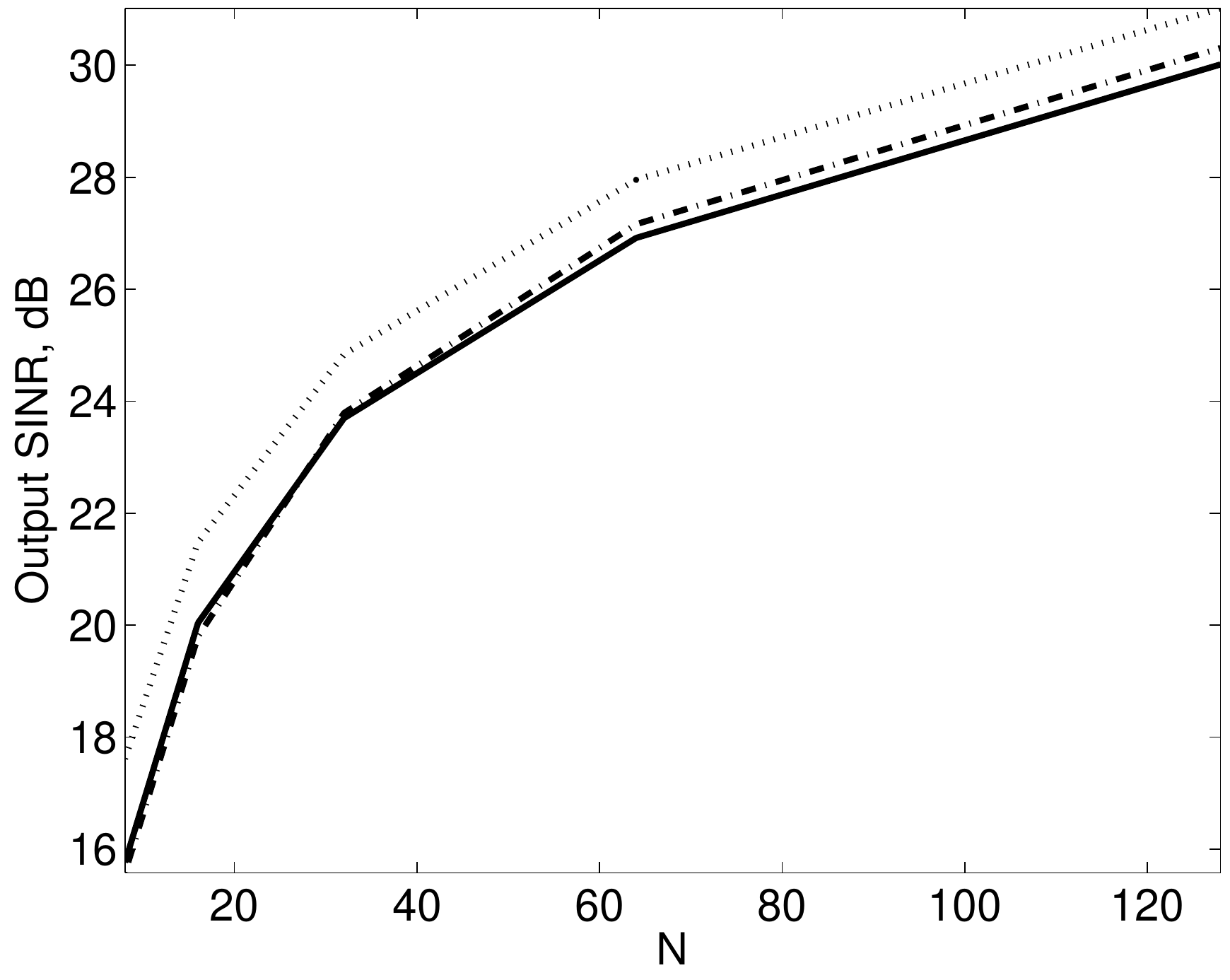}}
\hfill \subfigure[ОСПШ -20 дБ]{
\label{fig:reg_aut_1:-20} 
\includegraphics[width = 5.2cm]{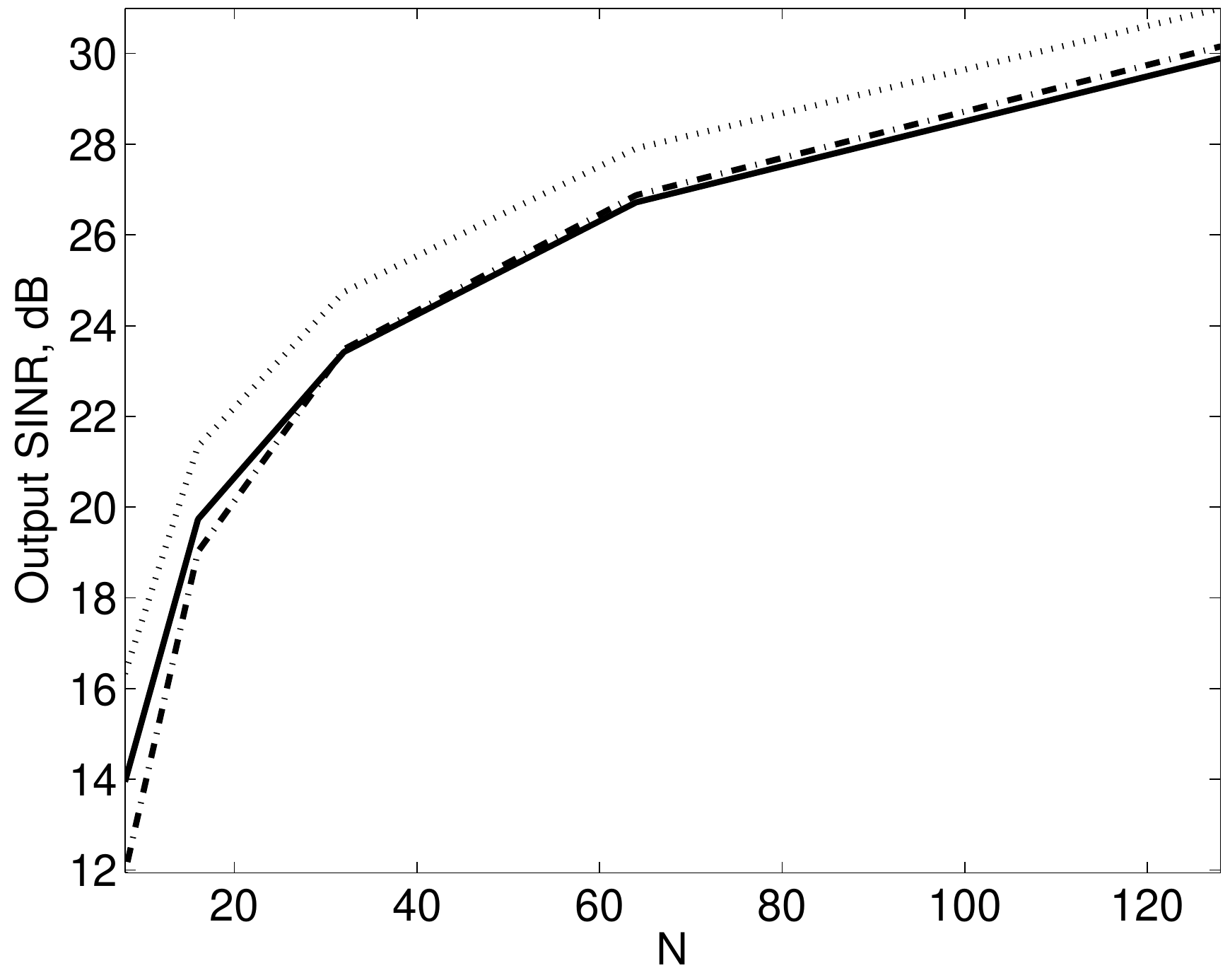}}
\hfill \subfigure[ОСПШ -40 дБ]{
\label{fig:reg_aut_1:-40} 
\includegraphics[width = 5.2cm]{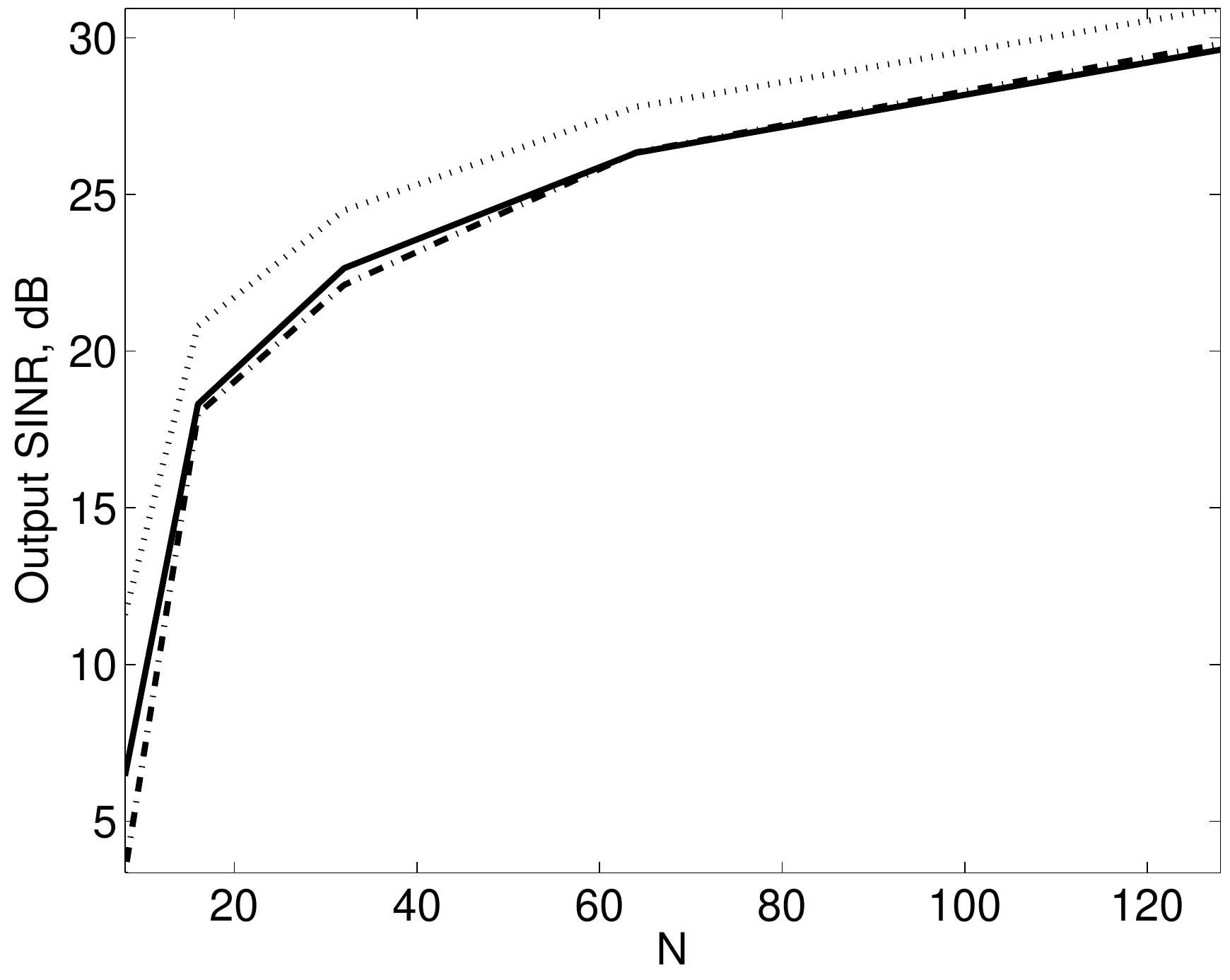}}
\hfill \subfigure[ОСПШ -60 дБ]{
\label{fig:reg_aut_1:-60} 
\includegraphics[width = 5.2cm]{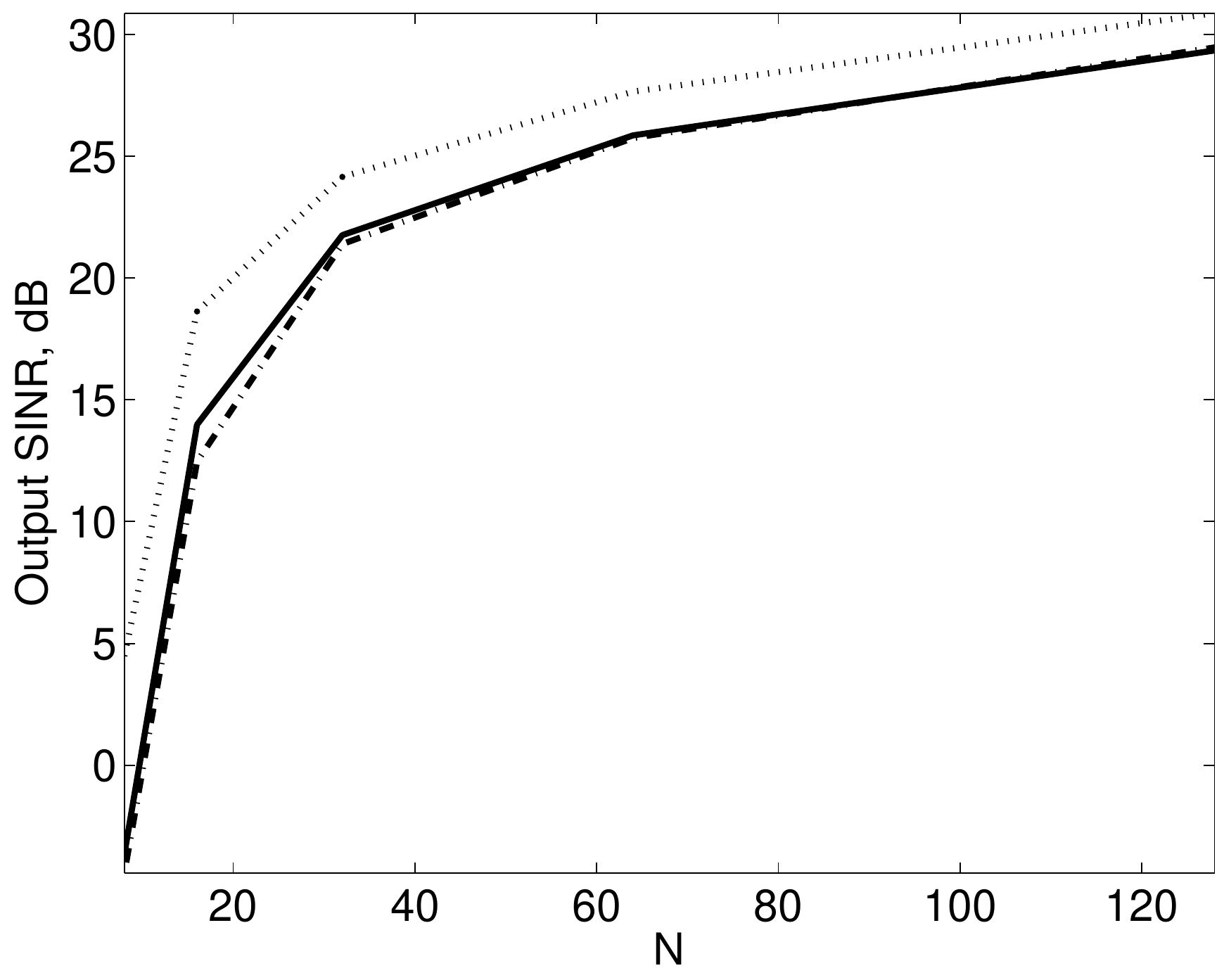}}
\hfill \subfigure[ОСПШ -80 дБ]{
\label{fig:reg_aut_1:-80} 
\includegraphics[width = 5.2cm]{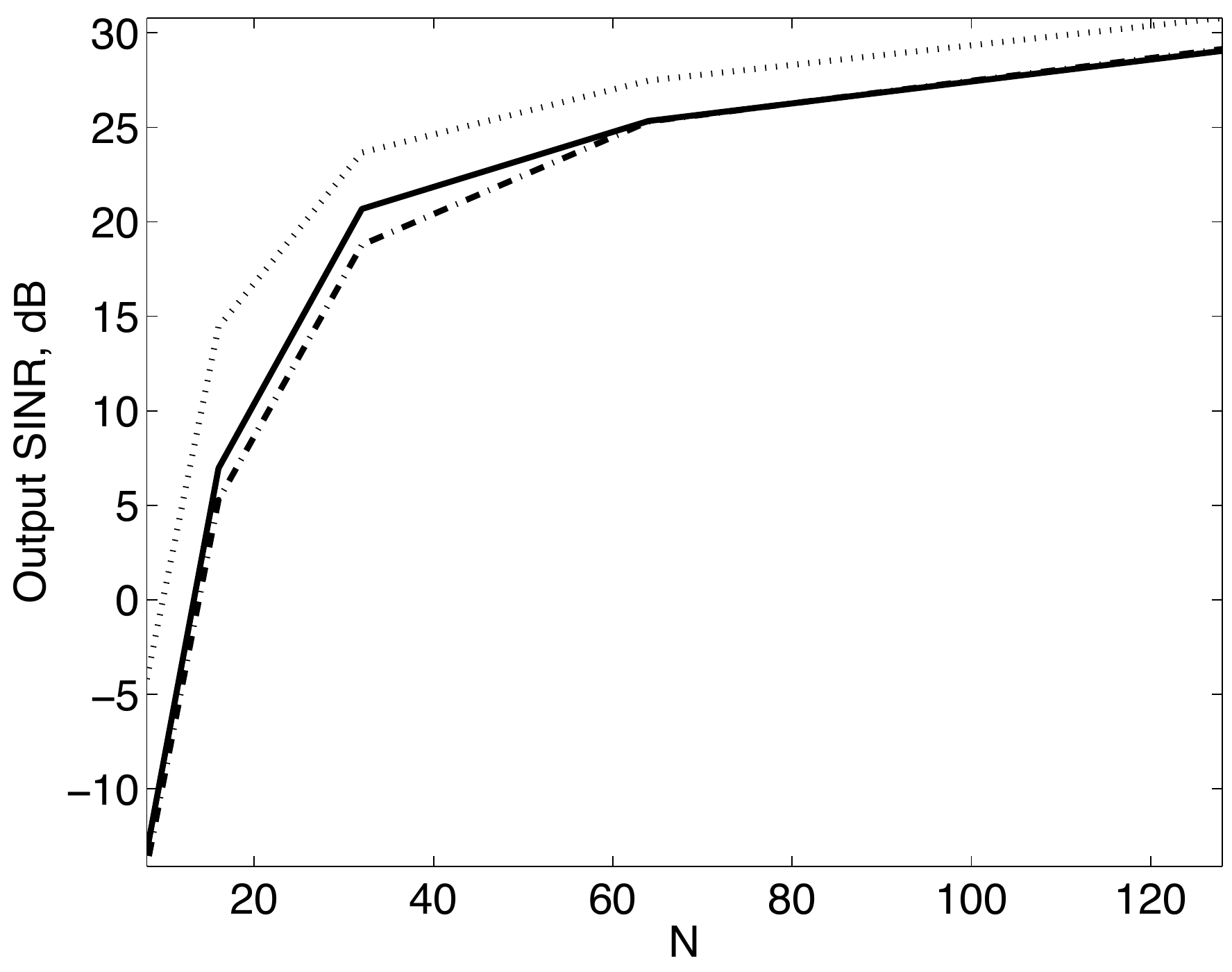}}
\caption{Зависимость ОСПШ на выходе формирователя достаточной
статистики от количества импульсов в пачке $N$, при различном
количестве усредняемых элементов разрешения $M$ и различных
значениях ОСПШ на входе. Усреднение по 500 реализаций. Пунктирная
линия обозначает идеальный случай с полностью известной
корреляционной матрицей помехи, сплошная линия обозначает
предлагаемый метод, штрих--пунктирная линия обозначает метод
\cite{Car88}. ОСПШ на входе составляет 10 дБ, $M = 0,5N$, а сигнал
от цели отсутствует в обучающей выборке.}
\label{fig:reg_aut_1} 
\end{figure*}

\begin{figure*}[tbp]
\centering \subfigure[ОСПШ 20 дБ]{
\label{fig:reg_aut_2:20} 
\includegraphics[width = 5.2cm]{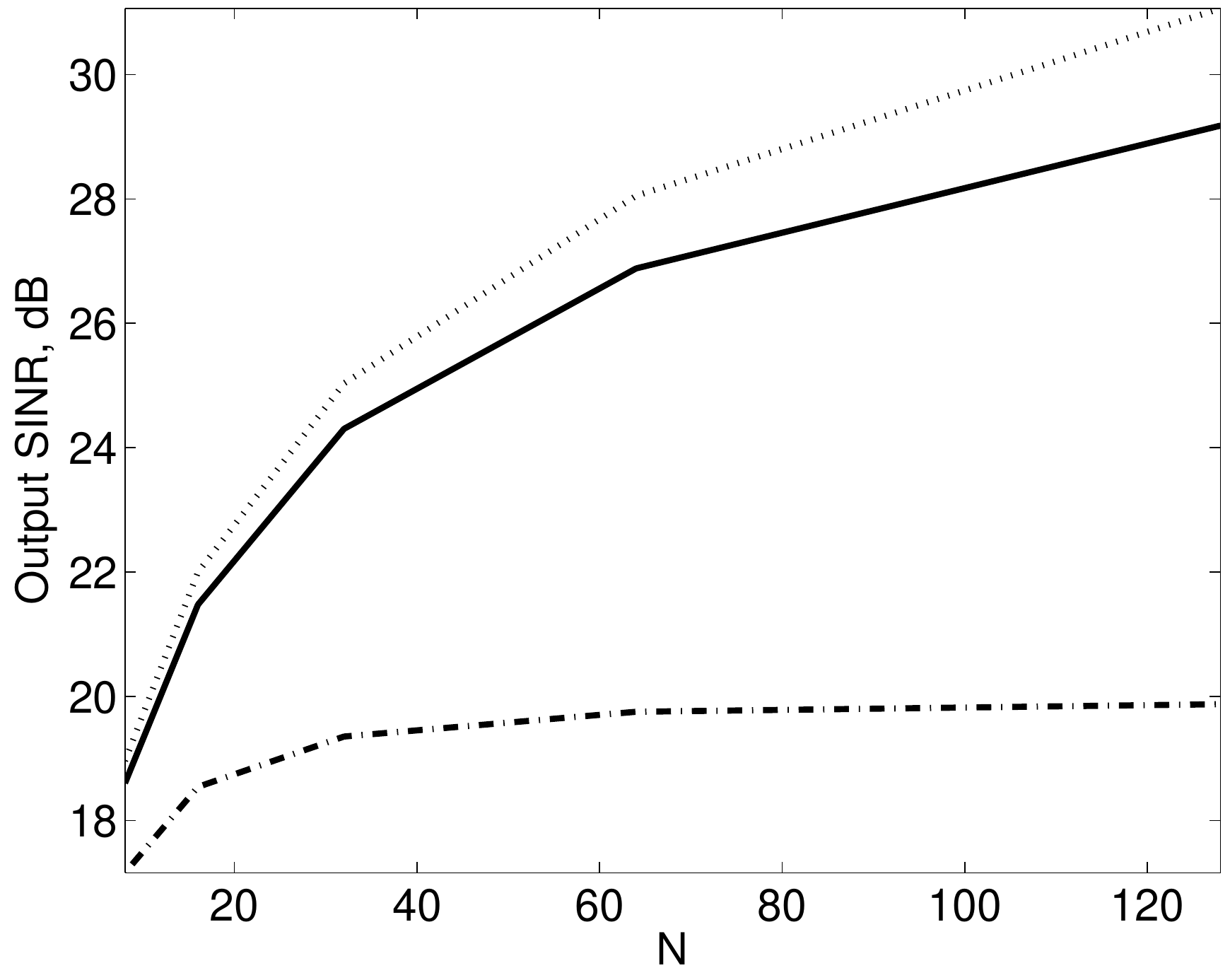}}
\hfill \subfigure[ОСПШ 10 дБ]{
\label{fig:reg_aut_2:10} 
\includegraphics[width = 5.2cm]{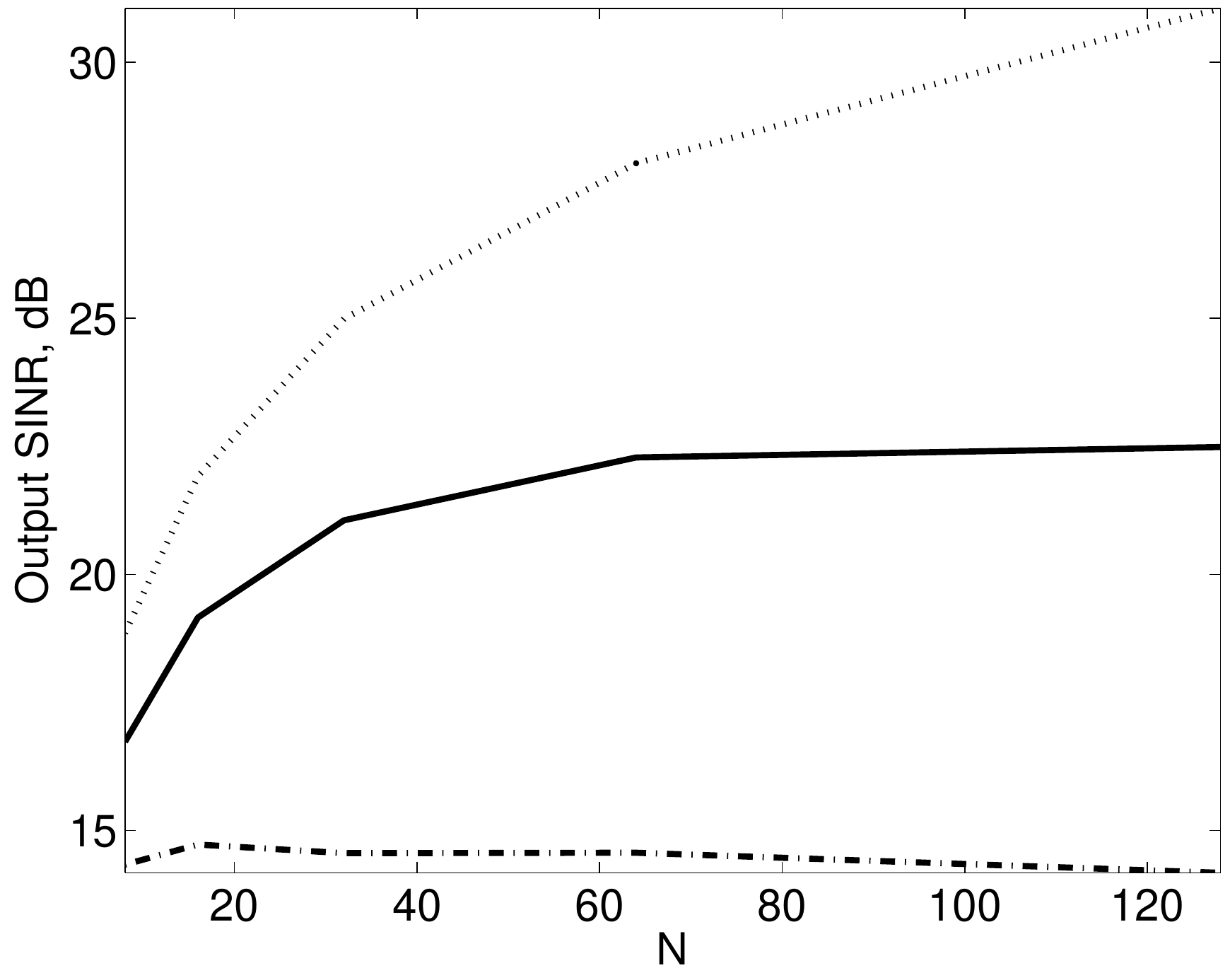}}
\hfill \subfigure[ОСПШ 0 дБ]{
\label{fig:reg_aut_2:0} 
\includegraphics[width = 5.2cm]{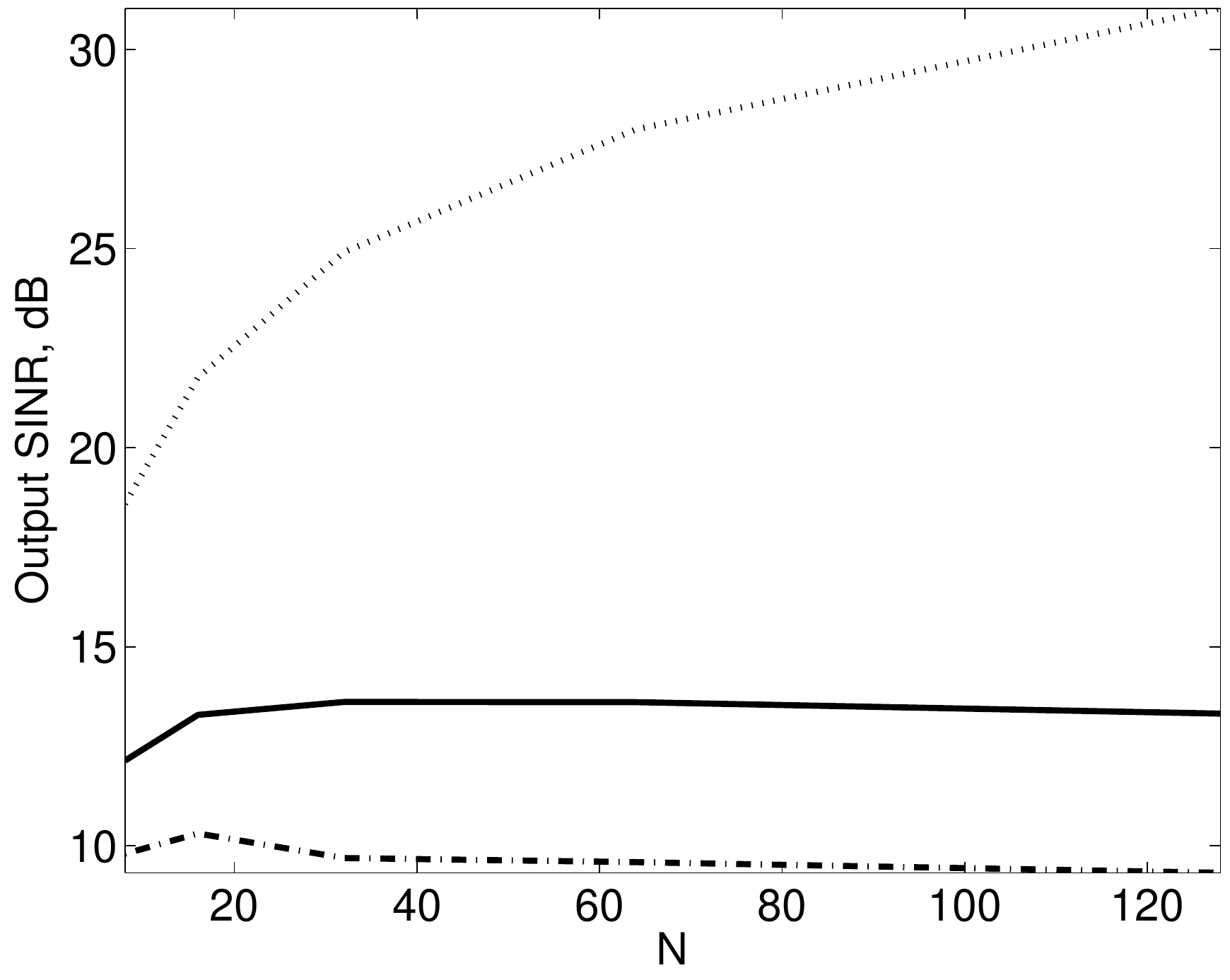}}
\hfill \subfigure[ОСПШ -10 дБ]{
\label{fig:reg_aut_2:-10} 
\includegraphics[width = 5.2cm]{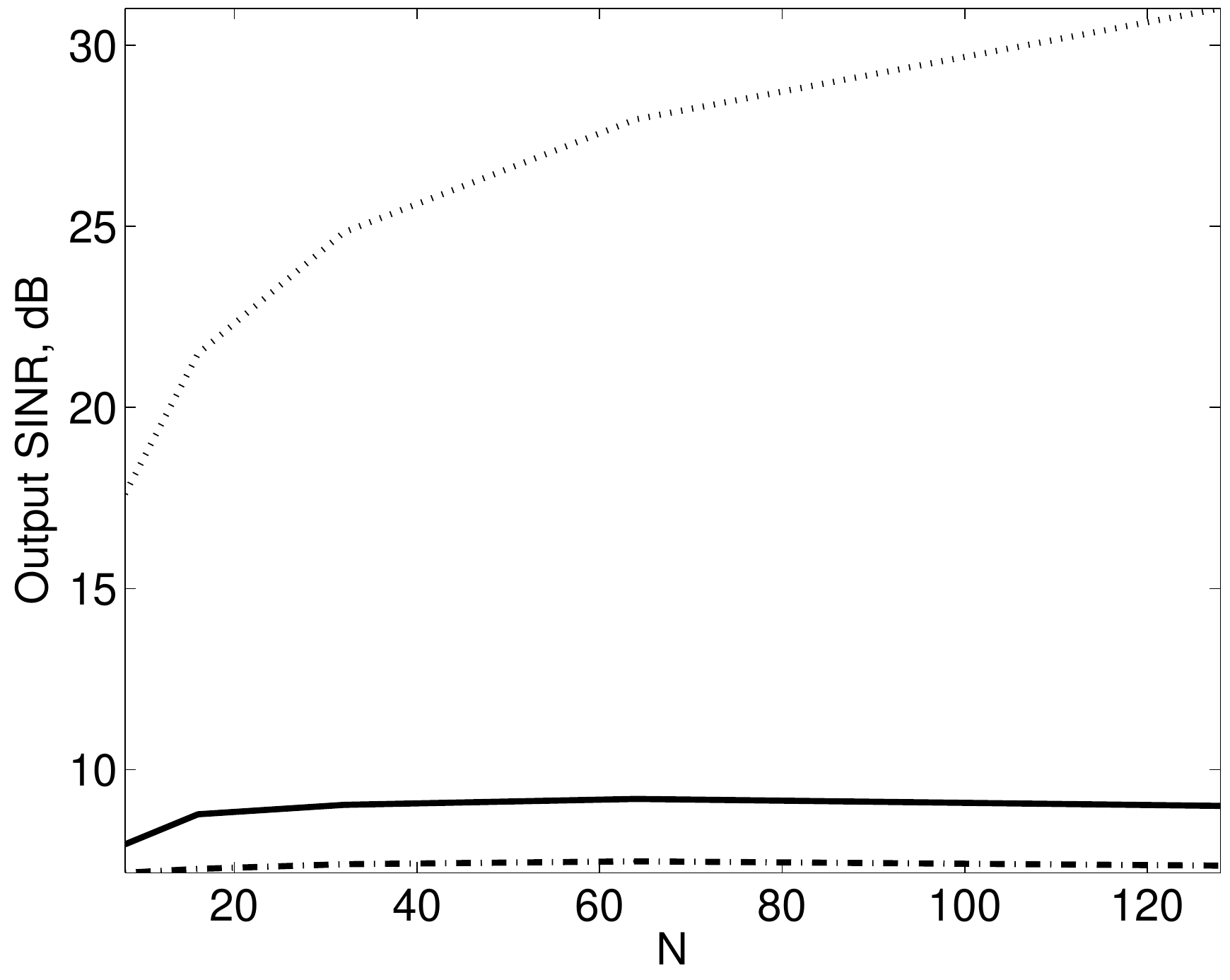}}
\hfill \subfigure[ОСПШ -20 дБ]{
\label{fig:reg_aut_2:-20} 
\includegraphics[width = 5.2cm]{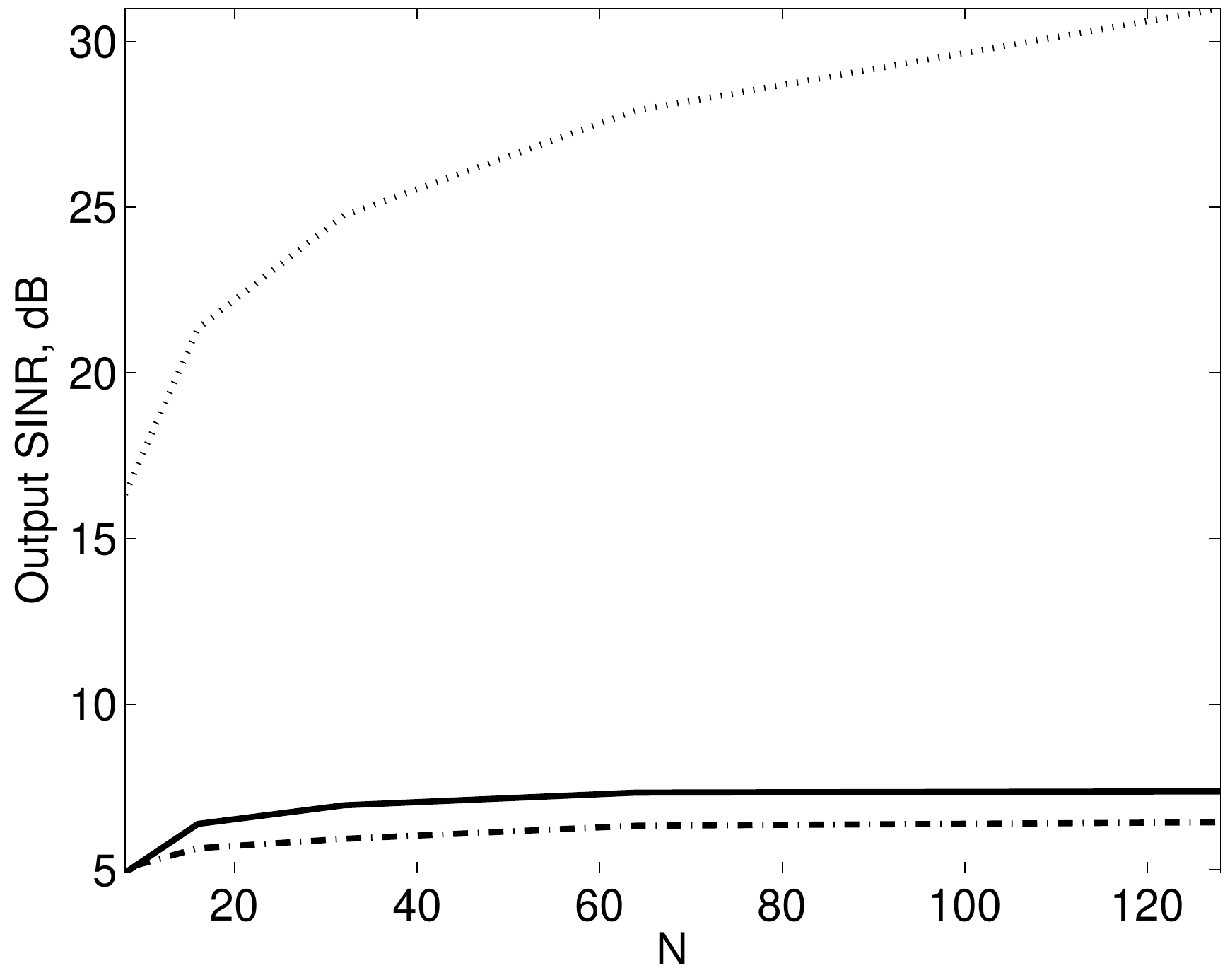}}
\hfill \subfigure[ОСПШ -40 дБ]{
\label{fig:reg_aut_2:-40} 
\includegraphics[width = 5.2cm]{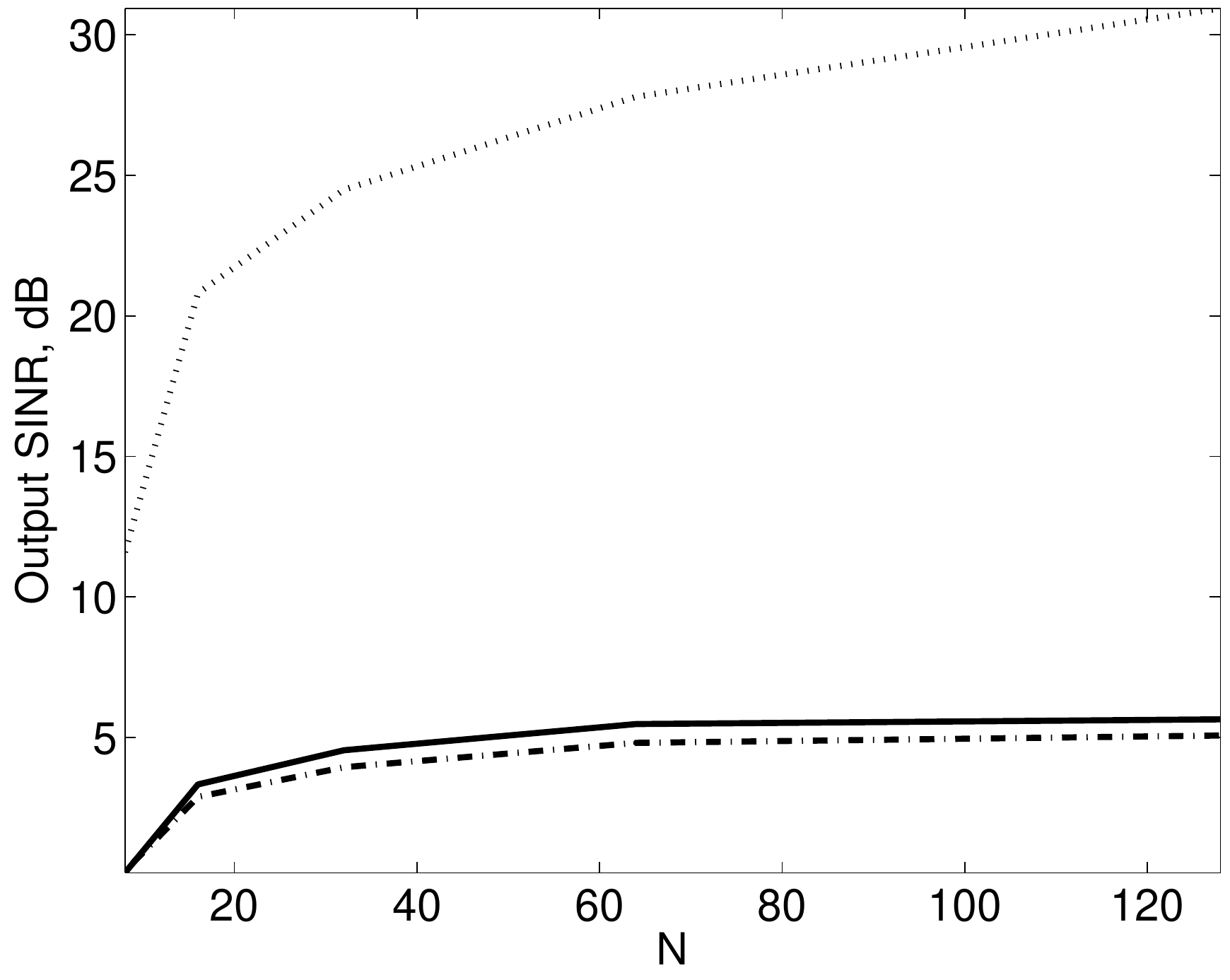}}
\hfill \subfigure[ОСПШ -60 дБ]{
\label{fig:reg_aut_2:-60} 
\includegraphics[width = 5.2cm]{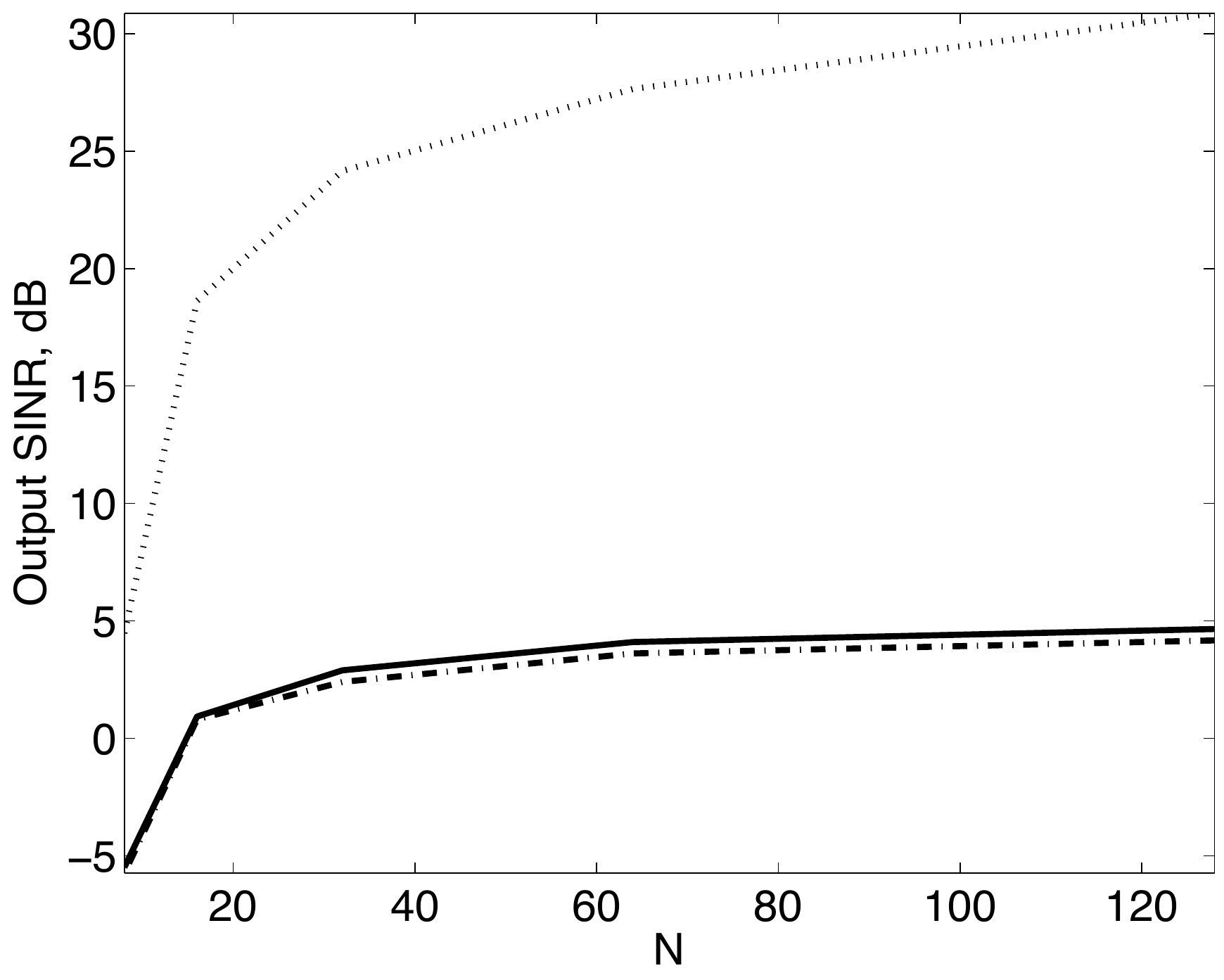}}
\hfill \subfigure[ОСПШ -80 дБ]{
\label{fig:reg_aut_2:-80} 
\includegraphics[width = 5.2cm]{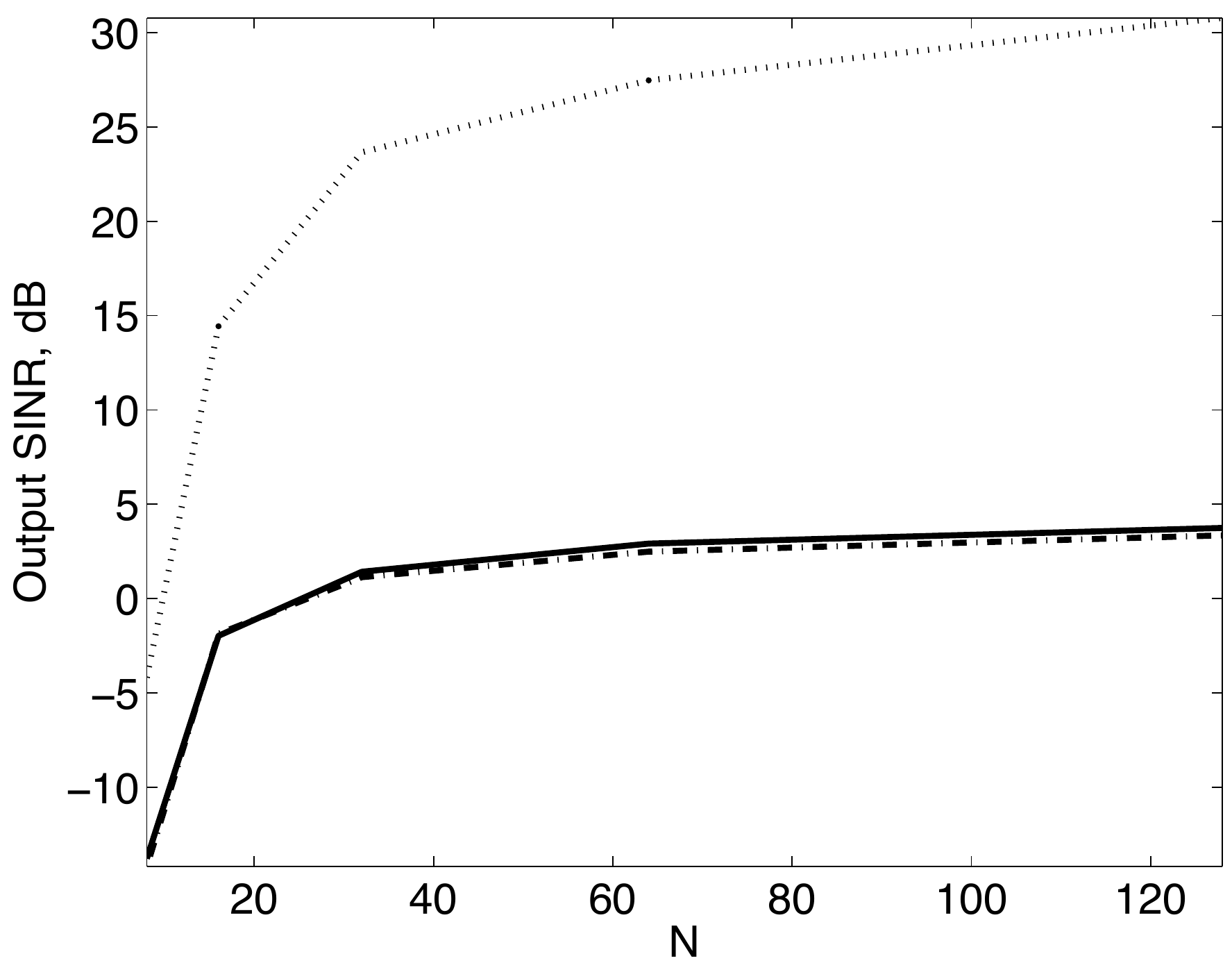}}
\caption{Зависимость ОСПШ на выходе формирователя достаточной
статистики от количества импульсов в пачке $N$, при различном
количестве усредняемых элементов разрешения $M$ и различных
значениях ОСПШ на входе. Усреднение по 500 реализаций. Пунктирная
линия обозначает идеальный случай с полностью известной
корреляционной матрицей помехи, сплошная линия обозначает
предлагаемый метод, штрих--пунктирная линия обозначает метод
\cite{Car88}. ОСШ на входе составляет 10 дБ, $M = 0,5N$, а сигнал от
цели присутствует в обучающей выборке.}
\label{fig:reg_aut_2} 
\end{figure*}

\begin{figure*}[tbp]
\centering \subfigure[ОСПШ 20 дБ]{
\label{fig:reg_aut_3:20} 
\includegraphics[width = 5.2cm]{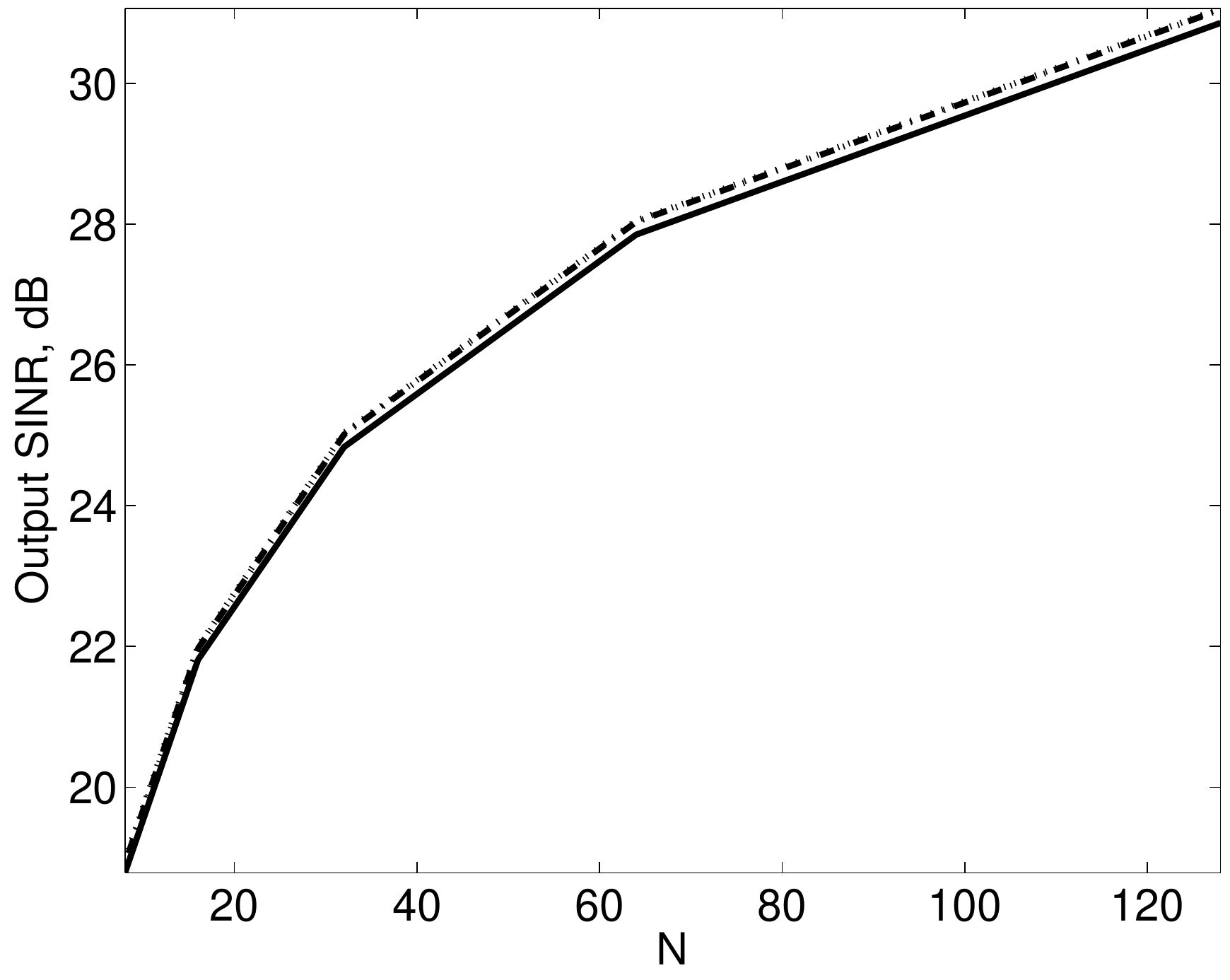}}
\hfill \subfigure[ОСПШ 10 дБ]{
\label{fig:reg_aut_3:10} 
\includegraphics[width = 5.2cm]{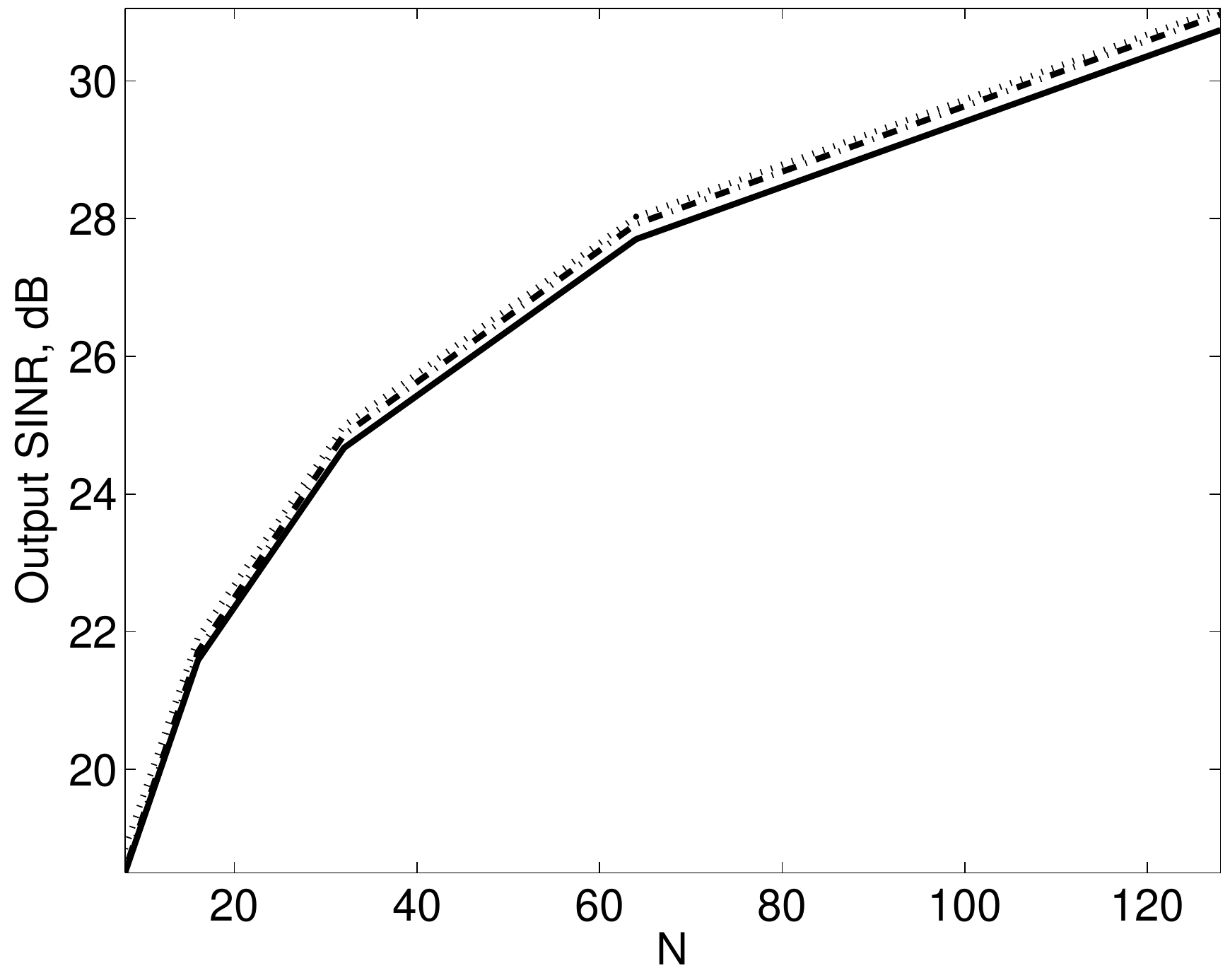}}
\hfill \subfigure[ОСПШ 0 дБ]{
\label{fig:reg_aut_3:0} 
\includegraphics[width = 5.2cm]{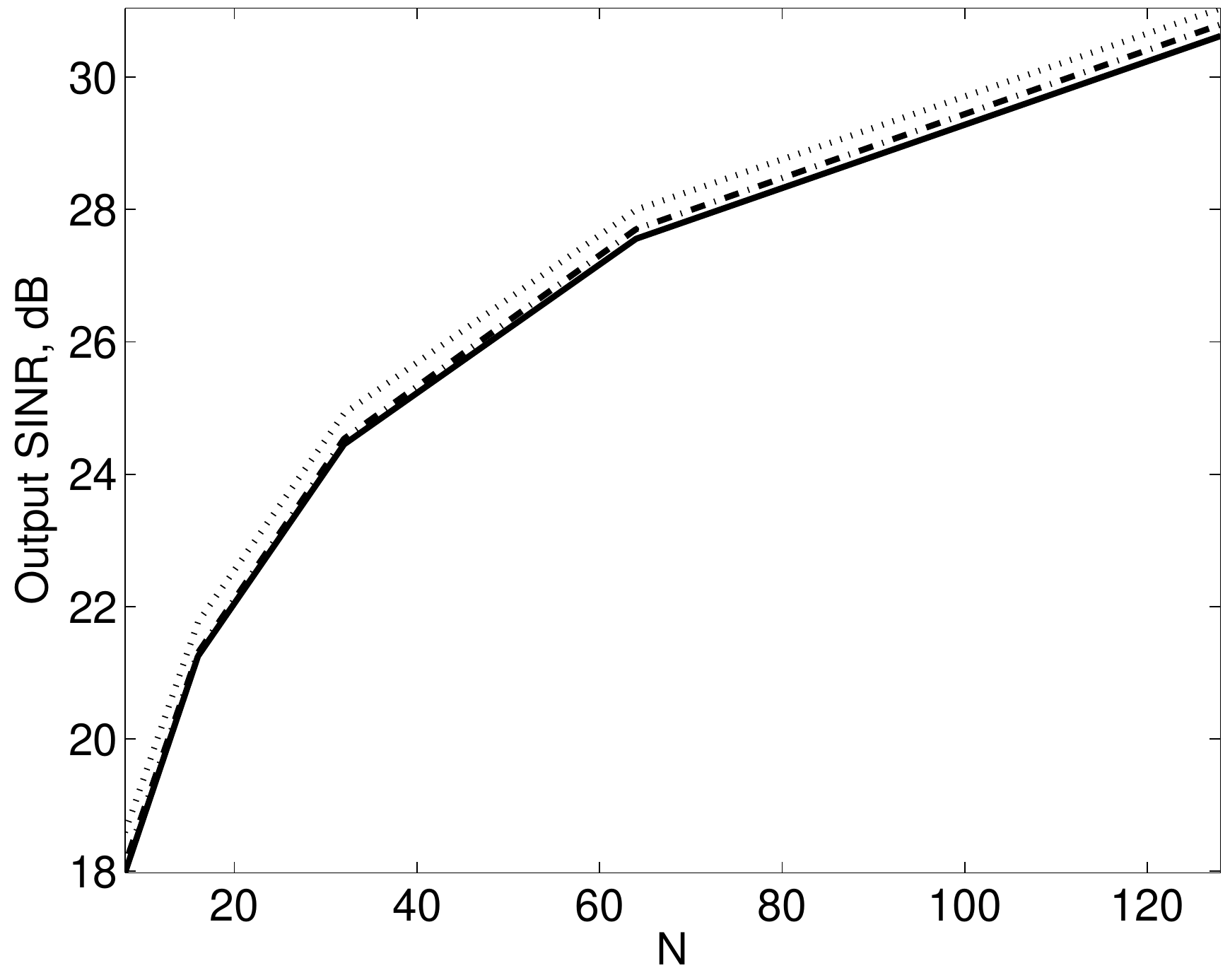}}
\hfill \subfigure[ОСПШ -10 дБ]{
\label{fig:reg_aut_3:-10} 
\includegraphics[width = 5.2cm]{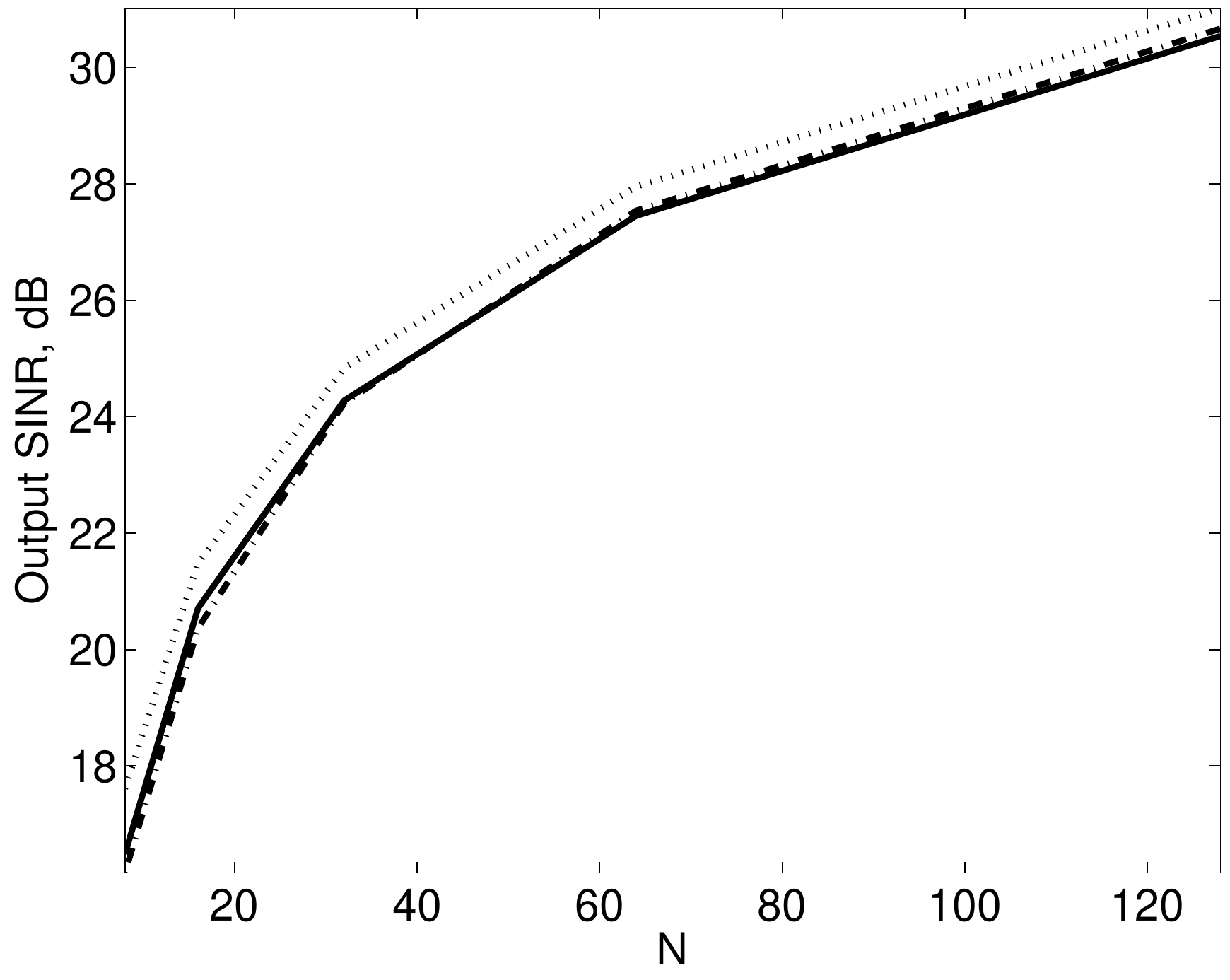}}
\hfill \subfigure[ОСПШ -20 дБ]{
\label{fig:reg_aut_3:-20} 
\includegraphics[width = 5.2cm]{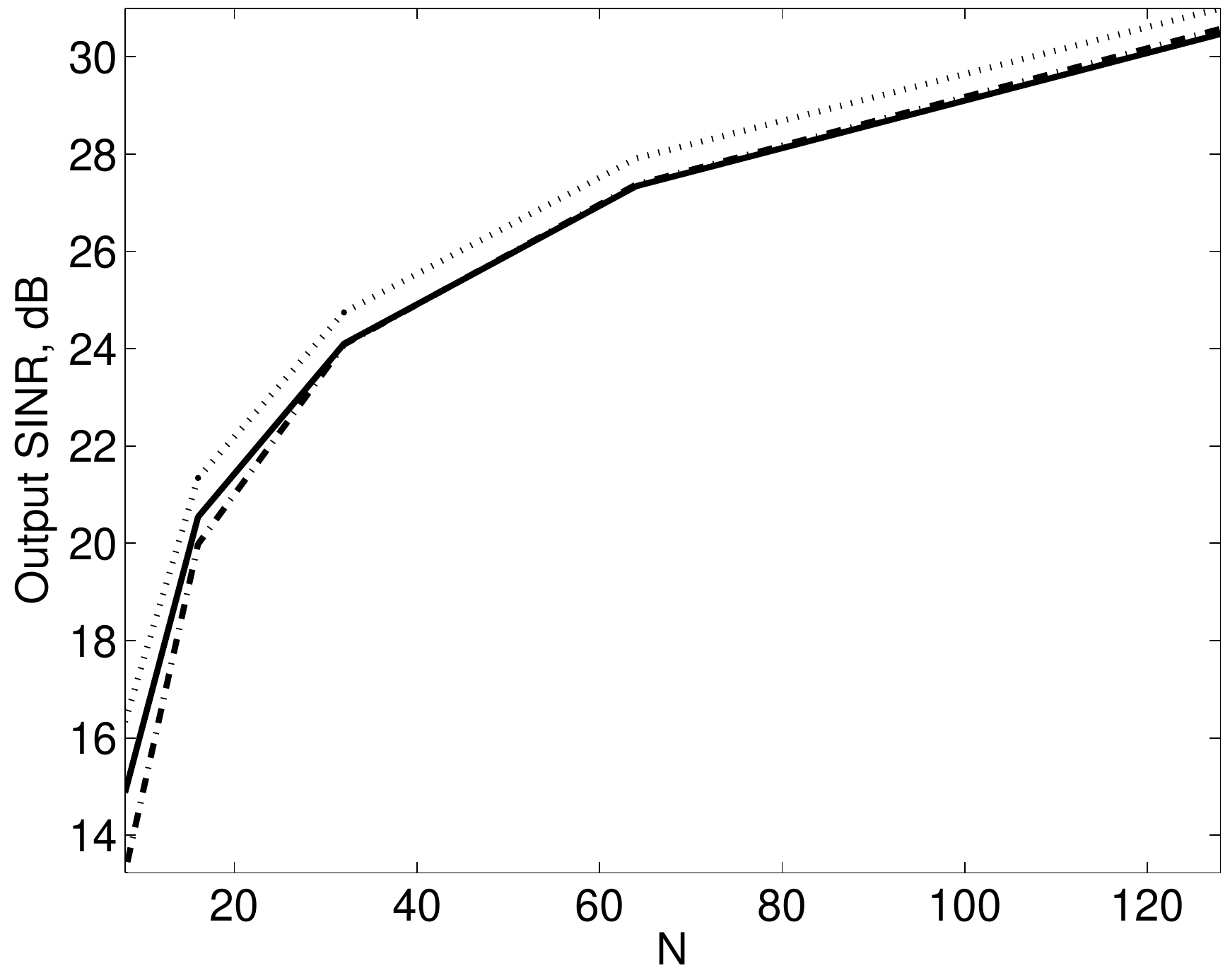}}
\hfill \subfigure[ОСПШ -40 дБ]{
\label{fig:reg_aut_3:-40} 
\includegraphics[width = 5.2cm]{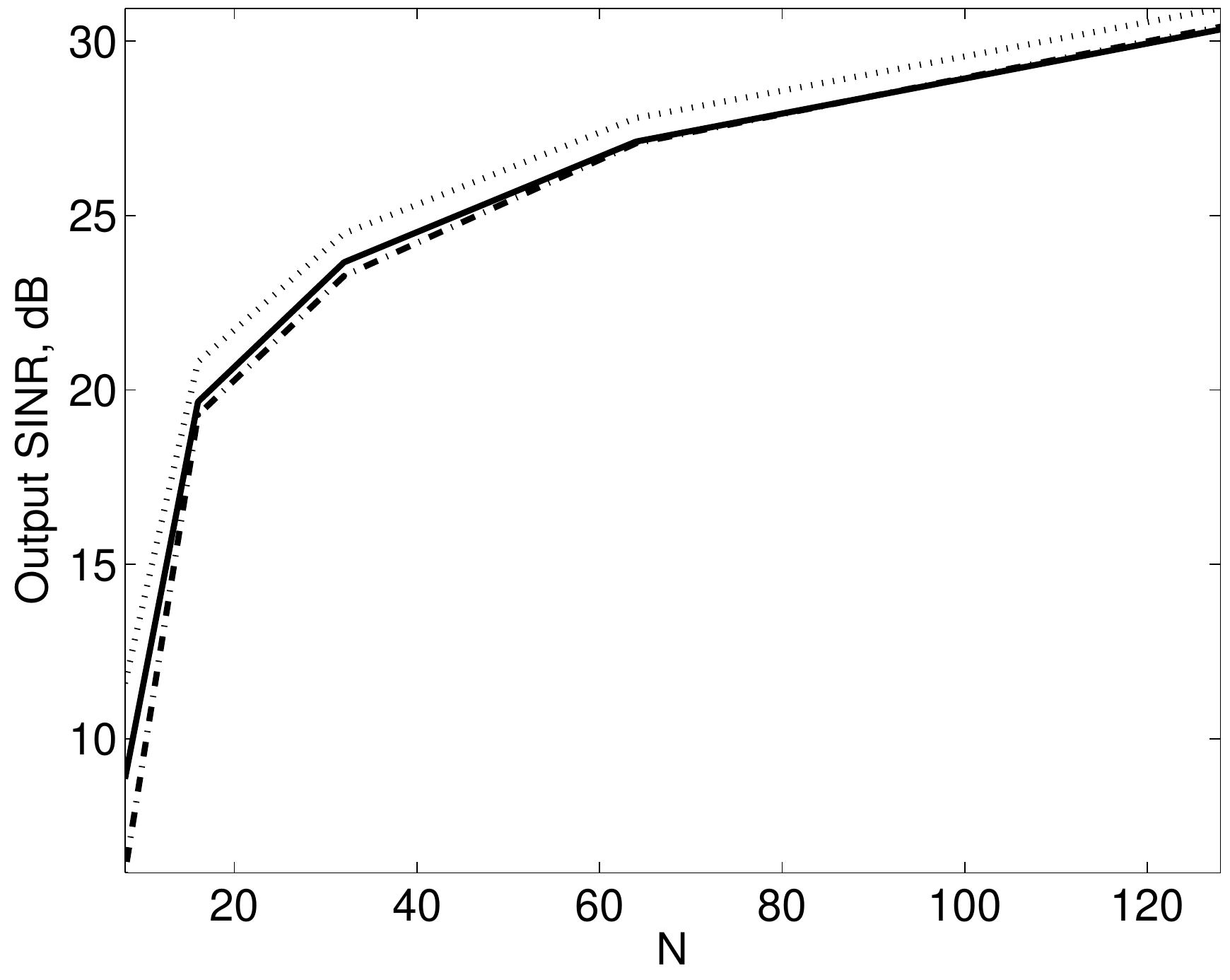}}
\hfill \subfigure[ОСПШ -60 дБ]{
\label{fig:reg_aut_3:-60} 
\includegraphics[width = 5.2cm]{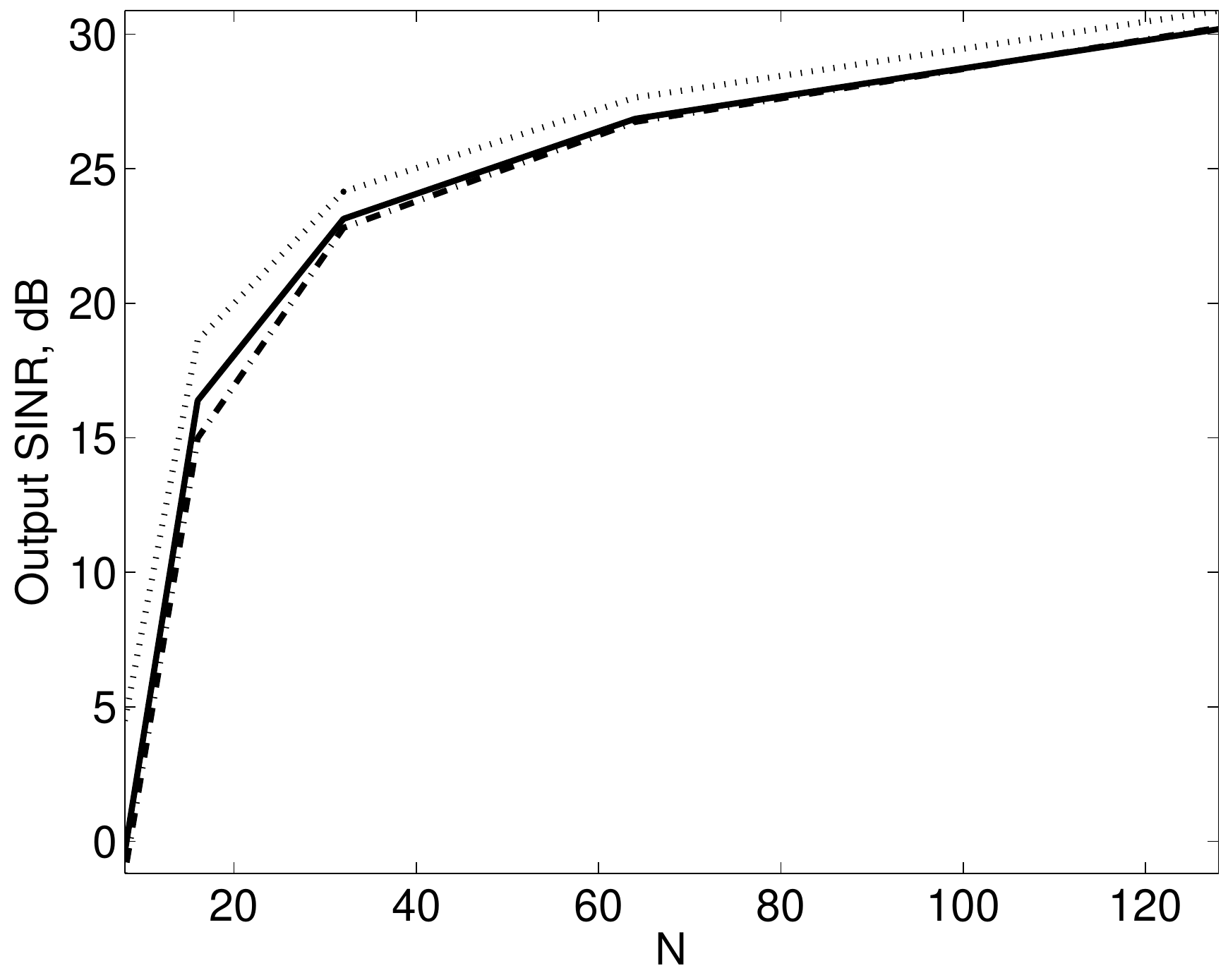}}
\hfill \subfigure[ОСПШ -80 дБ]{
\label{fig:reg_aut_3:-80} 
\includegraphics[width = 5.2cm]{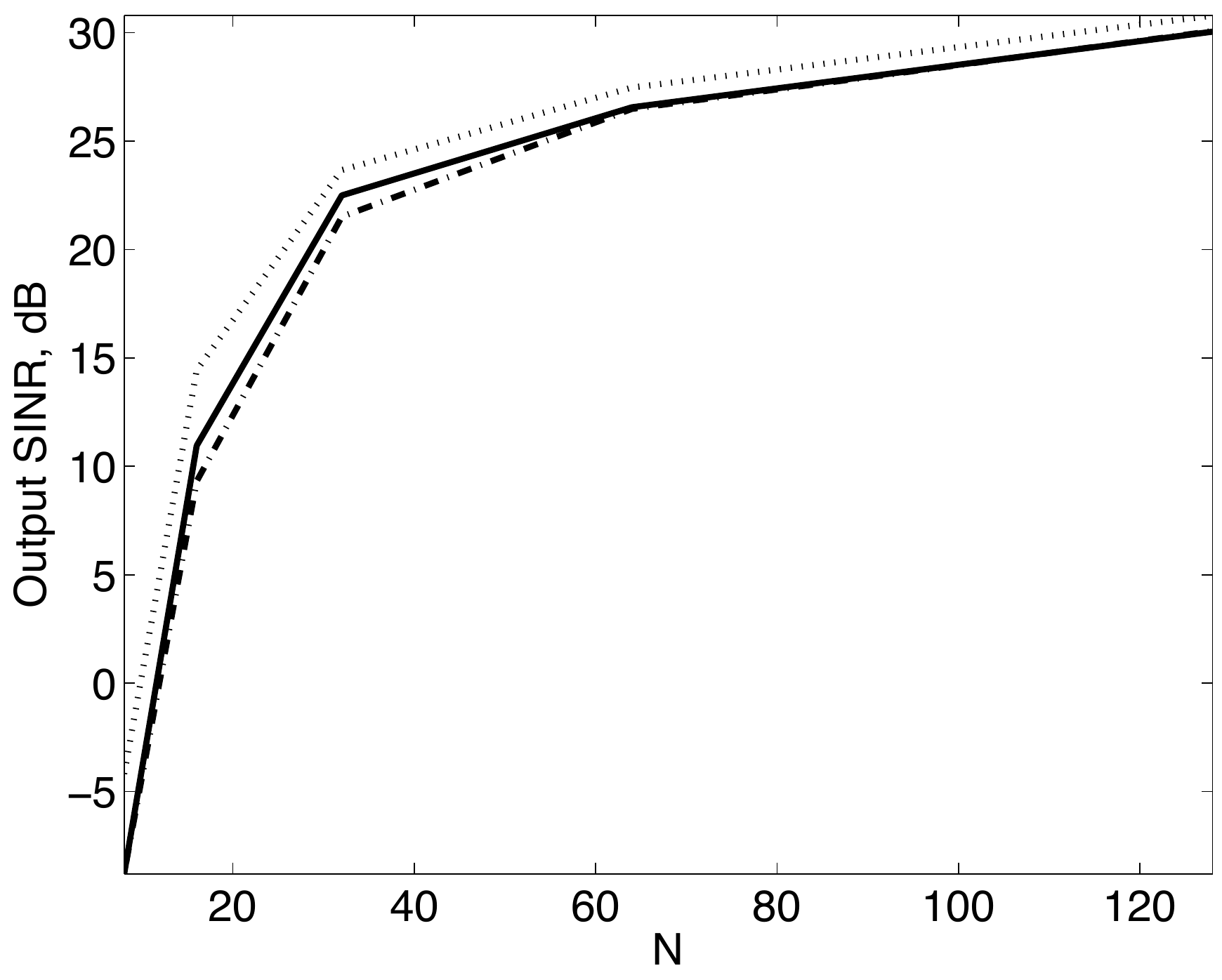}}
\caption{Зависимость ОСПШ на выходе формирователя достаточной
статистики от количества импульсов в пачке $N$, при различном
количестве усредняемых элементов разрешения $M$ и различных
значениях ОСПШ на входе. Усреднение по 500 реализаций. Пунктирная
линия обозначает идеальный случай с полностью известной
корреляционной матрицей помехи, сплошная линия обозначает
предлагаемый метод, штрих--пунктирная линия обозначает метод
\cite{Car88}. ОСШ на входе составляет 10 дБ, $M = N$, а сигнал от
цели отсутствует в обучающей выборке.}
\label{fig:reg_aut_3} 
\end{figure*}

\begin{figure*}[tbp]
\centering \subfigure[ОСПШ 20 дБ]{
\label{fig:reg_aut_4:20} 
\includegraphics[width = 5.2cm]{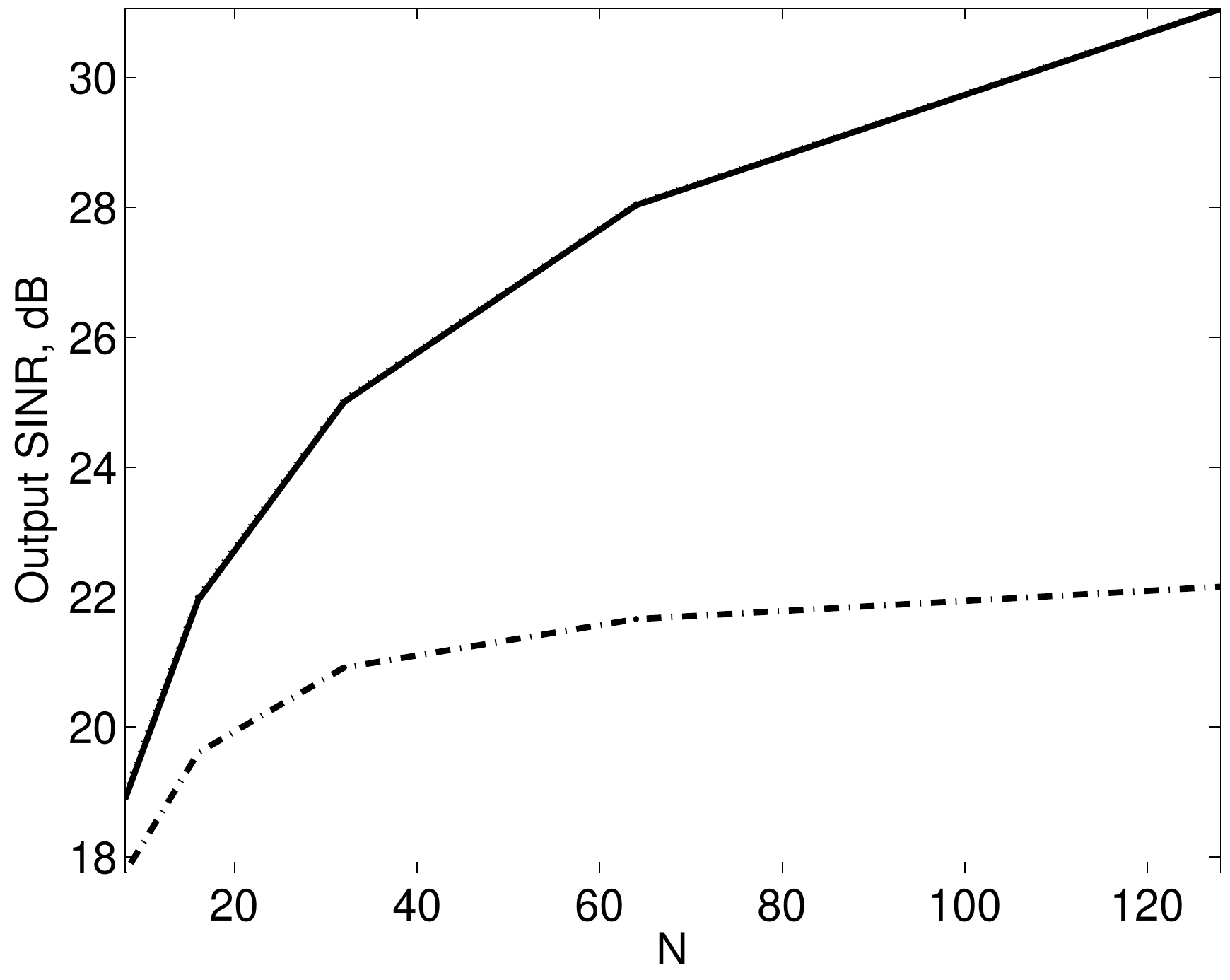}}
\hfill \subfigure[ОСПШ 10 дБ]{
\label{fig:reg_aut_4:10} 
\includegraphics[width = 5.2cm]{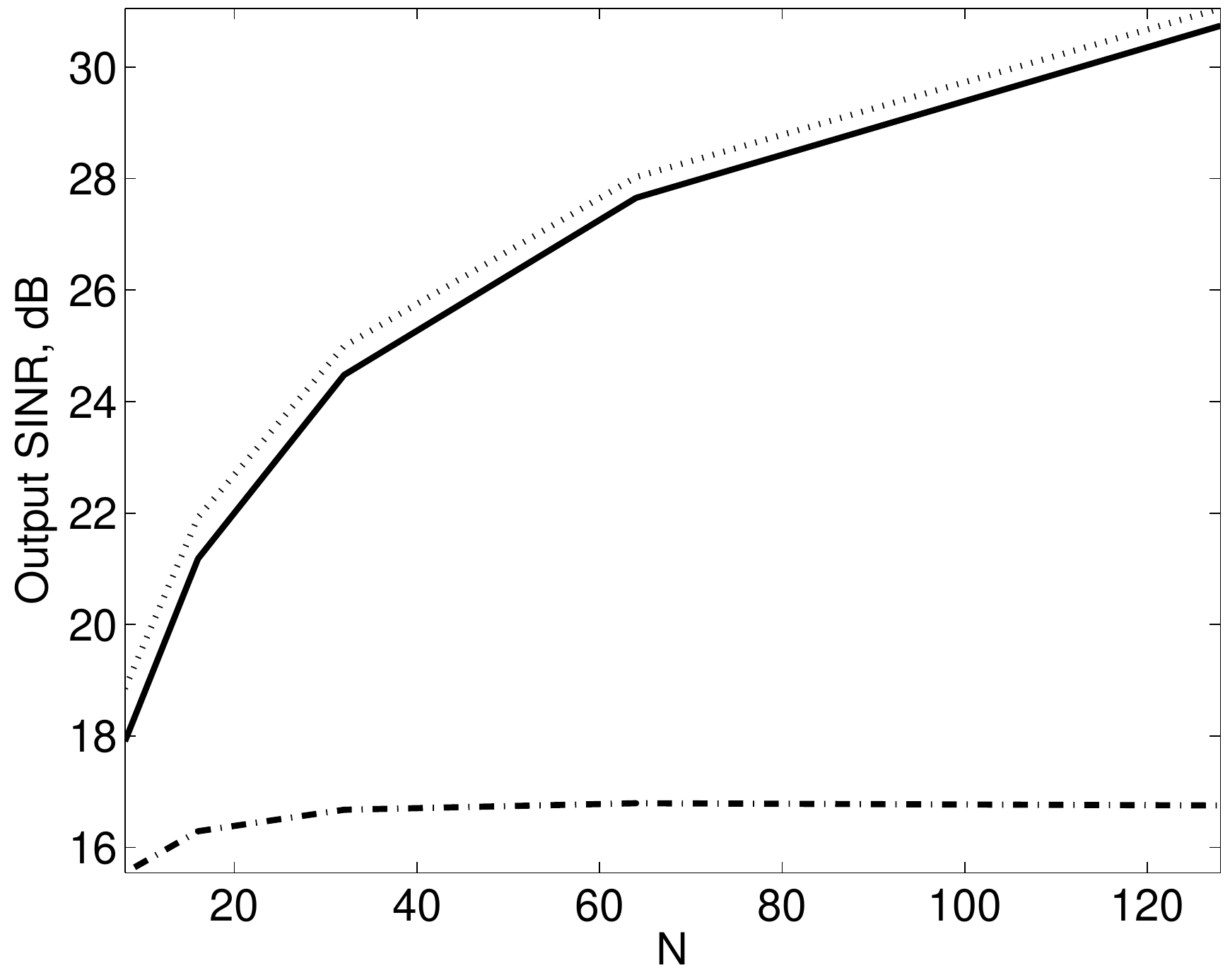}}
\hfill \subfigure[ОСПШ 0 дБ]{
\label{fig:reg_aut_4:0} 
\includegraphics[width = 5.2cm]{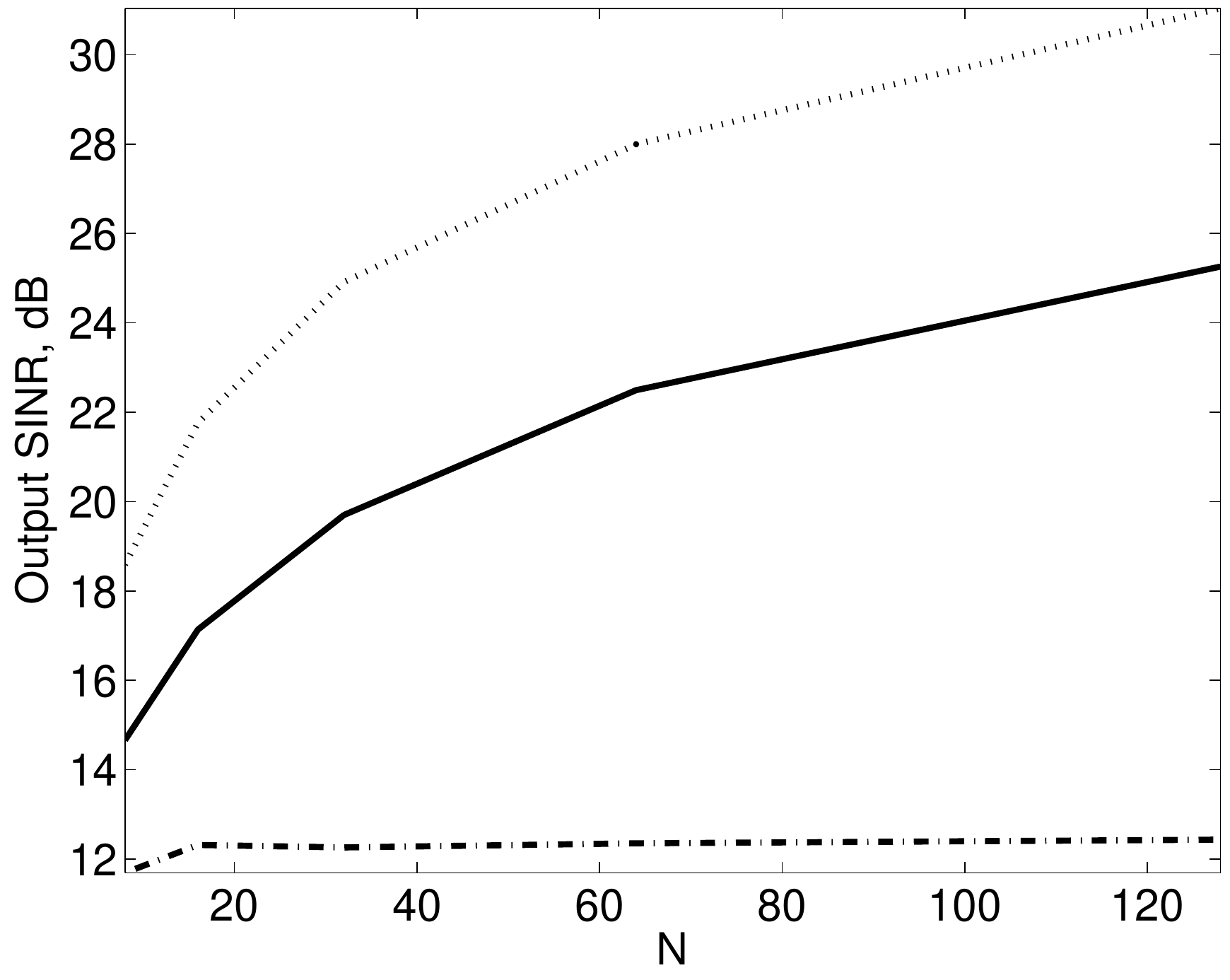}}
\hfill \subfigure[ОСПШ -10 дБ]{
\label{fig:reg_aut_4:-10} 
\includegraphics[width = 5.2cm]{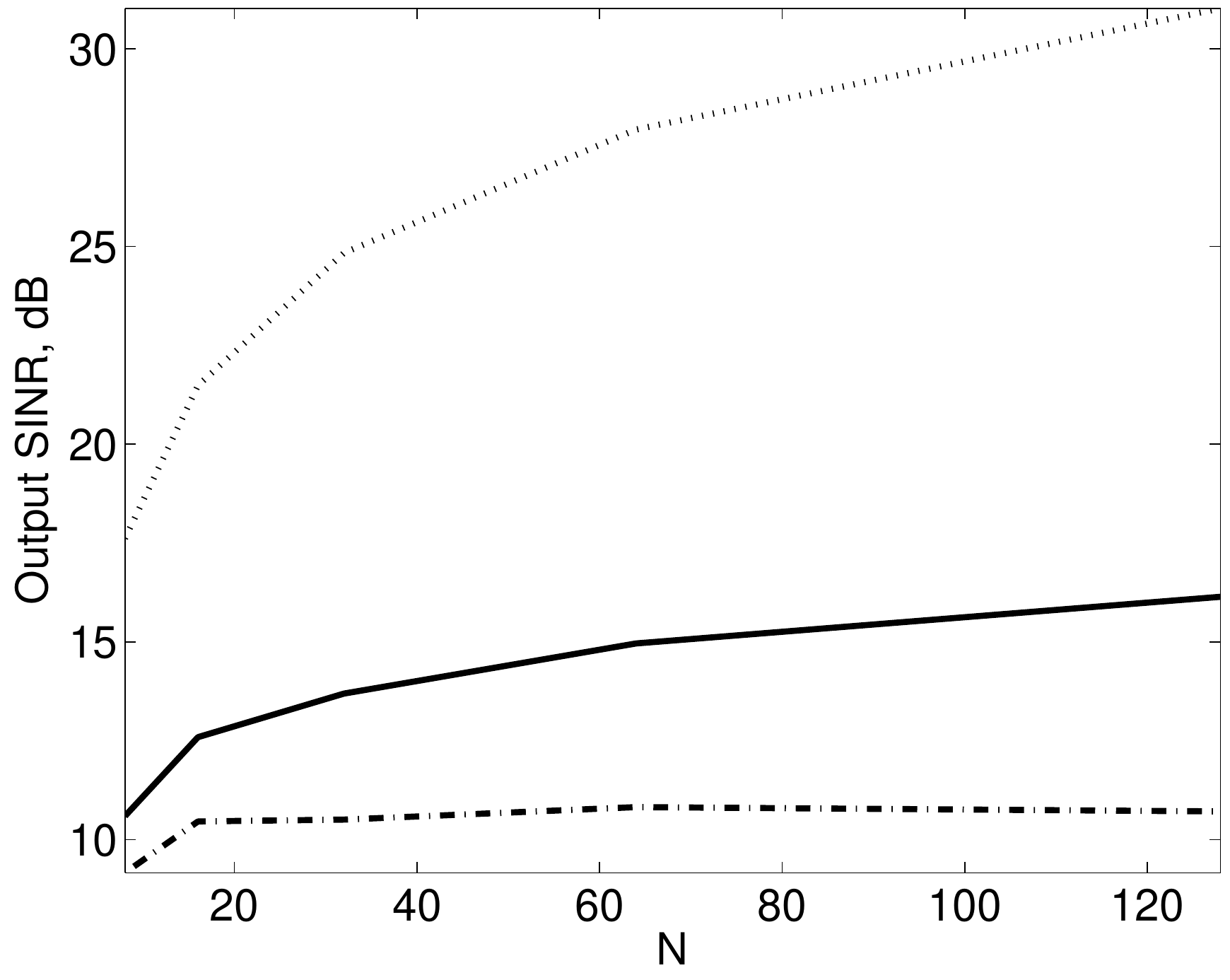}}
\hfill \subfigure[ОСПШ -20 дБ]{
\label{fig:reg_aut_4:-20} 
\includegraphics[width = 5.2cm]{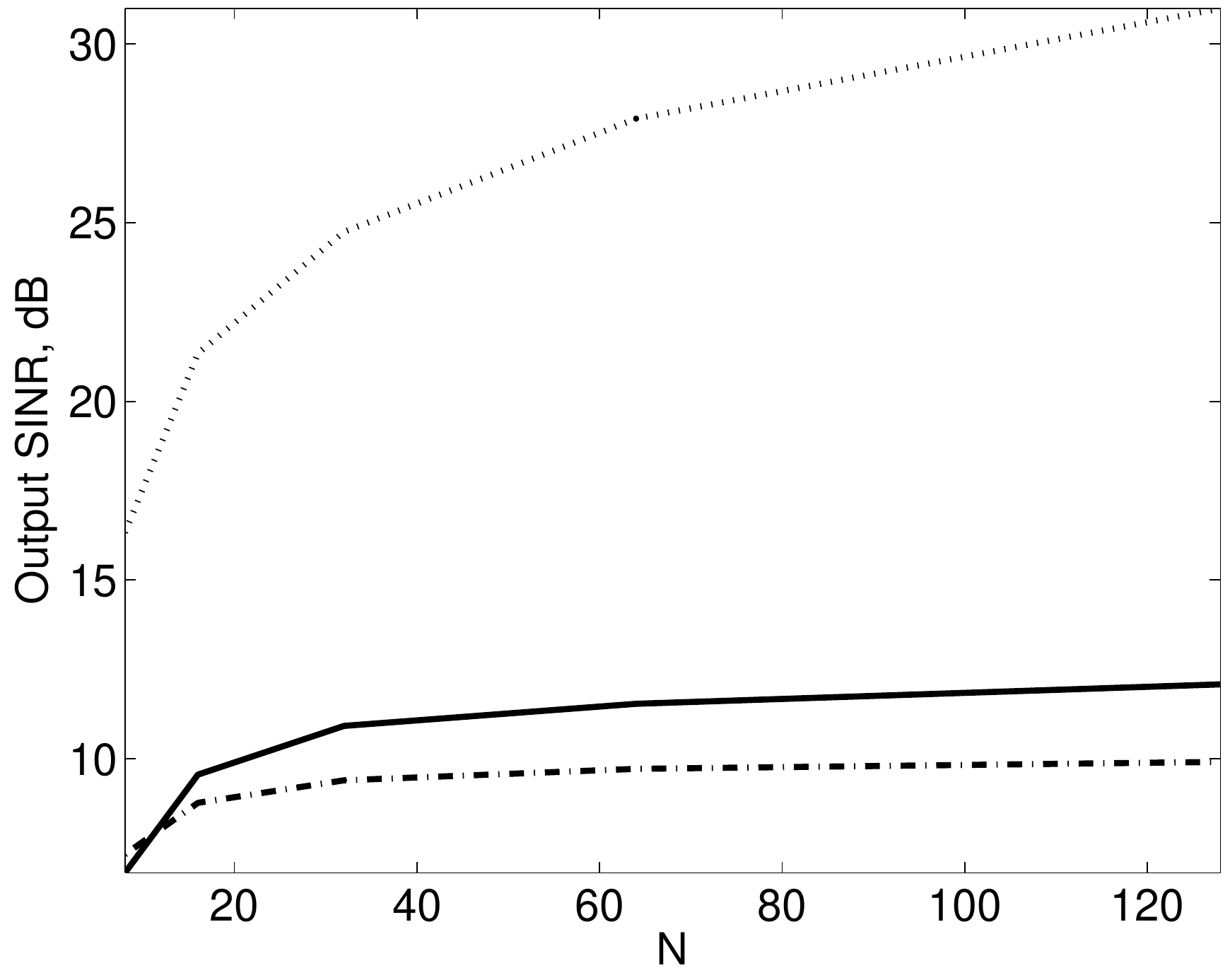}}
\hfill \subfigure[ОСПШ -40 дБ]{
\label{fig:reg_aut_4:-40} 
\includegraphics[width = 5.2cm]{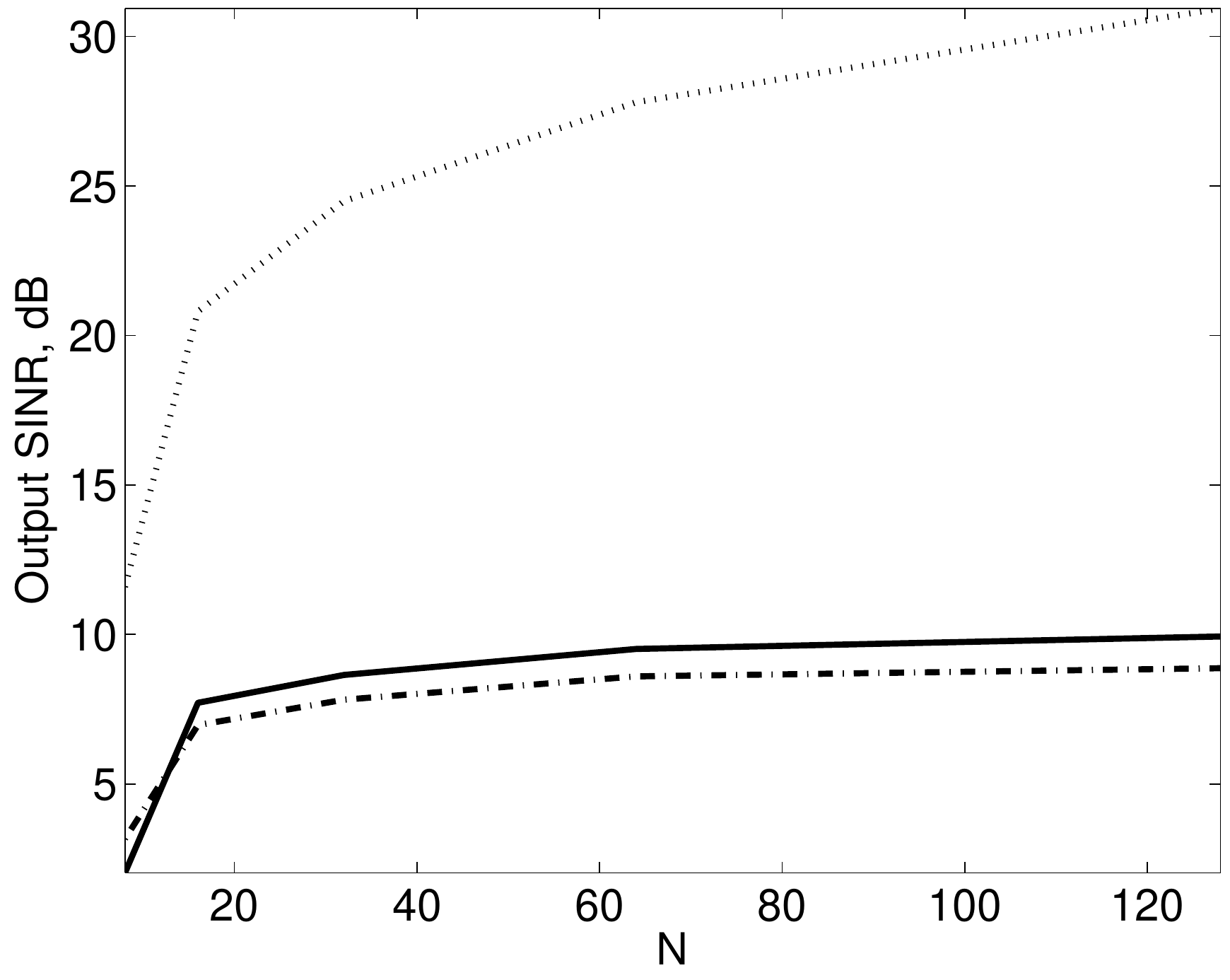}}
\hfill \subfigure[ОСПШ -60 дБ]{
\label{fig:reg_aut_4:-60} 
\includegraphics[width = 5.2cm]{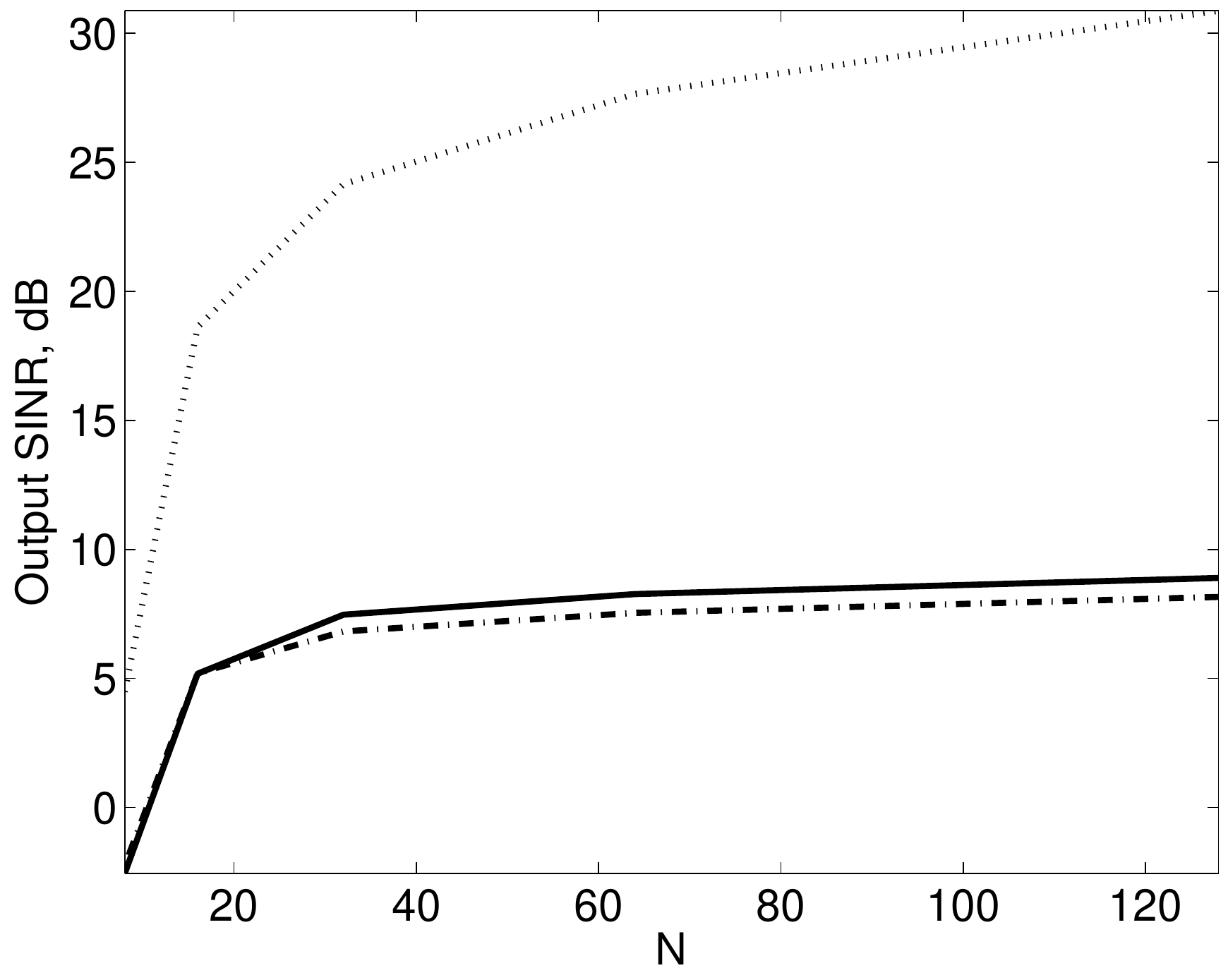}}
\hfill \subfigure[ОСПШ -80 дБ]{
\label{fig:reg_aut_4:-80} 
\includegraphics[width = 5.2cm]{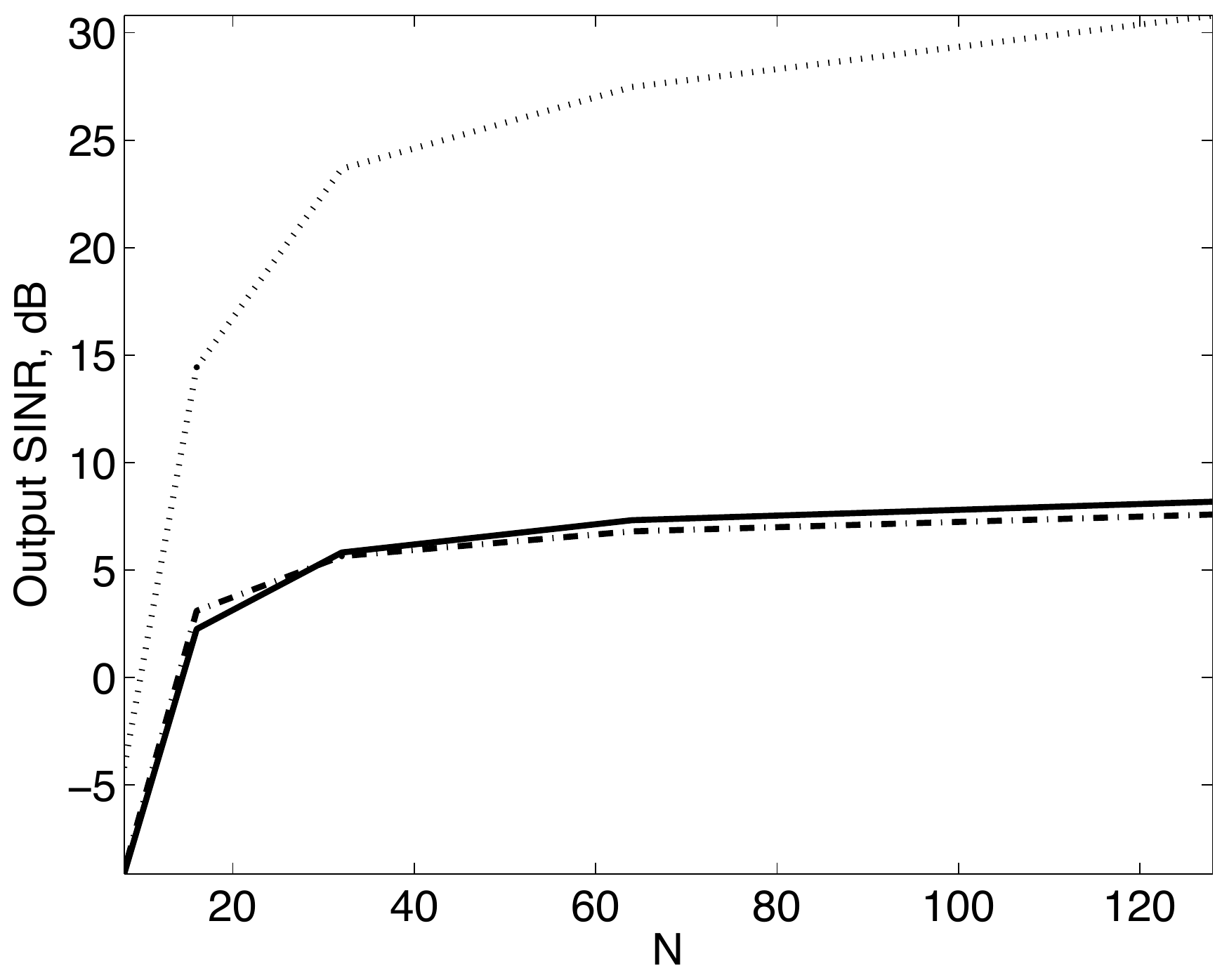}}
\caption{Зависимость ОСПШ на выходе формирователя достаточной
статистики от количества импульсов в пачке $N$, при различном
количестве усредняемых элементов разрешения $M$ и различных
значениях ОСПШ на входе. Усреднение по 500 реализаций. Пунктирная
линия обозначает идеальный случай с полностью известной
корреляционной матрицей помехи, сплошная линия обозначает
предлагаемый метод, штрих--пунктирная линия обозначает метод
\cite{Car88}. ОСШ на входе составляет 10 дБ, $M = N$, а сигнал от
цели присутствует в обучающей выборке.}
\label{fig:reg_aut_4} 
\end{figure*}

Результаты моделирования в виде зависимостей ОСПШ на выходе при
различных значениях параметров помеховой обстановки показаны на Рис.
\ref{fig:reg_aut_1}--\ref{fig:reg_aut_4}. Из рисунков следует, что
применение предлагаемого метода оценивания оптимальной величины
$\upalpha$ позволяет в значительной мере ослаблять эффект подавления
полезного сигнала, когда он присутствует в обучающей выборке. В
частности из Рис. \ref{fig:reg_aut_2:20}--Рис.
\ref{fig:reg_aut_2:-10}, Рис. \ref{fig:reg_aut_4:20}--Рис.
\ref{fig:reg_aut_4:-10} следует, что в ситуации, когда средняя
мощность сигнала от цели в обучающей выборке сравнительно велика,
применение предлагаемого метода определения уровня регуляризации
приводит к значительному выигрышу в выходном ОСПШ по сравнению со
случаем, когда $\upalpha$ фиксировано на уровне 10 дБ выше уровня
шума \cite{Car88}. Более того, моделирование показывает также, что
увеличение количества итераций $T$ например до 5 приводит в случаях
Рис. \ref{fig:reg_aut_2:10}--Рис. \ref{fig:reg_aut_2:-10}, Рис.
\ref{fig:reg_aut_4:0}--Рис. \ref{fig:reg_aut_4:-20} к
дополнительному выигрышу, составляющему 5--10 дБ. Появление выигрыша
при применении предлагаемого метода объясняется тем, что в
результате подавления полезного сигнала, возникающего при его
просачивании в обучающую выборку, средний уровень боковых лепестков
адаптивного фильтра подавления помехи возрастает и помеха
просачивается через них на выход обнаружителя. Более того, если
средняя мощность помехи в обучающей выборке меньше, чем средняя
мощность полезного сигнала, алгоритм РОКМ перестает подавлять помеху
совсем и подавляет только полезный сигнал. Таким образом,
предлагаемый алгоритм автоматически регулирует уровень регуляризации
и пытается найти баланс между подавлением помехи автоматически
возникающим при применении согласованного фильтра на фоне белого
шума $\w = \s$ и подавлением помехи и сигнала, возникающим при
применении согласованного фильтра на фоне коррелированного шума $\w
= \widehat\R^{-1}\s$ в ситуации, когда полезный сигнал присутствует
в оценке корреляционной матрицы помехи. Т.е. целью предлагаемого
алгоритма по сути является автоматическое определение такой степени
влияния адаптивной части весов обнаружителя (\ref{eqn:lsmi_w}), что
выигрыш от адаптивной обработки максимален.

Следует отметить также, что применение предлагаемого алгоритма в
некоторых ситуациях, например Рис. \ref{fig:reg_aut_1:20}--Рис.
\ref{fig:reg_aut_1:-10}, Рис. \ref{fig:reg_aut_3:20}--Рис.
\ref{fig:reg_aut_3:0} приводит к незначительным потерям по сравнению
с алгоритмом \cite{Car88}. Эти потери не превышают долей децибела и
могут быть объяснены применением линейной аппроксимации
(\ref{eqn:lsmi_w_not_approx}) для выражения весов адаптивного
фильтра (\ref{eqn:lsmi_w}). Другим недостатком предлагаемого метода
оценивания $\upalpha$ является то, что он не решает проблемы резкого
снижения эффективности алгоритма РОКМ при наличии сигнала от цели в
обучающей выборке при низких входных ОСПШ. Однако этот недостаток
скорее относится к методу регуляризации в целом. В ситуации, когда
ОСПШ на входе не превышает -40 дБ, а оценка корреляционной матрицы
помехи загрязнена сигналом от цели, дополнительное подавление в
участках спектра, где сконцетрирована мощность помехи становится
необходимым и добиться выигрыша простым балансированием между
обработкой на фоне белого шума и адаптивной обработкой очень сложно
или даже невозможно. Поэтому увеличение эффективности адаптивных
алгоритмов в таких условиях и при малых размерах обучающей выборки
является перспективным направлением дальнейших исследований.

\section{Моделирование алгоритма LMS с квадратичным ограничением} \label{sec:LMS_sq:mod}

В данном разделе представлены результаты компьютерного моделирования
алгоритма LMS с квадратичным ограничением синтезированного в разделе
\ref{sec:LMS_sq}. Промоделированы две ситуации. Во--первых, в
обучающей выборке присутствует несколько помех с Релеевской
огибающей, приходящих под различными углами из дальней зоны, а
полезный сигнал отсутствует. Во--вторых, в обучающей выборке
присутствует и полезный сигнал и помехи. Полученные результаты
подтверждают тот факт, что алгоритм LMS с квадратичным условием
обладает лучшими характеристиками, чем алгоритм LMS с линейным
условием, как в условиях отсутствия, так и присутствия полезного
сигнала в обучающей выборке.

\begin{figure*}[tbp]
\centering \subfigure[ОСПШ = -60 дБ, $N$ = 256]{
\label{fig:quadconst_1:1} 
\includegraphics[width = 7cm]{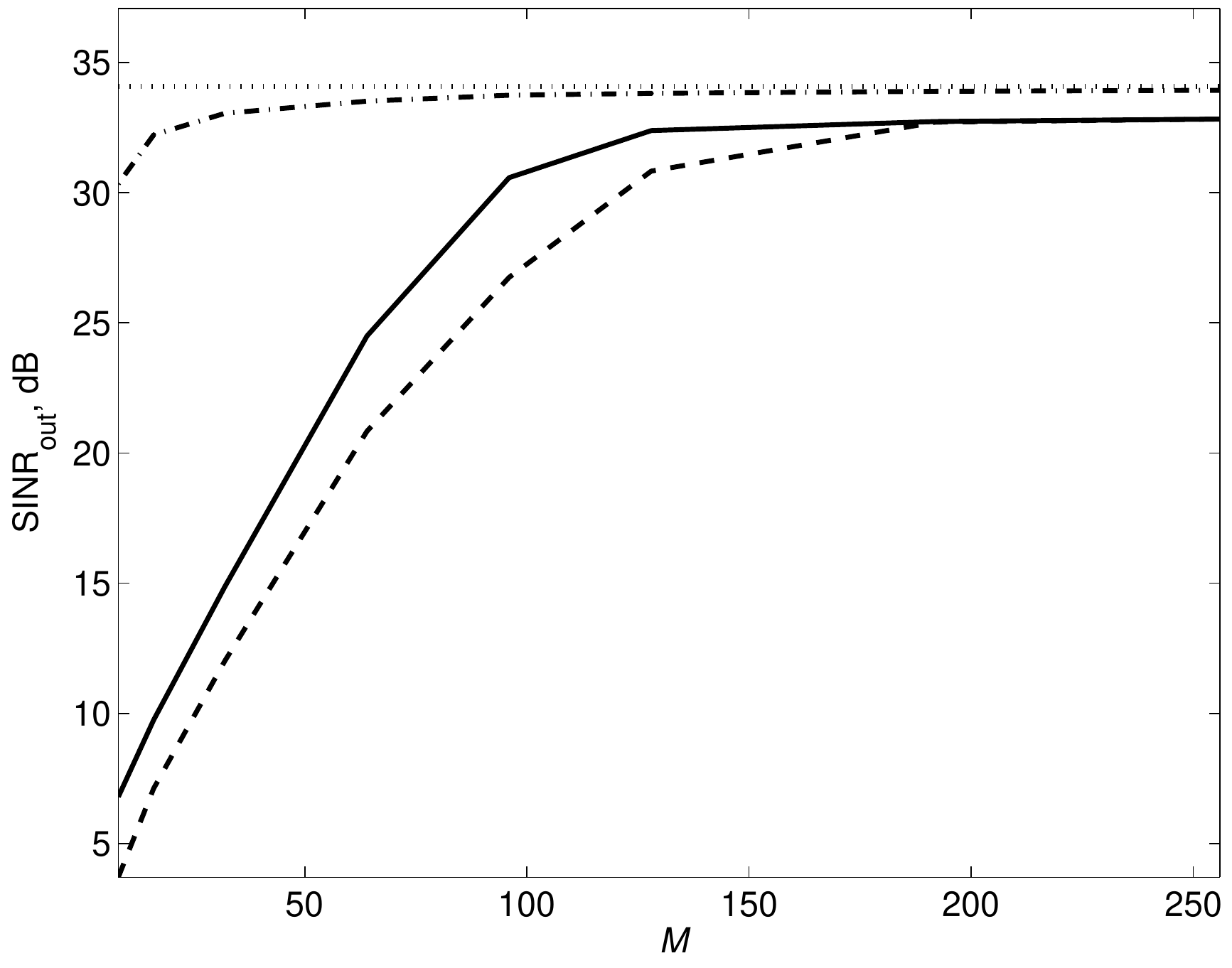}}
\hfill \subfigure[ОСПШ = -80 дБ, $N$ = 256]{
\label{fig:quadconst_1:2} 
\includegraphics[width = 7cm]{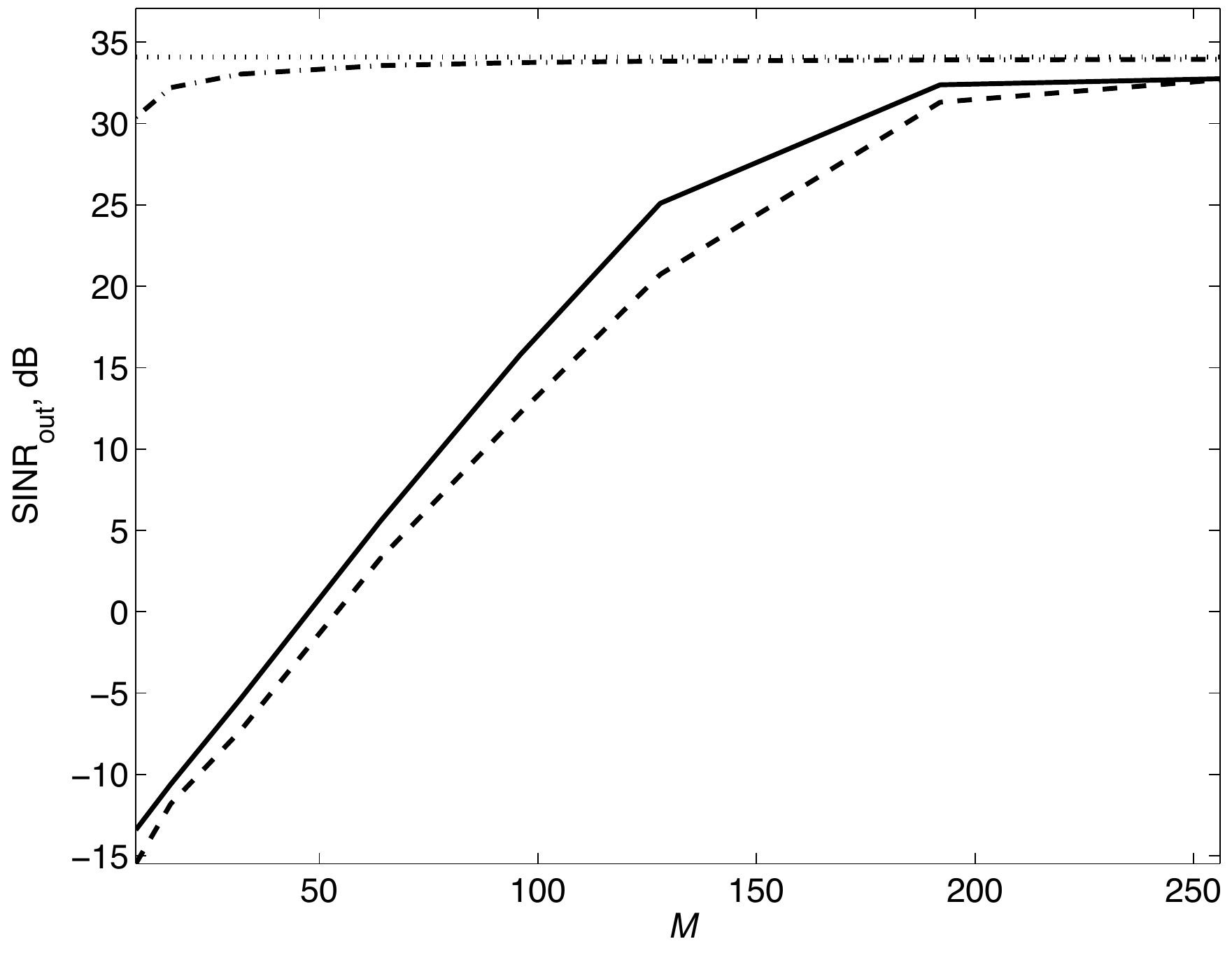}}
\hfill \subfigure[ОСПШ = -60 дБ, $N$ = 512]{
\label{fig:quadconst_1:3} 
\includegraphics[width = 7cm]{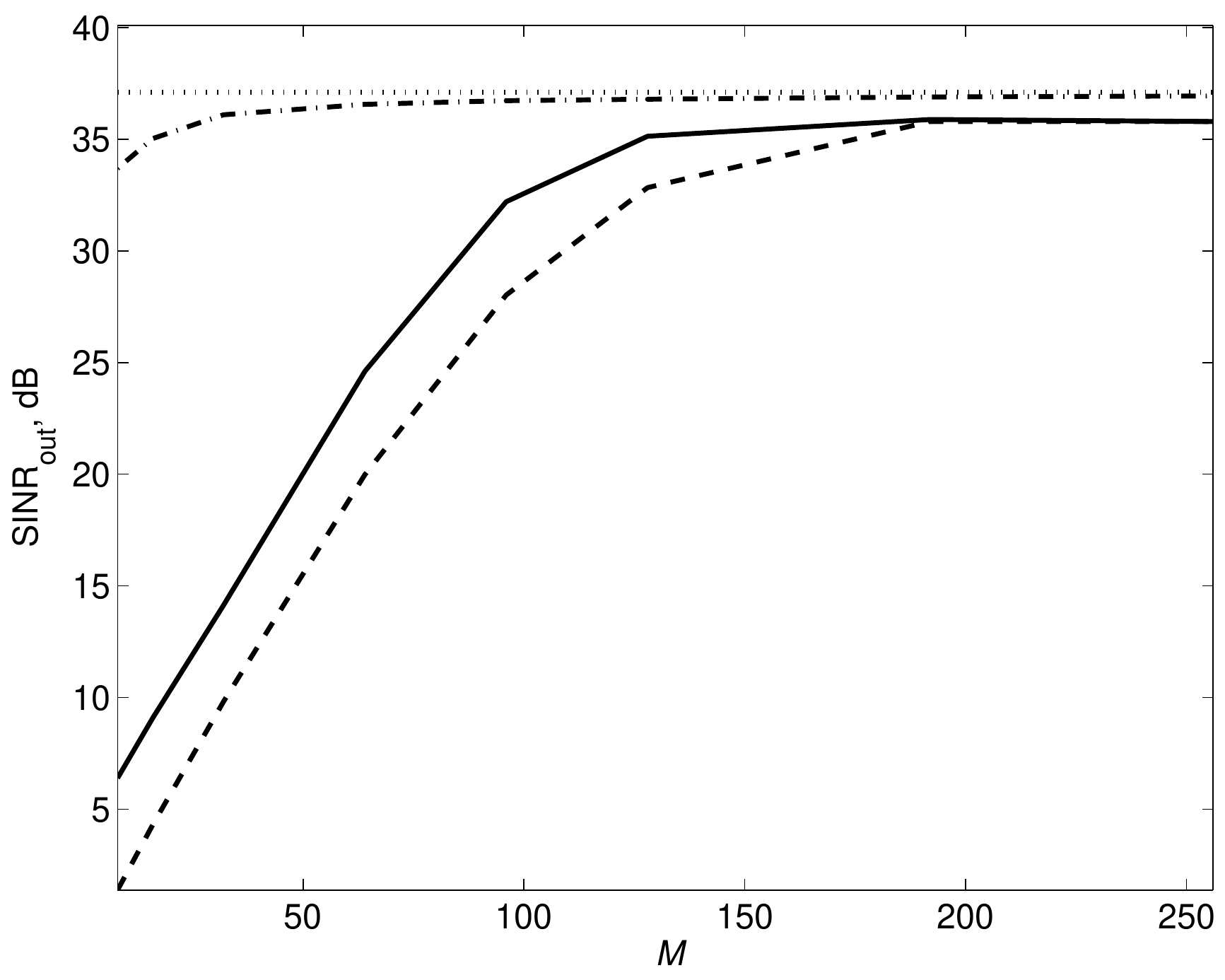}}
\hfill \subfigure[ОСПШ = -80 дБ, $N$ = 512]{
\label{fig:quadconst_1:4} 
\includegraphics[width = 7cm]{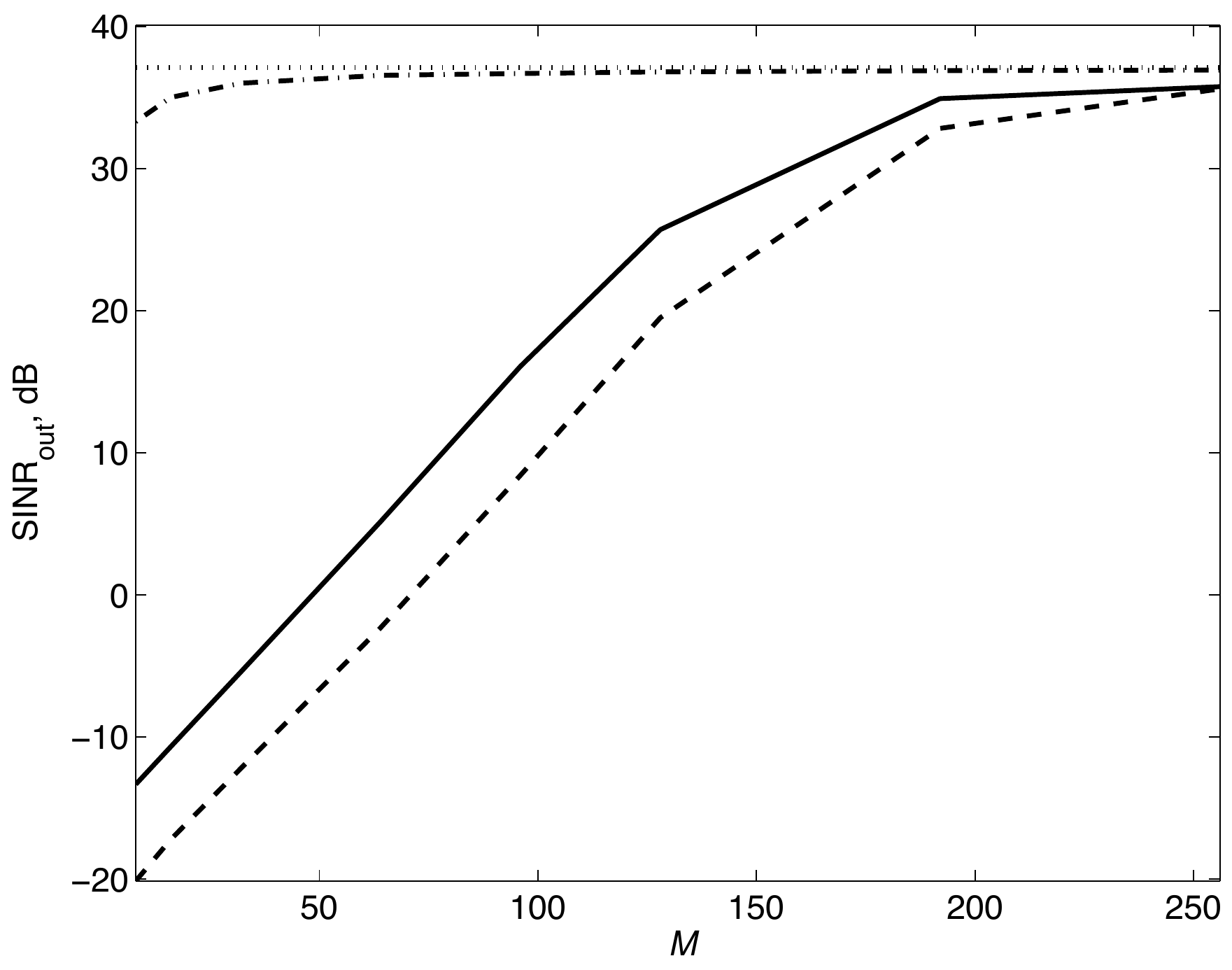}}
\hfill \subfigure[ОСПШ = -60 дБ, $N$ = 1024]{
\label{fig:quadconst_1:5} 
\includegraphics[width = 7cm]{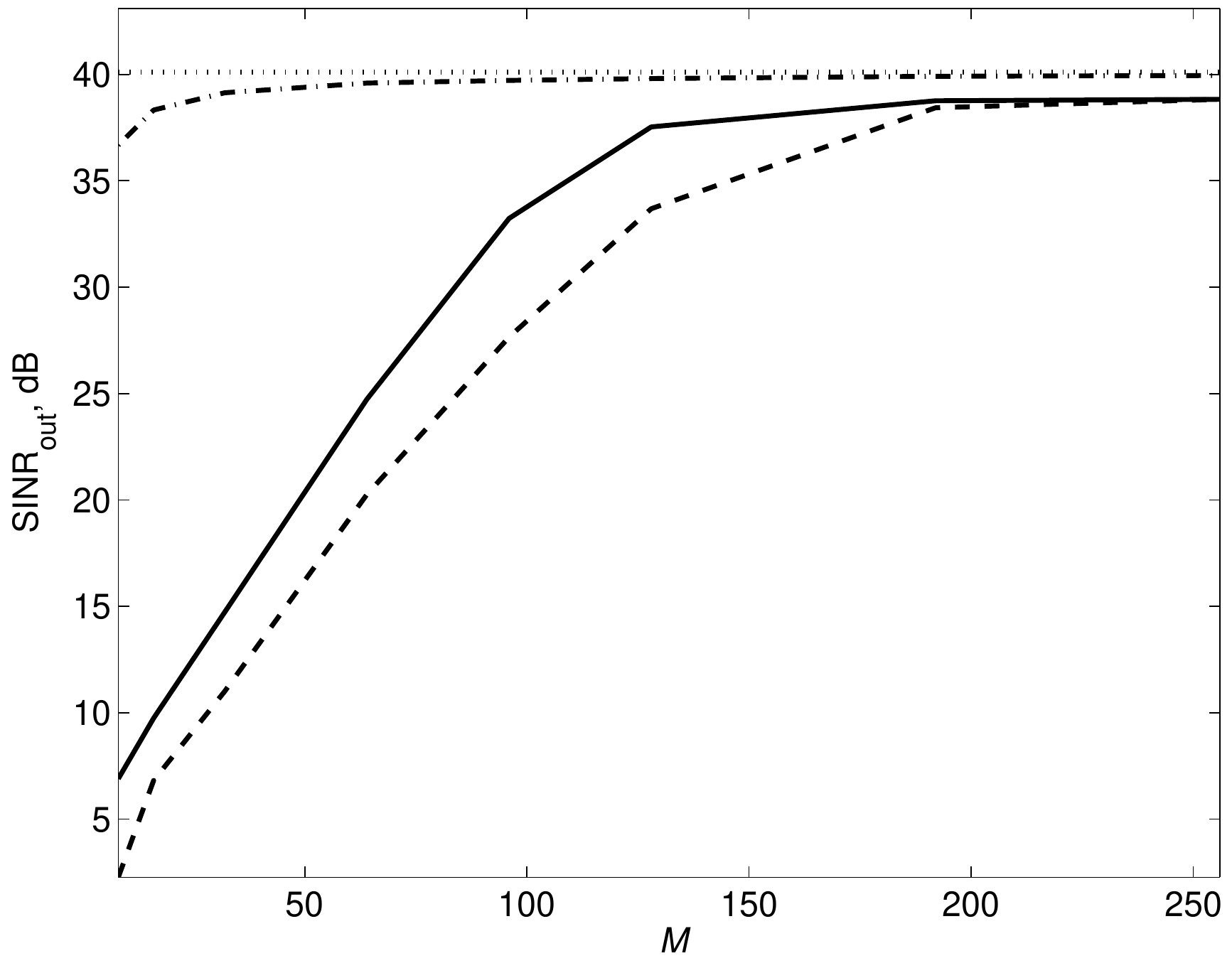}}
\hfill \subfigure[ОСПШ = -80 дБ, $N$ = 1024]{
\label{fig:quadconst_1:6} 
\includegraphics[width = 7cm]{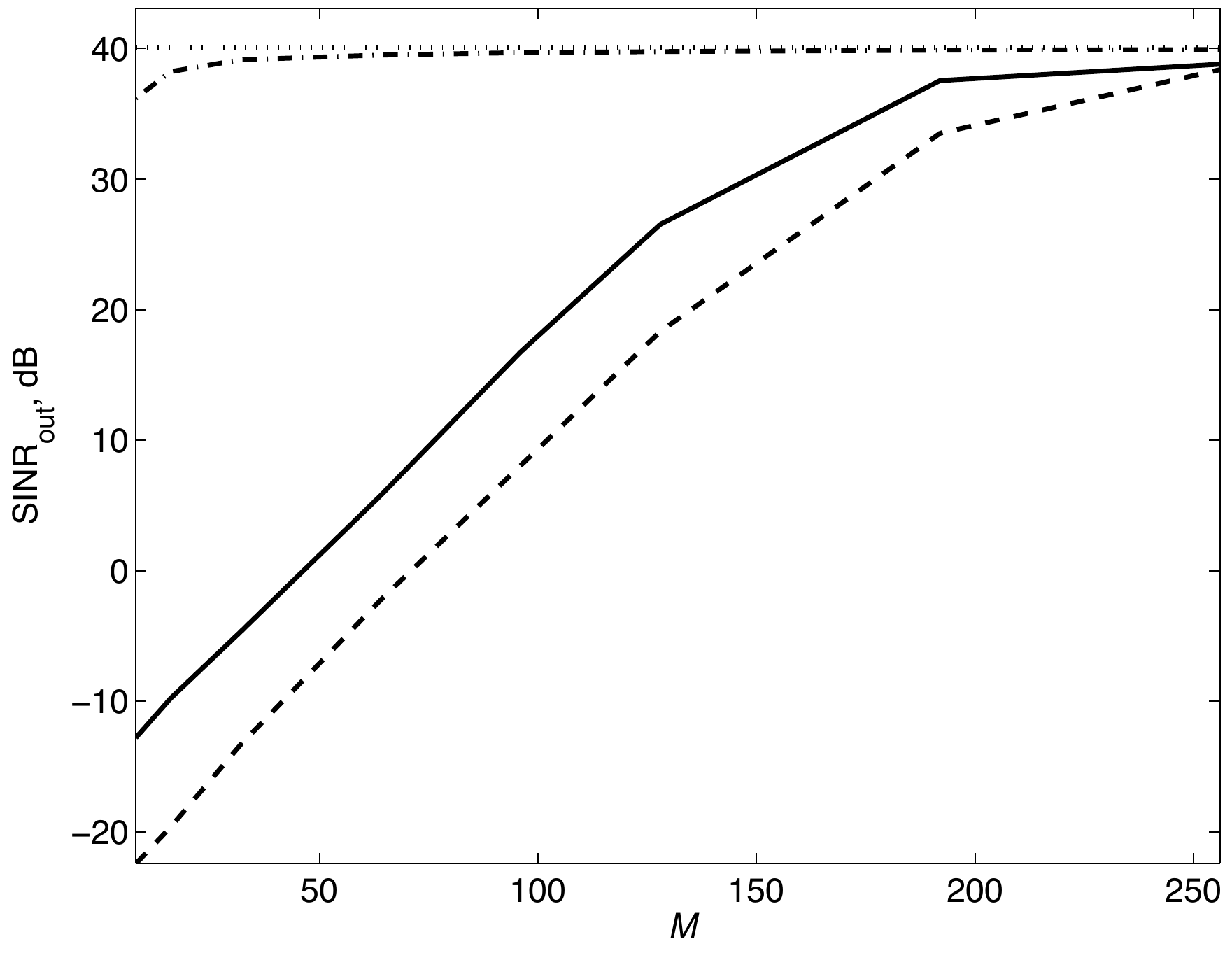}}
\caption{Кривые обучения алгоритмов РОКМ (штрих--пунктирная линия),
LMS с линейным ограничением (штриховая линия), LMS с квадратичным
ограничением (сплошная линия). Количество помех: 3, углы прихода
помех: -14, 71 и 66 градусов}
\label{fig:quadconst_1} 
\end{figure*}

Моделирование ситуации с отсутствием полезного сигнала в обучающей
выборке было проведено для следующих значений параметров помех и
антенны. Количество помех: 3, ОСПШ на входе: -60 и -80 дБ, отношение
сигнал/белый шум на входе 10 дБ, распределение мощности между
составляющими помехи: равномерное, огибающая помехи: Релеевская,
зона приема: дальняя, углы прихода помех по отношению к нормали
антенны: -14, 71 и 66 градусов, количество статистических испытаний:
100. Моделирование ситуации с присутствием полезного сигнала в
обучающей выборке было проведено для следующих значений параметров
(приведены только параметры со значениями отличными от первой
ситуации) ОСПШ в обучающей выборке: 10 дБ, $N = 128$, $M = 1024$.

Результаты моделирования первой ситуации показаны на Рис.
\ref{fig:quadconst_1}. Этот рисунок демонстрирует кривые обучения
алгоритмов РОКМ, LMS с линейным ограничением и LMS с квадратичным
ограничением. Он показывает насколько быстро сходится ОСПШ на выходе
адаптивных алгоритмов к ОСПШ на выходе идеального алгоритма ОКМ с
полностью известной ковариационной матрицей помехи при увеличении
объема обучающей выборки $M$. Видно, что при сравнительно большом
количестве элементов $N$ в антенной решетке адаптация алгоритма с
квадратичным ограничением проходит значительно быстрее, чем
адаптация алгоритма с линейным ограничением.

\begin{figure*}[tbp]
\centering \subfigure[LMS с линейным ограничением]{
\label{fig:fig:quadconst_2:1} 
\includegraphics[width = 5.2cm]{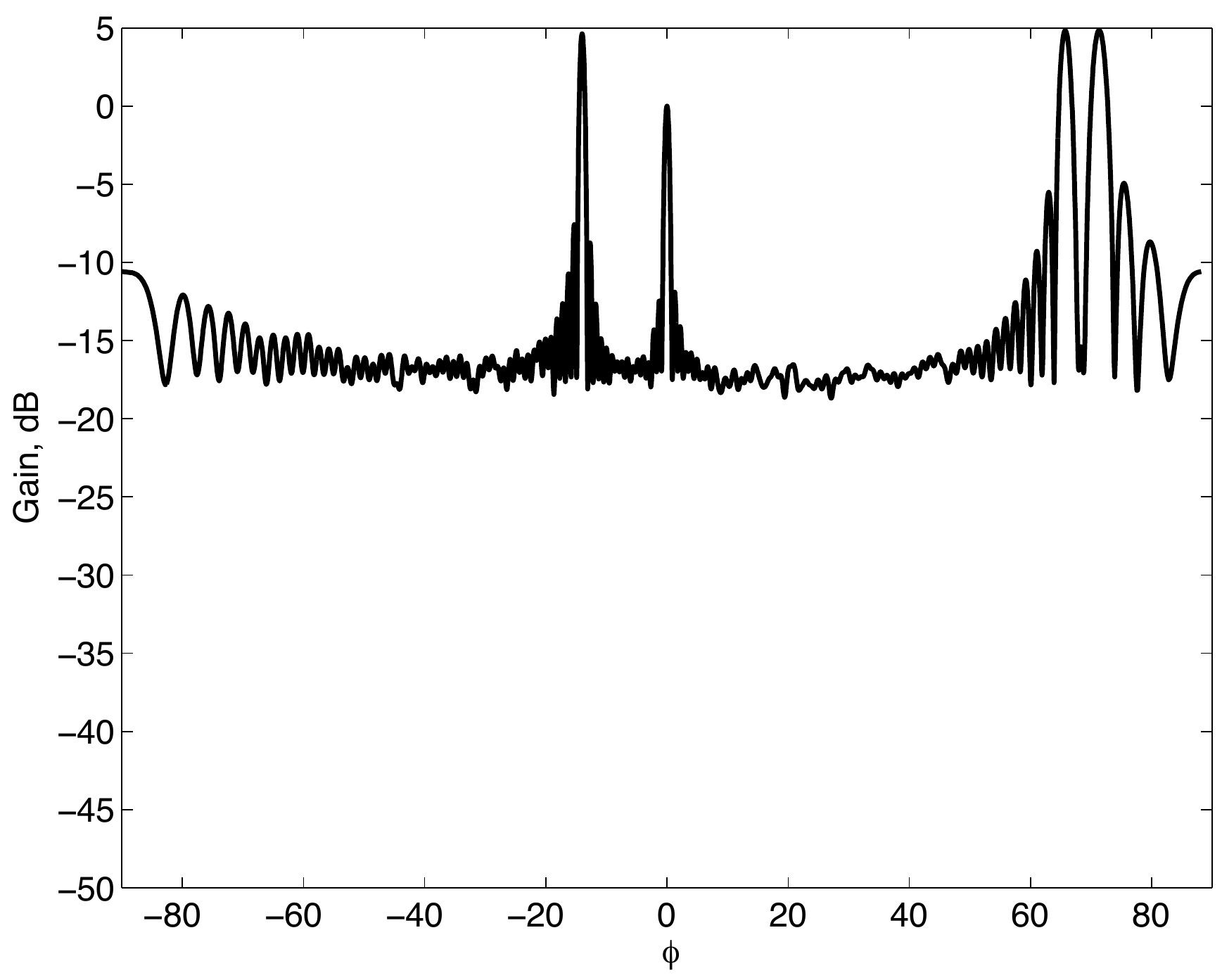}}
\hfill \subfigure[LMS с квадратичным ограничением]{
\label{fig:fig:quadconst_2:2} 
\includegraphics[width = 5.2cm]{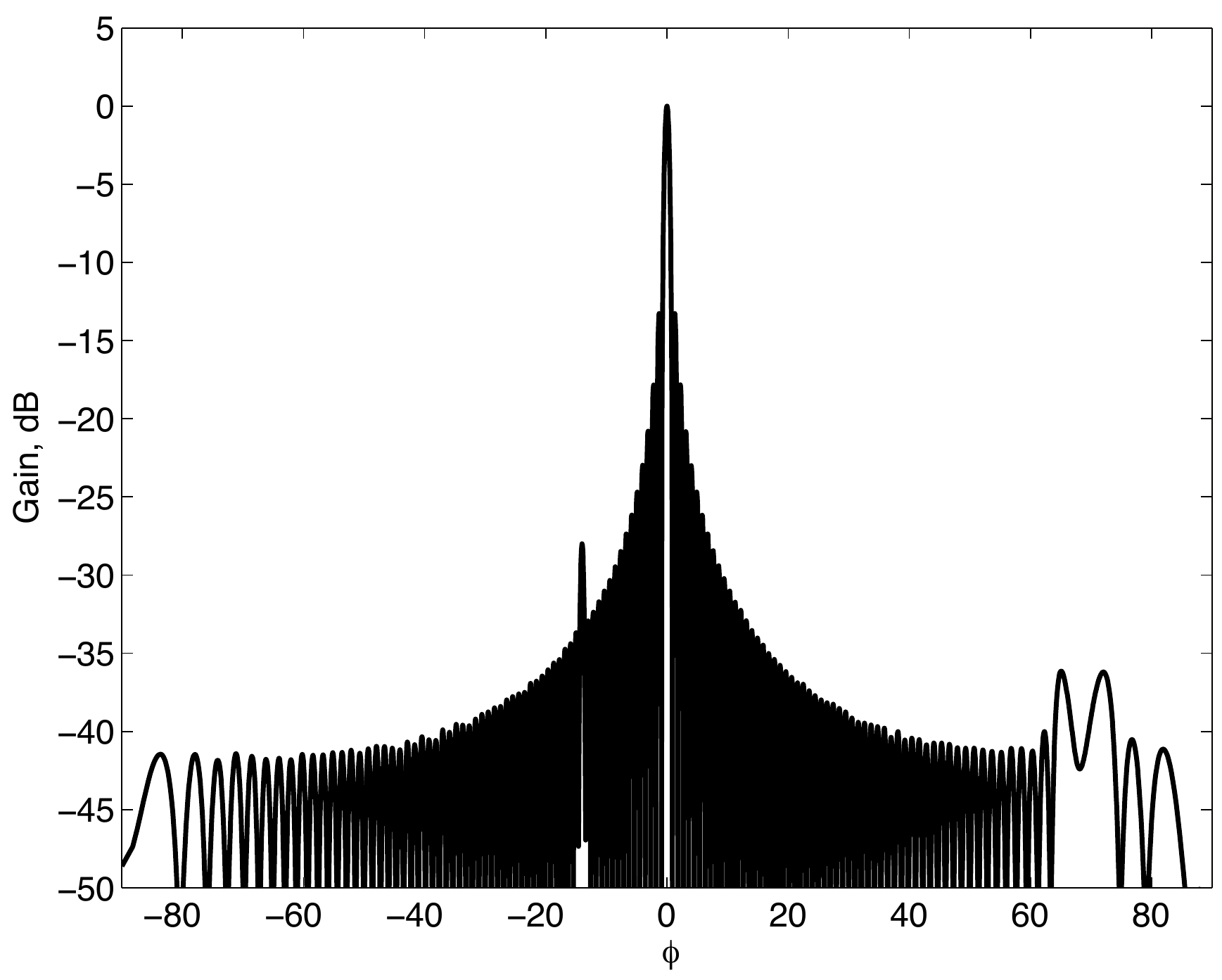}}
\hfill \subfigure[LMS без ограничений]{
\label{fig:fig:quadconst_2:3} 
\includegraphics[width = 5.2cm]{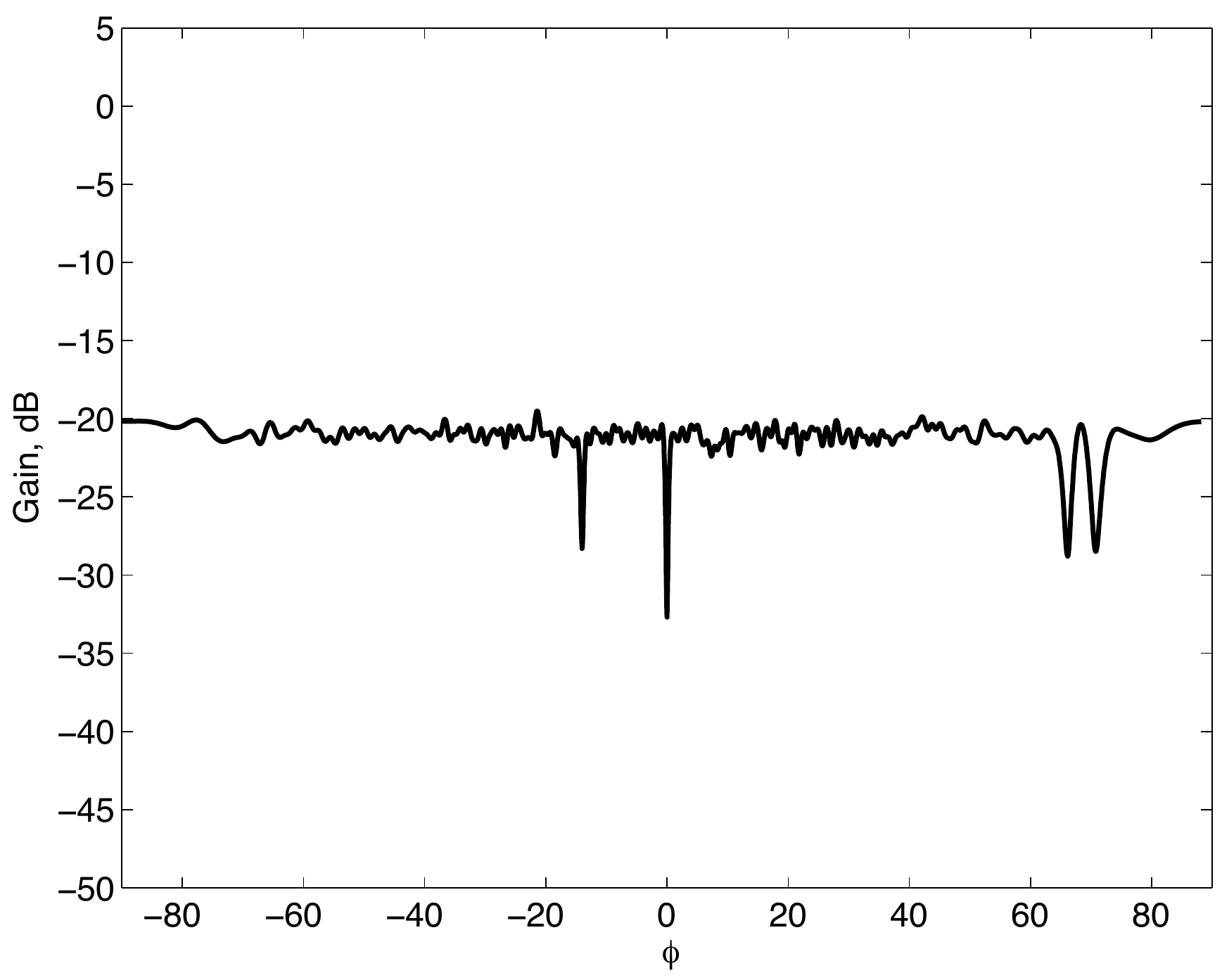}}
\caption{Вид диаграммы направленности антенны с количеством
элементов $N = 128$ после обучения по $M = 1024$ обучающим выборкам,
усредненный в результате 100 экспериментов. LMS с линейным
ограничением (слева), LMS с квадратичным ограничением (посередине) и
LMS без ограничений (справа). Количество помех: 3, углы прихода
помех: -14, 71 и 66 градусов, ОСПШ в обучающей выборке: 10 дБ}
\label{fig:quadconst_2} 
\end{figure*}

Результаты моделирования ситуации с присутствием полезного сигнала в
обучающей выборке представлены на Рис. \ref{fig:quadconst_2},
показывающем вид диаграммы направленности антенны с количеством
элементов $N = 128$ после обучения по $M = 1024$ обучающим выборкам,
усредненный в результате 100 экспериментов.
Для наглядности ОСПШ в обучающей выборке выбран равным 10 дБ, т. е.
на процесс адаптации влияет практически только полезный сигнал. Из
рисунка видно, что алгоритм LMS без ограничений подавляет и полезный
сигнал и помехи, алгоритм LMS с линейным ограничением выдерживает
условие единичного усиления в направлении полезного сигнала, однако
в то же время усиливает помехи так, что усиление в направлении помех
превосходит усиление в направлении полезного сигнала. В свою очередь
алгоритм LMS с квадратичным ограничением выдерживает условие
единичного усиления в направлении на полезный сигнал и в то же время
демпфирует влияние помеховых составляющих на формирование диаграммы
направленности, так что отклик в направлении на полезный сигнал на
30--35 дБ превышает отклик в направлении на помехи. Такая
значительная разница в свойствах алгоритмов LMS с различным видом
ограничения может быть объяснена тем, что линейное ограничение
обеспечивает небходимую, а не достаточную степень демпфирования
сигнала подстройки весов когда полезный сигнал присутствует в
обучающей выборке. В то же время алгоритм LMS с квадратичным
условием обеспечивает более высокую степень демпфирования и таким
образом предотвращает проникновение сигнала помехи в вектор весов в
указанной ситуации. Из результатов моделирования, представленных в
данном разделе можно сделать следующий вывод. Алгоритм LMS с
квадратичным условием обеспечивает более эффективное переключение
между режимами подстройки весов адаптивного фильтра, в одном из
которых блокировка сигнала подстройки весов адаптивного фильтра
необходима (полезный сигнал присутствует в обучающей выборке), а в
другой необходимости в такой блокировке нет (полезный сигнал
отсутствует в обучающей выборке).

\section{Заключение} \label{chap:sec:Ch4:concl}

В данной главе рассмотрены результаты моделирования адаптивных
алгоритмов АСДЦ синтезированных в главе \ref{chap:Ch3}.

Во--первых, в разделе \ref{sec:SMI_reg:mod} промоделирован
итеративный алгоритм оценивания оптимальной величины коэффициента
регуляризации алгоритма РОКМ, синтезированный в разделе
\ref{sec:SMI_reg} и предложенный в \cite{Ore07c,Ore08b,Ore08c}.
Данный алгоритм может быть применен совместно с алгоритмом РОКМ для
предотвращения или ослабления эффекта подавления сигнала от цели в
ситуации, когда этот сигнал проникает в обучающую выборку. Выявлены
положительные свойства алгоритма, такие как относительная простота,
значительный выигрыш в ситуации, когда полезный сигнал присутствует
в обучающей выборке помехи и практическое отсутствие потерь в
ситуации, когда полезного сигнала в ней нет. В частности, из анализа
Рис.~\ref{fig:reg_aut_1}--\ref{fig:reg_aut_4} следует, что
значительный выигрыш по сравнению с методом фиксирования
коэффициента регуляризации на уровне в 10 раз превышающем уровень
булого шума~\cite{Car88} при использовании данного алгоритма может
быть достигнут при использовании количества итераций $T$ равном 3.
Получаемый выигрыш зависит от ОСПШ на входе.

В ситуации, когда полезный сигнал присутствует в обучающей выборке
(см. Рис.~\ref{fig:reg_aut_2} и Рис.~\ref{fig:reg_aut_4}) и в
области положительных входных ОСПШ выигрыш составляет 3--8 дБ. В
области отрицательных ОСПШ выигрыш варьируется в пределах 0--3 дБ.
Предлагаемый алгоритм в рассматриваемой ситуации не приводит к
появлению потерь по отношению к алгоритму РОКМ с фиксированным
коэффициентом регуляризации~\cite{Car88}. Потери по отношению к
оптимальному алгоритму с полностью известной ковариационной матрицей
помехи увеличиваются с уменьшением входного ОСПШ, но никогда не
превышают потерь алгоритма РОКМ с фиксированным коэффициентом
регуляризации. Нами сделан также вывод о том, что этот недостаток
присущ методу регуляризации корреляционной матрицы помехи в целом.
Этот вывод основан на том факте, что регуляризация оценки
ковариационной матрицы приводит к тому, что результирующий фильтр
является комбинацией оптимального фильтра на фоне белого шума и
адаптивного не регуляризованного  фильтра на фоне неизвестной помехи
\cite{Sto05}. Изменение (или оптимизация) коэффициента регуляризации
приводят к варьированию вклада указанных фильтров в результирующий
регуляризованный фильтр. При низких значениях входного ОСПШ
получение выигрыша путем простого варьирования вклада указанных
фильтров невозможно в силу большой мощности помехи и ее просачивания
на выход регуляризованного фильтра через составляющую
соответствующую оптимальному фильтру на фоне белого шума.

В ситуации, когда полезный сигнал отсутствует в обучающей выборке
(см. Рис.~\ref{fig:reg_aut_1} и Рис.~\ref{fig:reg_aut_3})
предлагаемый алгоритм в некоторых случаях проигрывает алгоритму РОКМ
с фиксированным коэффициентом регуляризации~\cite{Car88}. Однако
соответствующие потери незначительны и составляют 0--0,5 дБ. Потери
предлагаемого алгоритма по сравнению с оптимальным алгоритмом,
использующем полностью известную матрицу помехи практически равны
потерям в алгоритме с фиксированным коэффициентом регуляризации.

Мы не проводили оптимизацию количества итераций $T$ предлагаемого
алгоритма, однако в соответствии с нашими наблюдениями, увеличение
$T$ приводит к росту выигрыша в ситуации, когда полезный сигнал
присутствует в обучающей выборке и росту проигрыша в ситуации, когда
полезный сигнал отсутствует в обучающей выборке. Поэтому $T$ может
быть увеличено если допустимо соответствующие увеличение
вычислительной нагрузки на процессор ПВАО и потерь во второй из
рассмотренных ситуаций.

Во--вторых, в разделе \ref{sec:LMS_sq:mod} промоделирован алгоритм
LMS с квадратичным ограничением, синтезированный в разделе
\ref{sec:LMS_sq} и предложенный в \cite{Ore07a,Ore08a}.
Промоделированы две ситуации, показывающие результаты адаптации
алгоритмов LMS с линейным условием, предлагаемого алгоритма LMS с
квадратичным условием, оптимального алгоритма с полностью известной
ковариационной матрицей помехи и алгоритма РОКМ с фиксированным
уровнем регуляризации. В первой из рассматриваемых ситуаций полезный
сигнал отсутствовал в обучающей выборке, во второй ---
присутствовал. Из анализа Рис.~\ref{fig:quadconst_1},
иллюстрирующего результаты моделирования первой ситуации, следует,
что применение алгоритма LMS с квадратичным условием позволяет
сократить объем обучающей выборки на 20--40\% по сравнению с
алгоритмом LMS с линейным ограниченем. Данный вывод следует из
анализа количества отсчетов в обучающей выборке, необходимого для
достижения максимального ОСПШ на выходе указанных алгоритмов. Анализ
результатов моделирования второй ситуации с загрязнением обучающей
выборки полезным сигналом, показанной на Рис.~\ref{fig:quadconst_2}
позволяет сделать следующие выводы. Применение алгоритма LMS с
квадратичным условием позволяет избежать подавления полезного
сигнала, как это имеет место в алгоритме LMS без условий с одной
стороны, и просачивания помех на выход адаптивного алгоритма, как
это имеет место в алгоритме с линейным ограничением, --- с другой.
Действительно, из Рис.~\ref{fig:quadconst_2} очевидно, что
подавление полезного сигнала, расположенного в направлении $0^\circ$
отсутствует в алгоритмах LMS, использующих линейное и квадратичное
условия. Однако с другой стороны, вид диаграммы направленности
адаптированной с помощью алгоритма LMS с линейным условием
показывает наличие мощных лепестков в направлении помех. Более того,
усиление в направлении помех превышает усиление в направлении
сигнала на 4--5 дБ. В свою очередь это говорит о том, что выходное
ОСПШ такого алгоритма в рассматриваемой ситуации будет меньше ОСПШ
на входе. С другой стороны, анализ диаграммы направленности,
сформированной алгоритмом LMS с квадратичным условием, показывает,
что лепестки диаграммы в направлении помех имеют много меньшую
величину. А именно, применение алгоритма LMS с квадратичным условием
позволяет снизить уровень боковых лепестков в направлении помех при
загрязнении обучающей выборки полезным сигналом на 30--35 дБ. Данный
факт объясняется особенностями формирования члена, отвечающего за
выполнение ограничений в алгоритме LMS с квадратичным ограничением.

\conclusion

В настоящей работе рассмотрены основные свойства, принципы синтеза и
анализа адаптивных алгоритмов и особенности их применения в системе
СДЦ. В частности, в Главе~\ref{chap:Ch1} рассмотрены основные
принципы построения систем обработки радиолокационной информации на
фоне помех и приведены аргументы, показывающие необходимость
применения адаптивных систем обработки в реальной ситуации, когда
спектральные свойства помех могут быть неизвестными или
неопределенными. В Главе~\ref{chap:Ch2} представлен обзор подходов к
решению задачи АСДЦ. Приведено статистически оптимальное решение
задачи обнаружения полезного сигнала на фоне коррелированной помехи
с известными спектральными свойствами. Изложен субоптимальный подход
к решению задачи обнаружения полезного сигнала при априорной
неопределенности относительно спектральных свойств помех, основанный
на обобщенном отношении правдоподобия. После этого рассмотрен ряд
существующих алгоритмов ПВАО, в большой степени основанных на этом
подходе. В последующих двух главах приведены основные результаты
диссертационного исследования, заключающиеся в синтезе и анализе
трех адаптивных алгоритмов, предназначенных для борьбы с помехами в
системе АСДЦ. Основные результаты диссертации получены методом
математического и компьютерного моделирования и сводятся к
следующему:
\begin{enumerate}
\item Синтезирован непараметрический алгоритм итеративной оптимизации коэффициента регуляризации
алгоритма РОКМ. Показано методом компьютерного моделирования, что применение разработанного алгоритма оптимизации позволяет значительно снизить потери, возникающие при проникании полезного
сигнала в обучающую выборку. \item Синтезирован алгоритм LMS с квадратичным ограничением.
Проанализированны свойства разработанного алгоритма и  показано, что алгоритмы LMS с квадратичным и линейным ограничением при определенных условиях асимптотически эквивалентны алгоритму LMS без
ограничений. Методом компьютерного моделирования показано, что применение разработанного алгоритма позволяет увеличить скорость сходимости LMS при наложении квадратичного ограничения по сравнению
со случаем линейного ограничения. Кроме этого, при моделировании выяснено, что при проникновении полезного сигнала в обучающую выборку, LMS с квадратичным ограничением эффективнее предотвращает
просачивание помехи на выход адаптивного фильтра режекции, чем алгоритм LMS с линейным ограничением.
\end{enumerate}
Анализ результатов главы~\ref{chap:Ch3} и главы~\ref{chap:Ch4}
позволяет сделать следующие выводы.

Во--первых, применение алгоритма оптимизации величины коэффициента
регуляризации алгоритма РОКМ, предложенного в
\cite{Ore07c,Ore08b,Ore08c} позволяет получить выигрыш в выходном
ОСПШ по сравнению с алгоритмом РОКМ, использующем фиксированный
коэффициент регуляризации. А именно, в ситуации, когда полезный
сигнал присутствует в обучающей выборке и в области положительных
входных ОСПШ выигрыш составляет 3--8 дБ. В области отрицательных
ОСПШ выигрыш варьируется в пределах 0--3 дБ. Потери по отношению к
оптимальному алгоритму с полностью известной ковариационной матрицей
помехи не превышают потерь алгоритма РОКМ с фиксированным
коэффициентом регуляризации. В ситуации, когда полезный сигнал
отсутствует в обучающей выборке предлагаемый алгоритм в некоторых
случаях проигрывает алгоритму РОКМ с фиксированным коэффициентом
регуляризации. Однако соответствующие потери незначительны и
составляют 0--0,5 дБ. Потери предлагаемого алгоритма по сравнению с
оптимальным алгоритмом, использующем полностью известную матрицу
помехи практически равны потерям в алгоритме с фиксированным
коэффициентом регуляризации. Увеличение количества итераций $T$
предлагаемого алгоритма приводит к росту выигрыша в ситуации, когда
полезный сигнал присутствует в обучающей выборке и росту проигрыша в
ситуации, когда полезный сигнал отсутствует в обучающей выборке.
Поэтому $T$ может быть увеличено если допустимо соответствующие
увеличение вычислительной нагрузки на процессор ПВАО и потерь во
второй из рассмотренных ситуаций. Предлагаемый алгоритм может быть
применен в адаптивной системе СДЦ, когда существует опасность
загрязнения обучающей выборки отраженным сигналом от цели.

Во--вторых, применение алгоритма LMS с квадратичным ограничением,
предложенного в \cite{Ore07a,Ore08a} позволяет сократить объем
обучающей выборки на 20--40\% по сравнению с алгоритмом LMS с
линейным ограниченем. Данный вывод следует из анализа количества
отсчетов в обучающей выборке, необходимого для достижения
максимального ОСПШ на выходе указанных алгоритмов. Кроме этого,
применение алгоритма LMS с квадратичным условием позволяет
значительно ослабить просачивание помех на выход адаптивного
алгоритма, как это имеет место в алгоритме LMS с линейным
ограничением. А именно, применение алгоритма LMS с квадратичным
условием позволяет снизить уровень боковых лепестков в направлении
помех при загрязнении обучающей выборки полезным сигналом на 30--35
дБ. Данный алгоритм может быть применен для подстройки весов
адаптивной антенной решетки в ситуации, когда объем обучающей
выборки ограничен неоднородностью помехи и существует опасность
проникновения отражений от цели в обучающую выборку.

\appendix
\chapter{Листинги программ, использованных при моделировании} \label{app:a}

\setlength{\parindent}{0pt}

\section{подпрограммы общего назначения} \label{app:a:gen_purp}

\subsection{get\_coef.m} \label{app:a:gen_purp:get_coef}

function [M, U] = get\_coef(Interf, SimParam)

\% подпрограмма вычисления коэффициентов матричного фильтра\\
\% для генерации случайного процесса с заданным количеством\\
\% и шириной спектральных мод

sigmaSQ = (Interf.df(:, SimParam.SigChan).\textasciicircum
2)/2.77/2;

rho = exp(-2*(pi*sqrt(sigmaSQ)/SimParam.PRF).\textasciicircum 2);

R = zeros(SimParam.N);

R\_comp = zeros(SimParam.N, SimParam.N, Interf.N);

for l = 1:Interf.N

\hspace{1cm} om = 2*i*pi*Interf.freq(l,
SimParam.SigChan)/SimParam.PRF;

\hspace{1cm}    for j = 1:SimParam.N

\hspace{1cm}\hspace{1cm} R\_comp(j, 1, l) =
rho(l)\textasciicircum((j-1)\textasciicircum 2)*exp(-(1-j)*om);

\hspace{1cm}    end;

\hspace{1cm}    R\_comp(:, :, l) = R\_comp(:, :,
l)*Interf.IntPwrComp(l, SimParam.SigChan);

end;

R = sum(R\_comp, 3);

for j = 2:SimParam.N

\hspace{1cm}R(j:SimParam.N, j) = R(1:SimParam.N-j+1, 1);

\hspace{1cm}R(1:j-1, j) = flipud(conj(R(2:j, 1)));

end;

CorMtxIdeal = R;

M = CorMtxIdeal + 1e-12*eye(SimParam.N);

U = chol(M);

M = CorMtxIdeal + SimParam.NoisePWR*eye(SimParam.N);

U = U\textprimstress; 

return;

\subsection{my\_fft.m} \label{app:a:gen_purp:my_fft}

function [fft\_out freq] = my\_fft(U\_in,Points,Fs)

\% подпрограмма вычисления СПМ fft\_out входного вектора U\_in\\
\% и сетки чатот freq с количеством точек Points, соответствующей \\
\% частоте дискретизации Fs

S\_y = fft(U\_in, Points)/length(U\_in);

S\_y = S\_y.*conj(S\_y);

freq = Fs*(0:Points-1)/Points;

fft\_out = S\_y;

\section{программа моделирования оптимизации уровня регуляризации
для ослабления выбеливания цели в алгоритме РОКМ}
\label{app:a:reg_aut}

\subsection{reg\_aut\_main.m} \label{app:a:reg_aut:main}

\% программа моделирования оптимизации уровня регуляризации\\
\% для ослабления выбеливания цели в алгоритме РОКМ\\
\% см. раздел \ref{sec:SMI_reg}

clc;

\% number of pulses in a burst

NN = [8; 16; 32; 64; 128];

\% number of Monte-Carlo trials

num\_of\_trials = 1;

\% input signal to interference ratios

SIR = [20; 10; 0; -10; -20; -40; -60; -80];

\% input signal to noise ratio SNR = 10;

\% useful signal present in training sample (pres = 1) or

\% not pres = 0)

pres = 1;

\% number of pulses in a training sample relative to the number of

\% pulses in a burst

MN = 0.5;

\% width of one spectrum mode, Hz

df = 500;

\% initialize output variables

SINR\_ideal = zeros(length(NN), length(SIR));

SINR\_10dB = zeros(length(NN), length(SIR));

SINR\_est = zeros(length(NN), length(SIR));

for s\_SIR = 1:length(SIR)

\hspace{1cm}    clc;

\hspace{1cm}    fprintf(1, \textprimstress Run \%d of \%d
\textbackslash n\textprimstress , s\_SIR, length(SIR));

\hspace{1cm}   for p = 1:length(NN)

\hspace{2cm}        \% SIMULATION PARAMETERS SPECIFICATIONS

\hspace{2cm}        \%==============================

\hspace{2cm}        \% Number of training samples

\hspace{2cm}        SimParam.NumOfChan = MN*NN(p) + 1;

\hspace{2cm}        SimParam.SigChan = (SimParam.NumOfChan + 1)/2;

\hspace{2cm}        \% Noise Level, dB

\hspace{2cm}        SimParam.NoiseLevel = -70;

\hspace{2cm}        \% Signal to White Noise Ratio, dB

\hspace{2cm}        SimParam.SNR = SNR;

\hspace{2cm}        \% Signal Level, dB

\hspace{2cm}        SimParam.SigLevel = SimParam.NoiseLevel +
SimParam.SNR;

\hspace{2cm}        \% Signal to total Interference power Ratio, dB

\hspace{2cm}        SimParam.SIR = SIR(s\_SIR);

\hspace{2cm}        \% Total Interference Level

\hspace{2cm}        SimParam.IntLevel = SimParam.SigLevel -
SimParam.SIR;

\hspace{2cm}        \% Number of samples to process

\hspace{2cm}        SimParam.N = NN(p);

\hspace{2cm}        \% Sample Length (FFT length)

\hspace{2cm}        SimParam.fftN = 8*SimParam.N;

\hspace{2cm}        \% INTERFERENCE PARAMETERS SPECIFICATIONS

\hspace{2cm}        \%================================

\hspace{2cm}        a = 2;

\hspace{2cm}        \% number of components

\hspace{2cm}        Interf.N = 2;

\hspace{2cm}        \% code for interference type: 1 - gauss, 2 -
resonans

\hspace{2cm}        Interf.code = zeros(Interf.N,
SimParam.NumOfChan);

\hspace{2cm}        \% central frequency for each of interference
modes (components)

\hspace{2cm}        Interf.freq = zeros(Interf.N,
SimParam.NumOfChan);

\hspace{2cm}        \% spectrum peak width for each of interference
modes (components)

\hspace{2cm}        Interf.df = zeros(Interf.N, SimParam.NumOfChan);

\hspace{2cm}        \% modeling filter coefficients for components

\hspace{2cm}        Interf.w(Interf.N, SimParam.NumOfChan).coef = 0;

\hspace{2cm}        \% interference levels in percent from the total
interference power

\hspace{2cm}        Interf.IntPwrPercent = zeros(Interf.N,
SimParam.NumOfChan);

\hspace{2cm}        \% power of each frequency component

\hspace{2cm}        Interf.IntPwrComp = zeros(Interf.N,
SimParam.NumOfChan);

\hspace{2cm}        \% interference components themselves

\hspace{2cm}        Interf.y = zeros(SimParam.N, 2,
SimParam.NumOfChan);

\hspace{2cm}        \% Interference doppler shift transform
components

\hspace{2cm}        Interf.yDopplerI = ...

\hspace{2cm}        zeros(SimParam.N, nterf.N, SimParam.NumOfChan);

\hspace{2cm}        Interf.yDopplerQ = ...

\hspace{2cm}        zeros(SimParam.N, Interf.N, SimParam.NumOfChan);

\hspace{2cm}        Interf.Signal = zeros(SimParam.N,
SimParam.NumOfChan);

\hspace{2cm}        \% SIGNAL PARAMETERS SPECIFICATIONS

\hspace{2cm}        \%================================

\hspace{2cm}        \% Number of signal components

\hspace{2cm}        Signal.N = 1;

\hspace{2cm}        \% Doppler Frequency

\hspace{2cm}        Signal.freq = zeros(Signal.N, 1);

\hspace{2cm}        \% Initial Phase

\hspace{2cm}        Signal.phase = zeros(Signal.N, 1);

\hspace{2cm}        \% Initialize signal vector

\hspace{2cm}        Signal.x = zeros(SimParam.N, Signal.N);

\hspace{2cm}        \% SIMULATION SETUP

\hspace{2cm}        \%================================

\hspace{2cm}        \% Pulse repetition frequency (Sampling
Frequency)

\hspace{2cm}        SimParam.PRF = 20000;

\hspace{2cm}        \% Pulse Repetition Interval

\hspace{2cm}        SimParam.T = 1/SimParam.PRF;

\hspace{2cm}        \% noise power

\hspace{2cm}        SimParam.NoisePWR =
10\textasciicircum(SimParam.NoiseLevel/10);

\hspace{2cm}        \% signal power

\hspace{2cm}        SimParam.SigPWR =
10\textasciicircum((SimParam.SigLevel)/10);

\hspace{2cm}        \% total interference power

\hspace{2cm}        SimParam.IntPWR =
10\textasciicircum((SimParam.IntLevel)/10);

\hspace{2cm}        Signal.t = (0:SimParam.N-1)\textprimstress
/SimParam.PRF;

\hspace{2cm}        Signal.freq = SimParam.PRF/5;

\hspace{2cm}        Interf.f0 = 1003;

\hspace{2cm}        \% setup frequency parameters

\hspace{2cm}        for j = 1:SimParam.NumOfChan

\hspace{3cm}            Interf.IntPwrPercent(:, j) = [0.5; 0.5];

\hspace{3cm}            Interf.IntPwrComp(:, j) = ...

\hspace{3cm}            Interf.IntPwrPercent(:, j)*SimParam.IntPWR;

\hspace{3cm}            \% frequencies of interference components

\hspace{3cm}            Interf.freq(:, j) = [0; Interf.f0];

\hspace{3cm}            \% spectrum peak widths

\hspace{3cm}            Interf.df(:, j) = [df; df];

\hspace{2cm}        end;

\hspace{2cm}        \% get matrix filter coefficients

\hspace{2cm}        [Processing.CorMtxIdeal, Interf.w(1, 1).coef] =
get\_coef(Interf, SimParam);

\hspace{2cm}        N = SimParam.N;

\hspace{2cm}        M = SimParam.NumOfChan;

\hspace{2cm}       for n = 1:num\_of\_trials

\hspace{3cm}            \% Apply amplitude distribution to channels

\hspace{3cm}            for j = 1:SimParam.NumOfChan

\hspace{4cm}                Interf.white\_noiseI = ...

\hspace{4cm} sqrt(SimParam.NoisePWR)*randn(SimParam.N, 1)/sqrt(2);

\hspace{4cm}                Interf.white\_noiseQ = ...

\hspace{4cm} sqrt(SimParam.NoisePWR)*randn(SimParam.N, 1)/sqrt(2);

\hspace{4cm}                x = randn(N, 1) + i*randn(N, 1);

\hspace{4cm}                Interf.Signal(:, j) = Interf.w(1,
1).coef*x;

\hspace{4cm}                Interf.Signal(:, j) = Interf.Signal(:,
j) + ...

\hspace{4cm}                Interf.white\_noiseI +
i*Interf.white\_noiseQ;

\hspace{4cm}                phi = pi*(rand(1, 1) - 1/2)*2;

\hspace{4cm}                signal\_us =
cos(2*pi*Signal.freq*Signal.t + phi) + ...

\hspace{4cm}                    i*sin(2*pi*Signal.freq*Signal.t +
phi);

\hspace{4cm}                signal\_us =
signal\_us/sqrt(signal\_us\textprimstress *signal\_us/N);

\hspace{4cm}                signal\_us =
sqrt(SimParam.SigPWR)*(signal\_us);

\hspace{4cm}                A = sqrt(2/pi)*raylrnd(1, 1, 1);

\hspace{4cm}                Interf.Signal(:, j) = Interf.Signal(:,
j) + ...

\hspace{4cm}                    sqrt(pres)*A*signal\_us;

\hspace{3cm}            end;

\hspace{3cm}            Processing.CorMtxUni = zeros(SimParam.N);

\hspace{3cm}            for j = 1:SimParam.NumOfChan

\hspace{4cm}               \% Estimate Correlation matrix

\hspace{4cm}                Processing.CorMtxUni =
Processing.CorMtxUni + ...

\hspace{4cm}                    Interf.Signal(:, j)*Interf.Signal(:,
j)\textprimstress ;

\hspace{3cm}           end;

\hspace{3cm}            Processing.CorMtxUni = ...

\hspace{3cm}            Processing.CorMtxUni/(SimParam.NumOfChan);

\hspace{3cm}            Signal.x = cos(2*pi*Signal.freq*Signal.t) +
...

\hspace{3cm} i*sin(2*pi*Signal.freq*Signal.t);

\hspace{3cm}            Signal.x =
Signal.x/sqrt(Signal.x\textprimstress *Signal.x/length(Signal.x));

\hspace{3cm}            Signal.x = sqrt(SimParam.SigPWR)*(Signal.x);

\hspace{3cm}            Load10dB = 16*SimParam.NoisePWR;

\hspace{3cm}            s = Signal.x/(sqrt(Signal.x\textprimstress
*Signal.x));

\hspace{3cm}            w\_ideal = ((Processing.CorMtxIdeal + ...

\hspace{3cm}1e-99*eye(SimParam.N))\textbackslash Signal.x);

\hspace{3cm}            w\_10dB = ((Processing.CorMtxUni + ...

\hspace{3cm}Load10dB*eye(SimParam.N))\textbackslash Signal.x);

\hspace{3cm}            lam\_init = 0.1*Load10dB;

\hspace{3cm}            w = ((Processing.CorMtxUni +
lam\_init*eye(SimParam.N))\textbackslash s);

\hspace{3cm}            v = ((Processing.CorMtxUni +
lam\_init*eye(SimParam.N))\textbackslash w);

\hspace{3cm}            mu = 1 + real(2*(1 - w\textprimstress
*s)*w\textprimstress *v + (w\textprimstress *w)*(v\textprimstress *s))/(abs(s\textprimstress *v)\textasciicircum2);

\hspace{3cm}            lam = real(w\textprimstress *w +
v\textprimstress *s - mu*v\textprimstress
*s)/real(2*v\textprimstress *w);

\hspace{3cm}            w\_til = w - lam*v;

\hspace{3cm}            w = ((Processing.CorMtxUni +
lam*eye(SimParam.N))\textbackslash s);

\hspace{3cm}            v = ((Processing.CorMtxUni +
lam*eye(SimParam.N))\textbackslash w);

\hspace{3cm}            mu = 1 + real(2*(1 - w\textprimstress
*s)*w\textprimstress *v + (w\textprimstress *w)*(v\textprimstress *s))/(abs(s\textprimstress *v)\textasciicircum2);

\hspace{3cm}            lam = real(w\textprimstress *w +
v\textprimstress *s - mu*v\textprimstress
*s)/real(2*v\textprimstress *w);

\hspace{3cm}            w = ((Processing.CorMtxUni +
lam*eye(SimParam.N))\textbackslash s);

\hspace{3cm}            v = ((Processing.CorMtxUni +
lam*eye(SimParam.N))\textbackslash w);

\hspace{3cm}            mu = 1 + real(2*(1 - w\textprimstress
*s)*w\textprimstress *v + (w\textprimstress *w)*(v\textprimstress *s))/(abs(s\textprimstress *v)\textasciicircum2);

\hspace{3cm}            lam = real(w\textprimstress *w +
v\textprimstress *s - mu*v\textprimstress
*s)/real(2*v\textprimstress *w);

\hspace{3cm}            w\_est = ((Processing.CorMtxUni + ...

\hspace{3cm}                    (lam)*eye(SimParam.N))\textbackslash
Signal.x);

\hspace{3cm}            SINR\_ideal(p, s\_SIR) = SINR\_ideal(p,
s\_SIR) + ...

\hspace{3cm} (SimParam.SigPWR)*abs(w\_ideal\textprimstress
*Signal.x/...

\hspace{3cm}           sqrt(SimParam.SigPWR))\textasciicircum2/...

\hspace{3cm} abs(w\_ideal\textprimstress
*Processing.CorMtxIdeal*w\_ideal);

\hspace{3cm}            SINR\_10dB(p, s\_SIR) = SINR\_10dB(p,
s\_SIR) + ...

\hspace{3cm} (SimParam.SigPWR)*abs(w\_10dB\textprimstress
*Signal.x/...

\hspace{3cm}           sqrt(SimParam.SigPWR))\textasciicircum2/...

\hspace{3cm} abs(w\_10dB\textprimstress
*(Processing.CorMtxIdeal)*w\_10dB);

\hspace{3cm}            SINR\_est(p, s\_SIR) = SINR\_est(p, s\_SIR)
+ ...

\hspace{3cm}           (SimParam.SigPWR)*abs(w\_est\textprimstress
*Signal.x/...

\hspace{3cm}           sqrt(SimParam.SigPWR))\textasciicircum2/...

\hspace{3cm}           abs(w\_est\textprimstress
*(Processing.CorMtxIdeal)*w\_est);

\hspace{2cm}        end;

\hspace{1cm}    end;

end;

\% Scale the result

SINR\_idealdB = 10*log10(SINR\_ideal/num\_of\_trials);

SINR\_10dBdB = 10*log10(SINR\_10dB/num\_of\_trials);

SINR\_estdB = 10*log10(SINR\_est/num\_of\_trials);

\% Save the results

save(strcat(\textprimstress results/M\textprimstress , num2str(MN),
\textprimstress N\textprimstress ,...

\hspace{5cm}                     \textprimstress
\_sig\textprimstress , num2str(pres),...

\hspace{5cm}                     \textprimstress
\_SNR\textprimstress , num2str(SNR),...

\hspace{5cm}                     \textprimstress \_tr\textprimstress
, num2str(num\_of\_trials),...

\hspace{5cm}                     \textprimstress \_fs\textprimstress
, num2str(Signal.freq),...

\hspace{5cm}                     \textprimstress \_df\textprimstress
, num2str(df),...

\hspace{5cm}                     \textprimstress .mat\textprimstress
));

newdir = strcat(\textprimstress M\textprimstress , num2str(MN),
\textprimstress N\textprimstress ,...

\hspace{5cm}                     \textprimstress
\_sig\textprimstress , num2str(pres),...

\hspace{5cm}                     \textprimstress
\_SNR\textprimstress , num2str(SNR),...

\hspace{5cm}                     \textprimstress \_tr\textprimstress
, num2str(num\_of\_trials),...

\hspace{5cm}                     \textprimstress \_fs\textprimstress
, num2str(Signal.freq),...

\hspace{5cm}                     \textprimstress \_df\textprimstress
, num2str(df));

mkdir(strcat(\textprimstress fig/\textprimstress , newdir));

\% plot and save the figures

for j = 1:length(SIR)

\hspace{1cm}    h = plot(NN, SINR\_idealdB(:, j), \textprimstress
:k\textprimstress , \textprimstress linewidth\textprimstress , 2.5);

\hspace{1cm}    hold on;

\hspace{1cm}    plot(NN, SINR\_10dBdB(:, j), \textprimstress
-.k\textprimstress , \textprimstress linewidth\textprimstress ,
2.5);

\hspace{1cm}    plot(NN, SINR\_estdB(:, j), \textprimstress
-k\textprimstress , \textprimstress linewidth\textprimstress , 2.5);

\hspace{1cm}    xlabel(\textprimstress N\textprimstress ,
\textprimstress FontSize\textprimstress , 18);

\hspace{1cm}    ylabel(\textprimstress Output SINR,
dB\textprimstress , \textprimstress FontSize\textprimstress , 18);

\hspace{1cm}    set(gca, \textprimstress FontSize\textprimstress ,
18);

\hspace{1cm}    ylim([min([SINR\_10dBdB(:, j); SINR\_estdB(:, j)]);
max(SINR\_idealdB(:, j))]);

\hspace{1cm}    xlim([min(NN); max(NN)]);

\hspace{1cm}    saveas(h, strcat(\textprimstress fig/\textprimstress
, newdir, \textprimstress /\textprimstress , newdir,...

\hspace{5cm}                     \textprimstress
\_SIR\textprimstress , num2str(SIR(j)),...

\hspace{5cm}                     \textprimstress .eps\textprimstress
));

\hspace{1cm}    hold off;

\hspace{1cm}    delete(gcf);

end;

\section{программы моделирования алгоритма LMS с квадратичным ограничением}
\label{app:a:LMS_sq}

\subsection{quad\_LMS\_PSD.m} \label{app:a:LMS_sq:PSD}

\% программа моделирования вида диаграммы направленности антенны\\
\% после подстройки коэффициентов при помощи алгоритмов LMS без \\
\% ограничений, LMS с линейным ограничением (алгоритм Фроста) \\
\% и предлагаемый алгоритм LMS с квадратичным ограничением \\
\% см. раздел \ref{sec:LMS_sq} и Рис. \ref{fig:quadconst_2}
\\

clc;

\% Number of antenna elements

NN = [8; 16; 32; 64; 128];

\% NN = [256; 512; 1024];

NN = 128;

\% Number of Monte-Carlo Trials

num\_of\_trials = 100;

\% useful signal present in training sample (pres = 1) or \% not
(pres = 0)

sig\_pres = 1;

\% Number of samples in FFT

Nf = 64*64;

\% initialize output vectors

PSD\_My = zeros(Nf, length(NN));

PSD\_Frost = zeros(Nf, length(NN));

PSD\_LMS = zeros(Nf, length(NN));

for p = 1:length(NN)

\hspace{1cm}    clc;

\hspace{1cm}    fprintf(1, \textprimstress Run \%d of \%d
\textbackslash n\textprimstress , p, length(NN));

\hspace{1cm}    \% SIMULATION PARAMETERS SPECIFICATIONS

\hspace{1cm}    \%================================

\hspace{1cm}    \% Number of channels to process

\hspace{1cm}    SimParam.NumOfChan = 1024; \% 1*NN(p) + 1; \% 256

\hspace{1cm}    \% Number of channel in which signal is

\hspace{1cm}    SimParam.SigChan = (SimParam.NumOfChan + 1)/2;

\hspace{1cm}    \% Noise Level, dB

\hspace{1cm}    SimParam.NoiseLevel = -60;

\hspace{1cm}    \% Signal to White Noise Ratio, dB

\hspace{1cm}    SimParam.SNR = 10;

\hspace{1cm}    \% Signal Level, dB

\hspace{1cm}    SimParam.SigLevel = SimParam.NoiseLevel +
SimParam.SNR;

\hspace{1cm}    \% Signal to total Interference power Ratio, dB

\hspace{1cm}    SimParam.SIR = 10;

\hspace{1cm}    \% Total Interference Level

\hspace{1cm}    SimParam.IntLevel = SimParam.SigLevel -
SimParam.SIR;

\hspace{1cm}    \% Num of samples to process1

\hspace{1cm}    SimParam.N = NN(p);

\hspace{1cm}    \% Sample Length (FFT length)

\hspace{1cm}    SimParam.fftN = 8*SimParam.N; \% 2*1024

\hspace{1cm}    N = SimParam.N;

\hspace{1cm}    \% noise power

\hspace{1cm}    SimParam.NoisePWR =
10\textasciicircum(SimParam.NoiseLevel/10);

\hspace{1cm}    \% signal power

\hspace{1cm}    SimParam.SigPWR =
10\textasciicircum((SimParam.SigLevel)/10);

\hspace{1cm}    \% total interference power

\hspace{1cm}    SimParam.IntPWR =
10\textasciicircum((SimParam.IntLevel)/10);

\hspace{1cm}    \% INTERFERENCE PARAMETERS SPECIFICATIONS

\hspace{1cm}    \%==================================

\hspace{1cm}    \% angle for each of interference modes (components)

\hspace{1cm}    Interf.th = pi*[-50; -25; 25]/180;

\hspace{1cm}    Interf.th = pi*[-14; 20; 71]/180;

\hspace{1cm}    \%     Interf.th = pi*[-33; 14; 71; -60]/180;

\hspace{1cm}    Interf.th = pi*[-14; 71; 66]/180;

\hspace{1cm}    \% number of components

\hspace{1cm}    Interf.N = length(Interf.th);

\hspace{1cm}    \% power distribution among components

 \hspace{1cm}   Interf.PwrPerc(1:Interf.N, 1) = 1/Interf.N;

\hspace{1cm}    \% interference components

\hspace{1cm}    Interf.y = zeros(SimParam.N, SimParam.NumOfChan);

\hspace{1cm}    \% interference steering vectors

\hspace{1cm}    Interf.e = zeros(N, Interf.N);

\hspace{1cm}    Interf.Signal = zeros(SimParam.N,
SimParam.NumOfChan);

\hspace{1cm}    Interf.PwrComp = Interf.PwrPerc*SimParam.IntPWR;

\hspace{1cm}    \% SIGNAL PARAMETERS SPECIFICATIONS

\hspace{1cm}    \%=============================

\hspace{1cm}    \% Number of signal components

\hspace{1cm}    Signal.N = 1;

\hspace{1cm}    \% Angle

\hspace{1cm}    Signal.th = 0;

\hspace{1cm}    \% initialize signal vector

\hspace{1cm}    Signal.x = zeros(SimParam.N, Signal.N);

\hspace{1cm}    \% SIMULATION SETUP

\hspace{1cm}    \%==============================

\hspace{1cm}    \% Pulse repetition frequency (Sampling Frequency)

\hspace{1cm}    SimParam.PRF = 1;

\hspace{1cm}    \% Pulse Repetition Interval

\hspace{1cm}    SimParam.T = 1/SimParam.PRF;

\hspace{1cm}    Signal.t = (0:SimParam.N-1)\textprimstress
/SimParam.PRF;

\hspace{1cm}    for n = 1:num\_of\_trials

\hspace{2cm}        \% generate interference steering angles

\hspace{2cm}        k = (0:N-1)\textprimstress  - (N-1)/2;

\hspace{2cm}        Processing.CorMtxIdeal = zeros(N);

\hspace{2cm}        for j = 1:Interf.N

\hspace{3cm}            phi = pi*sin(Interf.th(j));

\hspace{3cm}            Interf.e(:, j) = exp(i*phi*k);

\hspace{3cm}            Interf.e(:, j) = sqrt(N)*Interf.e(:, j)/...

\hspace{3cm}                sqrt(Interf.e(:, j)\textprimstress
*Interf.e(:, j));

\hspace{3cm}            Processing.CorMtxIdeal =
Processing.CorMtxIdeal + ...

\hspace{3cm}                (Interf.PwrComp(j))*(Interf.e(:,
j)*Interf.e(:, j)\textprimstress );

\hspace{2cm}        end;

\hspace{2cm}      Processing.CorMtxIdeal = Processing.CorMtxIdeal +
...

\hspace{2cm}            SimParam.NoisePWR*eye(N);

\hspace{2cm}        phi = pi*sin(Signal.th);

\hspace{2cm}        Signal.x = exp(i*phi*k);

\hspace{2cm}        Signal.x = Signal.x/sqrt(Signal.x\textprimstress
*Signal.x/length(Signal.x));

\hspace{2cm}        Signal.x = sqrt(SimParam.SigPWR)*(Signal.x);

\hspace{2cm}        s = Signal.x/sqrt(Signal.x\textprimstress
*Signal.x);

\hspace{2cm}        \% generate training sample

\hspace{2cm}        Interf.Signal = zeros(SimParam.N,
SimParam.NumOfChan);

\hspace{2cm}        for j = 1:SimParam.NumOfChan

\hspace{3cm}            rayl = raylrnd(sqrt(2/pi), Interf.N, 1);

\hspace{3cm}            Interf.white\_noiseI =
sqrt(SimParam.NoisePWR)*...

\hspace{3cm}                randn(SimParam.N, 1)/sqrt(2);

\hspace{3cm}            Interf.white\_noiseQ =
sqrt(SimParam.NoisePWR)*...

\hspace{3cm}                randn(SimParam.N, 1)/sqrt(2);

\hspace{3cm}            for m = 1:Interf.N

\hspace{4cm}                Interf.Signal(:, j) = Interf.Signal(:,
j) ...

\hspace{4cm}                    + Interf.e(:,
m)*rayl(m)*sqrt(Interf.PwrComp(m));

\hspace{3cm}            end;

\hspace{3cm}            Interf.Signal(:, j) = Interf.Signal(:, j) +
...

\hspace{3cm}                Interf.white\_noiseI +
i*Interf.white\_noiseQ + ...

\hspace{3cm}                sqrt(sig\_pres)*Signal.x;

\hspace{2cm}        end;

\hspace{2cm}        \% Estimate Correlation matrix

\hspace{2cm}        Processing.CorMtxUni = zeros(SimParam.N);

\hspace{2cm}        X = zeros(N, SimParam.NumOfChan - 1);

\hspace{2cm}        count = 1;

\hspace{2cm}        for j = 1:SimParam.NumOfChan

\hspace{3cm}            if (j $\sim$= SimParam.SigChan)

\hspace{4cm}                Processing.CorMtxUni =
Processing.CorMtxUni + ...

\hspace{4cm}                    (Interf.Signal(:,
j))*(Interf.Signal(:, j))\textprimstress ;

\hspace{4cm}                X(:, count) = Interf.Signal(:, j);

\hspace{4cm}                count = count + 1;

\hspace{3cm}            end;

\hspace{2cm}        end;

\hspace{2cm}        Processing.CorMtxUni = Processing.CorMtxUni...

\hspace{2cm}            /(SimParam.NumOfChan - 1);

\hspace{2cm}        Load10dB = 16*SimParam.NoisePWR;

\hspace{2cm}        w\_ideal = ((Processing.CorMtxIdeal + ...

\hspace{2cm}            1e-99*eye(SimParam.N))\textbackslash
Signal.x);

\hspace{2cm}        w\_10dB = ((Processing.CorMtxUni + ...

\hspace{2cm}        Load10dB*eye(SimParam.N))\textbackslash
Signal.x);

\hspace{2cm}        mu\_0 = 0.25;

\hspace{2cm}        w = s;

\hspace{2cm}        for m = 1:SimParam.NumOfChan-1

\hspace{3cm}            x = X(:, m);

\hspace{3cm}            mu = mu\_0/(x\textprimstress *x);

\hspace{3cm}            l = abs(w\textprimstress
*x)\textasciicircum2/abs(w\textprimstress *s)\textasciicircum2;

\hspace{3cm}            w = w - x*2*mu*(x\textprimstress *w) +
l*2*mu*(s\textprimstress *w);

\hspace{3cm}            w = w/sqrt(w\textprimstress *w);

\hspace{2cm}        end;

\hspace{2cm}        w\_e = s;

\hspace{2cm}        l = 1;

\hspace{2cm}        for m = 1:SimParam.NumOfChan-1

\hspace{3cm}            x = X(:, m);

\hspace{3cm}            y = x\textprimstress *w\_e;

\hspace{3cm}            mu = mu\_0/(x\textprimstress *x);

\hspace{3cm}            e = w\_e - 2*mu*y*x;

\hspace{3cm}            w\_e = e - sum(e)/length(e) +
1/length(w\_e);

\hspace{2cm}        end;

\hspace{2cm}        w\_LMS = s;

\hspace{2cm}        l = 1;

\hspace{2cm}        for m = 1:SimParam.NumOfChan-1

\hspace{3cm}            x = X(:, m);

\hspace{3cm}            y = w\_LMS\textprimstress *x;

\hspace{3cm}            mu = mu\_0/(x\textprimstress *x);

\hspace{3cm}            w\_LMS = w\_LMS - x*2*mu*y\textprimstress ;

\hspace{3cm}            w\_LMS = w\_LMS/sqrt(w\_LMS\textprimstress
*w\_LMS);

\hspace{2cm}        end;

\hspace{2cm}        [pattern1 angle] = my\_fft(w, Nf, 1);

\hspace{2cm}        pattern1 = circshift(pattern1, Nf/2);

\hspace{2cm}        [pattern2 angle] = my\_fft(w\_e, Nf, 1);

\hspace{2cm}        pattern2 = circshift(pattern2, Nf/2);

\hspace{2cm}        [pattern3 angle] = my\_fft(w\_LMS, Nf, 1);

\hspace{2cm}        pattern3 = circshift(pattern3, Nf/2);

\hspace{2cm}        angle = 2*(angle - 0.5);

\hspace{2cm}        angle = 180*asin(angle)/pi;

\hspace{2cm}        PSD\_My = PSD\_My + pattern1;

\hspace{2cm}        PSD\_Frost = PSD\_Frost + pattern2;

\hspace{2cm}        PSD\_LMS = PSD\_LMS + pattern3;

\hspace{1cm}    end;

\hspace{1cm}    PSD\_My(:, p) = N*PSD\_My(:, p)/num\_of\_trials;

\hspace{1cm}    PSD\_Frost(:, p) = N\textasciicircum2*PSD\_Frost(:,
p)/num\_of\_trials;

\hspace{1cm}    PSD\_LMS(:, p) = N*PSD\_LMS(:, p)/num\_of\_trials;

\hspace{1cm}    PSD\_My(:, p) = 10*log10(PSD\_My(:, p));

\hspace{1cm}    PSD\_Frost(:, p) = 10*log10(PSD\_Frost(:, p));

\hspace{1cm}    PSD\_LMS(:, p) = 10*log10(PSD\_LMS(:, p));

end;

\% Save the results

save(strcat(\textprimstress results/PSD\textprimstress ,...

\hspace{4cm}                     \textprimstress \_M\textprimstress
, num2str(SimParam.NumOfChan),...

\hspace{4cm}                     \textprimstress \_N\textprimstress
, num2str(NN),...

\hspace{4cm}                     \textprimstress
\_sig\textprimstress , num2str(sig\_pres),...

\hspace{4cm}                     \textprimstress
\_SNR\textprimstress , num2str(SimParam.SNR),...

\hspace{4cm}                     \textprimstress
\_SIR\textprimstress , num2str(SimParam.SIR),...

\hspace{4cm}                     \textprimstress \_tr\textprimstress
, num2str(num\_of\_trials),...

\hspace{4cm}                     \textprimstress .mat\textprimstress
));

newdir = strcat(\textprimstress PSD\textprimstress ,...

\hspace{4cm}                     \textprimstress \_M\textprimstress
, num2str(SimParam.NumOfChan),...

\hspace{4cm}                     \textprimstress \_N\textprimstress
, num2str(NN),...

\hspace{4cm}                     \textprimstress
\_sig\textprimstress , num2str(sig\_pres),...

\hspace{4cm}                     \textprimstress
\_SNR\textprimstress , num2str(SimParam.SNR),...

\hspace{4cm}                     \textprimstress
\_SIR\textprimstress , num2str(SimParam.SIR),...

\hspace{4cm}                     \textprimstress \_tr\textprimstress
, num2str(num\_of\_trials),...

\hspace{4cm}                     \textprimstress .mat\textprimstress
);

mkdir(strcat(\textprimstress fig/\textprimstress , newdir));

\% plot and save the figures

h = plot(angle, PSD\_My, \textprimstress k\textprimstress ,
\textprimstress linewidth\textprimstress , 2);

ylim([-50; 5]);

xlim([-90; 90]);

xlabel(\textprimstress \textbackslash phi\textprimstress );

ylabel(\textprimstress Gain, dB\textprimstress );

saveas(h, strcat(\textprimstress fig/\textprimstress , newdir,
\textprimstress /\textprimstress , newdir,...

\hspace{4cm}                     \textprimstress
\_PSD\_My\textprimstress ,\textprimstress .eps\textprimstress  ));

delete(gcf);

h = plot(angle, PSD\_Frost, \textprimstress k\textprimstress ,
\textprimstress linewidth\textprimstress , 2);

ylim([-50; 5]);

xlim([-90; 90]);

xlabel(\textprimstress \textbackslash phi\textprimstress );

ylabel(\textprimstress Gain, dB\textprimstress );

saveas(h, strcat(\textprimstress fig/\textprimstress , newdir,
\textprimstress /\textprimstress , newdir,...

\hspace{4cm}                     \textprimstress
PSD\_Frost\textprimstress ,\textprimstress .eps\textprimstress  ));

delete(gcf);

h = plot(angle, PSD\_LMS, \textprimstress k\textprimstress ,
\textprimstress linewidth\textprimstress , 2);

ylim([-50; 5]);

xlim([-90; 90]);

xlabel(\textprimstress \textbackslash phi\textprimstress );

ylabel(\textprimstress Gain, dB\textprimstress );

saveas(h, strcat(\textprimstress fig/\textprimstress , newdir,
\textprimstress /\textprimstress , newdir,...

\hspace{4cm}         \textprimstress PSD\_LMS\textprimstress
,\textprimstress .eps\textprimstress  ));

\subsection{quad\_LMS\_learn\_curve.m} \label{app:a:LMS_sq:LearnCurve}

\% программа моделирования вида зависимости ОСПШ на выходе от\\
\% количества обучающих выборок (кривой обучения) для алгоритма  \\
\% LMS с линейным ограничением (алгоритма Фроста) \\
\% и предлагаемого алгоритма LMS с квадратичным ограничением \\
\% см. раздел \ref{sec:LMS_sq} и Рис. \ref{fig:quadconst_1}
\\

clc;

\% Number of antenna elements

NN = [8; 16; 32; 64; 128; 256; 512; 1024];

\% mu\_0

mu\_mu = 0.25;

\% Number of training samples

MM = [8; 16; 32; 64; 96; 128; 192; 256];

\% Number of Monte-Carlo Trials

num\_of\_trials = 100;

\% useful signal present in training sample (pres = 1) or

\% not (pres = 0)

sig\_pres = 0;

\% initialize output vectors

SINR\_ideal = zeros(length(NN), length(MM));

SINR\_10dB = zeros(length(NN), length(MM));

SINR\_est = zeros(length(NN), length(MM));

SINR\_estMy = zeros(length(NN), length(MM));

\% Number of samples in FFT

Nf = 64*64;

for p = 1:length(NN)

\hspace{1cm}    clc;

\hspace{1cm}    fprintf(1, \textprimstress Run \%d of \%d
\textbackslash n\textprimstress , p, length(NN));

\hspace{1cm}    \% SIMULATION PARAMETERS SPECIFICATIONS

\hspace{1cm}    \%=============================

\hspace{1cm}    \% Number of channels to process

\hspace{1cm}    SimParam.NumOfChan = max(MM);

\hspace{1cm}    \% Noise Level, dB

\hspace{1cm}    SimParam.NoiseLevel = -60;

\hspace{1cm}    \% Signal to White Noise Ratio, dB

\hspace{1cm}    SimParam.SNR = 10;

\hspace{1cm}    \% Signal Level, dB

\hspace{1cm}    SimParam.SigLevel = SimParam.NoiseLevel +
SimParam.SNR;

\hspace{1cm}    \% Signal to total Interference power Ratio, dB

\hspace{1cm}    SimParam.SIR = -60;

\hspace{1cm}    \% Total Interference Level

\hspace{1cm}    SimParam.IntLevel = SimParam.SigLevel -
SimParam.SIR;

\hspace{1cm}    \% Number of samples to process

\hspace{1cm}    SimParam.N = NN(p);

\hspace{1cm}    N = SimParam.N;

\hspace{1cm}    \% noise power

\hspace{1cm}    SimParam.NoisePWR =
10\textasciicircum(SimParam.NoiseLevel/10);

\hspace{1cm}    \% signal power

\hspace{1cm}    SimParam.SigPWR =
10\textasciicircum((SimParam.SigLevel)/10);

\hspace{1cm}    \% total interference power

\hspace{1cm}    SimParam.IntPWR =
10\textasciicircum((SimParam.IntLevel)/10);

\hspace{1cm}    \% INTERFERENCE PARAMETERS SPECIFICATIONS

\hspace{1cm}    \%====================================

\hspace{1cm}    \% arrival angles of interference modes (components)

\hspace{1cm}    Interf.th = pi*[-50; -25; 25]/180;

\hspace{1cm}    Interf.th = pi*[-14; 71; 66]/180; \% pi*[-14; 71;
60]/180;

\hspace{1cm}    \% number of components

\hspace{1cm}    Interf.N = length(Interf.th);

\hspace{1cm} \% power distribution among components

\hspace{1cm} Interf.PwrPerc(1:Interf.N, 1) = 1/Interf.N;

\hspace{1cm} \% interference components

\hspace{1cm} Interf.y = zeros(SimParam.N, SimParam.NumOfChan);

\hspace{1cm} \% interference steering vectors

\hspace{1cm} Interf.e = zeros(N, Interf.N);

\hspace{1cm} Interf.Signal = zeros(SimParam.N, SimParam.NumOfChan);

\hspace{1cm} Interf.PwrComp = Interf.PwrPerc*SimParam.IntPWR;

\hspace{1cm} \% SIGNAL PARAMETERS SPECIFICATIONS

\hspace{1cm} \%============================

\hspace{1cm} \% Number of signal components

\hspace{1cm} Signal.N = 1;

\hspace{1cm} \% Angle

\hspace{1cm} Signal.th = 0;

\hspace{1cm} \% initialize signal vector

\hspace{1cm} Signal.x = zeros(SimParam.N, Signal.N);

\hspace{1cm} \% SIMULATION SETUP

\hspace{1cm} \%=============================

\hspace{1cm} \% Pulse repetition frequency (Sampling Frequency)

\hspace{1cm} SimParam.PRF = 1;

\hspace{1cm} \% Pulse Repetition Interval

\hspace{1cm} SimParam.T = 1/SimParam.PRF;

\hspace{1cm} Signal.t = (0:SimParam.N-1)\textprimstress
/SimParam.PRF;

\hspace{1cm} for n = 1:num\_of\_trials

\hspace{2cm}        k = (0:N-1)\textprimstress  - (N-1)/2;

\hspace{2cm}        Processing.CorMtxIdeal = zeros(N);

\hspace{2cm}        for j = 1:Interf.N

\hspace{3cm}            phi = pi*sin(Interf.th(j));

\hspace{3cm}            Interf.e(:, j) = exp(i*phi*k);

\hspace{3cm}            Interf.e(:, j) = sqrt(N)*Interf.e(:, j)/...

\hspace{3cm}                sqrt(Interf.e(:, j)\textprimstress
*Interf.e(:, j));

\hspace{3cm}            Processing.CorMtxIdeal =

\hspace{3cm}Processing.CorMtxIdeal + ...

\hspace{3cm}                (Interf.PwrComp(j))*(Interf.e(:,
j)*Interf.e(:, j)\textprimstress );

\hspace{2cm}        end;

\hspace{2cm}        Processing.CorMtxIdeal = Processing.CorMtxIdeal
+ ...

\hspace{2cm}            SimParam.NoisePWR*eye(N);

\hspace{2cm}        phi = pi*sin(Signal.th);

\hspace{2cm}        Signal.x = exp(i*phi*k);

\hspace{2cm}        Signal.x = Signal.x/sqrt(Signal.x\textprimstress
*Signal.x/length(Signal.x));

\hspace{2cm}        Signal.x = sqrt(SimParam.SigPWR)*(Signal.x);

\hspace{2cm}        s = Signal.x/sqrt(Signal.x\textprimstress
*Signal.x);

\hspace{2cm}        \% Apply amplitude distribution to channels

\hspace{2cm}        Interf.Signal = zeros(SimParam.N,
SimParam.NumOfChan);

\hspace{2cm}        for j = 1:SimParam.NumOfChan

\hspace{3cm}            rayl = raylrnd(sqrt(2/pi), Interf.N, 1);

\hspace{3cm}            Interf.white\_noiseI =
sqrt(SimParam.NoisePWR)*...

\hspace{3cm}                randn(SimParam.N, 1)/sqrt(2);

\hspace{3cm}            Interf.white\_noiseQ =
sqrt(SimParam.NoisePWR)*...

\hspace{3cm}                randn(SimParam.N, 1)/sqrt(2);

\hspace{3cm}            for m = 1:Interf.N

\hspace{4cm}                Interf.Signal(:, j) = Interf.Signal(:,
j) ...

\hspace{4cm}                    + Interf.e(:,
m)*rayl(m)*sqrt(Interf.PwrComp(m));

\hspace{3cm}            end;

\hspace{3cm}            Interf.Signal(:, j) = Interf.Signal(:, j) +
...

\hspace{3cm}                Interf.white\_noiseI +
i*Interf.white\_noiseQ + ...

\hspace{3cm}                sqrt(sig\_pres)*Signal.x;

\hspace{2cm}        end;

\hspace{2cm}        m\_c = 1;

\hspace{2cm}        Processing.CorMtxUni = zeros(SimParam.N);

\hspace{2cm}        X = zeros(N, SimParam.NumOfChan - 1);

\hspace{2cm}        count = 1;

\hspace{2cm}        for j = 1:MM(m\_c)

\hspace{3cm}            \% Estimate Correlation matrix

\hspace{3cm}            Processing.CorMtxUni = Processing.CorMtxUni
+ ...

\hspace{3cm}                (Interf.Signal(:, j))*(Interf.Signal(:,
j))\textprimstress ;

\hspace{3cm}            X(:, count) = Interf.Signal(:, j);

\hspace{3cm}            count = count + 1;

\hspace{2cm}        end;

\hspace{2cm}        Processing.CorMtxUni =
Processing.CorMtxUni/(MM(m\_c));

\hspace{2cm}        Load10dB = 16*SimParam.NoisePWR;

\hspace{2cm}        w\_ideal = ((Processing.CorMtxIdeal + ...

\hspace{2cm}            1e-99*eye(SimParam.N))\textbackslash
Signal.x);

\hspace{2cm}        w\_10dB = ((Processing.CorMtxUni + ...

\hspace{2cm}            Load10dB*eye(SimParam.N))\textbackslash
Signal.x);

\hspace{2cm}        mu\_0 = mu\_mu;

\hspace{2cm}        l = 1;

\hspace{2cm}        w = s;

\hspace{2cm}        for m = 1:MM(m\_c)

\hspace{3cm}            x = X(:, m);

\hspace{3cm}            mu = 2*mu\_0/(x\textprimstress *x);

\hspace{3cm}            l = abs(w\textprimstress
*x)\textasciicircum2/abs(w\textprimstress *s)\textasciicircum2;

\hspace{3cm}            w = w - mu*(x*(x\textprimstress *w) -
l*s*(s\textprimstress *w));

\hspace{3cm}            ll(m) = l;

\hspace{2cm}        end;

\hspace{2cm}        mu\_0 = mu\_mu;

\hspace{2cm}        w\_e = s;

\hspace{2cm}        l = 1;

\hspace{2cm}        for m = 1:MM(m\_c)

\hspace{3cm}            x = X(:, m);

\hspace{3cm}            y = x\textprimstress *w\_e;

\hspace{3cm}            mu = 2*mu\_0/(x\textprimstress *x);

\hspace{3cm}            e = w\_e - mu*y*x;

\hspace{3cm}            w\_e = e - sum(e)/length(e) +
1/length(w\_e);

\hspace{2cm}        end;

\hspace{2cm}        SINR\_ideal(p, m\_c) = SINR\_ideal(p, m\_c) +
...

\hspace{2cm}            abs(w\_ideal\textprimstress
*Signal.x)\textasciicircum2/...

\hspace{2cm} abs(w\_ideal\textprimstress
*Processing.CorMtxIdeal*w\_ideal);

\hspace{2cm}        SINR\_10dB(p, m\_c) = SINR\_10dB(p, m\_c) + ...

\hspace{2cm}            abs(w\_10dB\textprimstress
*Signal.x)\textasciicircum2/...

\hspace{2cm} abs(w\_10dB\textprimstress
*(Processing.CorMtxIdeal)*w\_10dB);

\hspace{2cm}        SINR\_est(p, m\_c) = SINR\_est(p, m\_c) + ...

\hspace{2cm}        abs(w\textprimstress
*Signal.x)\textasciicircum2/...

\hspace{2cm}            abs(w\textprimstress
*(Processing.CorMtxIdeal)*w);

\hspace{2cm}        SINR\_estMy(p, m\_c) = SINR\_estMy(p, m\_c) +
...

\hspace{2cm}        abs(w\_e\textprimstress
*Signal.x)\textasciicircum2/...

\hspace{2cm}            abs(w\_e\textprimstress
*(Processing.CorMtxIdeal)*w\_e);

\hspace{2cm}        for m\_c = 2:length(MM)

\hspace{3cm}            Processing.CorMtxUni =
Processing.CorMtxUni*(MM(m\_c-1));

\hspace{3cm}            for j = MM(m\_c-1)+1:MM(m\_c)

\hspace{4cm}                \% Estimate Correlation matrix

\hspace{4cm}                Processing.CorMtxUni =
Processing.CorMtxUni + ...

\hspace{4cm}                    (Interf.Signal(:,
j))*(Interf.Signal(:, j))\textprimstress ;

\hspace{4cm}                X(:, count) = Interf.Signal(:, j);

\hspace{4cm}                count = count + 1;

\hspace{3cm}            end;

\hspace{3cm}            Processing.CorMtxUni =
Processing.CorMtxUni/MM(m\_c);

\hspace{3cm}            Load10dB = 16*SimParam.NoisePWR;

\hspace{3cm}            w\_ideal = ((Processing.CorMtxIdeal + ...

\hspace{3cm}                1e-99*eye(SimParam.N))\textbackslash
Signal.x);

\hspace{3cm}            w\_10dB = ((Processing.CorMtxUni + ...

\hspace{3cm}                Load10dB*eye(SimParam.N))\textbackslash
Signal.x);

\hspace{3cm}            mu\_0 = mu\_mu;

\hspace{3cm}            l = 1;

\hspace{3cm}            for m = MM(m\_c-1)+1:MM(m\_c)

\hspace{4cm}                x = X(:, m);

\hspace{4cm}                mu = 2*mu\_0/(x\textprimstress *x);

\hspace{4cm}                l = abs(w\textprimstress
*x)\textasciicircum2/abs(w\textprimstress *s)\textasciicircum2;

\hspace{4cm}                w = w - mu*(x*(x\textprimstress *w) -
l*s*(s\textprimstress *w));

\hspace{4cm}                ll(m) = l;

\hspace{3cm}            end;

\hspace{3cm}            mu\_0 = mu\_mu;

\hspace{3cm}            l = 1;

\hspace{3cm}            for m = MM(m\_c-1)+1:MM(m\_c)

\hspace{4cm}                x = X(:, m);

\hspace{4cm}                y = x\textprimstress *w\_e;

\hspace{4cm}                mu = 2*mu\_0/(x\textprimstress *x);

\hspace{4cm}                e = w\_e - mu*y*x;

\hspace{4cm}                w\_e = e - sum(e)/length(e) +
1/length(w\_e);

\hspace{3cm}            end;

\hspace{3cm}            SINR\_ideal(p, m\_c) = SINR\_ideal(p, m\_c)
+ ...

\hspace{3cm} abs(w\_ideal\textprimstress
*Signal.x)\textasciicircum2/...

\hspace{3cm} abs(w\_ideal\textprimstress
*Processing.CorMtxIdeal*w\_ideal);

\hspace{3cm}            SINR\_10dB(p, m\_c) = SINR\_10dB(p, m\_c) +
...

\hspace{3cm} abs(w\_10dB\textprimstress
*Signal.x)\textasciicircum2/...

\hspace{3cm} abs(w\_10dB\textprimstress
*(Processing.CorMtxIdeal)*w\_10dB);

\hspace{3cm}            SINR\_est(p, m\_c) = SINR\_est(p, m\_c) +
...

\hspace{3cm} abs(w\textprimstress *Signal.x)\textasciicircum2/...

\hspace{3cm}                abs(w\textprimstress
*(Processing.CorMtxIdeal)*w);

\hspace{3cm}            SINR\_estMy(p, m\_c) = SINR\_estMy(p, m\_c)
+ ...

\hspace{3cm}                abs(w\_e\textprimstress
*Signal.x)\textasciicircum2/...

\hspace{3cm} abs(w\_e\textprimstress
*(Processing.CorMtxIdeal)*w\_e);

\hspace{2cm}        end;

\hspace{1cm}    end;

end;

SINR\_idealdB = 10*log10(SINR\_ideal/num\_of\_trials);

SINR\_10dBdB = 10*log10(SINR\_10dB/num\_of\_trials);

SINR\_estdB = 10*log10(SINR\_est/num\_of\_trials);

SINR\_estMydB = 10*log10(SINR\_estMy/num\_of\_trials);

\% Save the results

save(strcat(\textprimstress results/LC\textprimstress ,...

\hspace{4cm}                     \textprimstress \_M\textprimstress
, num2str(SimParam.NumOfChan),...

\hspace{4cm}                     \textprimstress \_N\textprimstress
, num2str(NN(1)), \textprimstress \_\textprimstress ,
num2str(NN(length(NN))),...

\hspace{4cm}                     \textprimstress
\_sig\textprimstress , num2str(sig\_pres),...

\hspace{4cm}                     \textprimstress
\_SNR\textprimstress , num2str(SimParam.SNR),...

\hspace{4cm}                     \textprimstress
\_SIR\textprimstress , num2str(SimParam.SIR),...

\hspace{4cm}                     \textprimstress \_tr\textprimstress
, num2str(num\_of\_trials),...

\hspace{4cm}                     \textprimstress .mat\textprimstress
));

newdir = strcat(\textprimstress LC\textprimstress ,...

\hspace{4cm}                     \textprimstress \_M\textprimstress
, num2str(SimParam.NumOfChan),...

\hspace{4cm}                     \textprimstress \_N\textprimstress
, num2str(NN(1)), \textprimstress \_\textprimstress ,
num2str(NN(length(NN))),...

\hspace{4cm}                     \textprimstress
\_sig\textprimstress , num2str(sig\_pres),...

\hspace{4cm}                     \textprimstress
\_SNR\textprimstress , num2str(SimParam.SNR),...

\hspace{4cm}                     \textprimstress
\_SIR\textprimstress , num2str(SimParam.SIR),...

\hspace{4cm}                     \textprimstress \_tr\textprimstress
, num2str(num\_of\_trials),...

\hspace{4cm}                     \textprimstress .mat\textprimstress
);

mkdir(strcat(\textprimstress fig/\textprimstress , newdir));

\% plot and save the figures

for pl\_col = 1:length(NN)

\hspace{1cm}    h = plot(MM, SINR\_idealdB(pl\_col, :),
\textprimstress :k\textprimstress , \textprimstress
linewidth\textprimstress , 2);

\hspace{1cm}    hold on;

\hspace{1cm}    plot(MM, SINR\_10dBdB(pl\_col, :), \textprimstress
-.k\textprimstress , \textprimstress linewidth\textprimstress , 2);

\hspace{1cm}    plot(MM, SINR\_estdB(pl\_col, :), \textprimstress
k\textprimstress ,  \textprimstress linewidth\textprimstress , 2);

\hspace{1cm}    plot(MM, SINR\_estMydB(pl\_col, :), \textprimstress
--k\textprimstress , \textprimstress linewidth\textprimstress , 2);

\hspace{1cm}    ylim([min(SINR\_estMydB(pl\_col, :));
max(SINR\_idealdB(pl\_col, :))+3]);

\hspace{1cm}    xlim([min(MM); max(MM)]);

\hspace{1cm}    xlabel(\textprimstress {\it M}\textprimstress ,
\textprimstress FontSize\textprimstress , 14);

\hspace{1cm}    ylabel(\textprimstress SINR\_{out},
dB\textprimstress , \textprimstress FontSize\textprimstress , 14);

\hspace{1cm}    set(gca, \textprimstress FontSize\textprimstress ,
14);

\hspace{1cm}    drawnow;

\hspace{1cm}    filename = strcat(\textprimstress LC\textprimstress
,...

\hspace{5cm}                     \textprimstress \_M\textprimstress
, num2str(SimParam.NumOfChan),...

\hspace{5cm}                     \textprimstress \_N\textprimstress
, num2str(NN(pl\_col)),...

\hspace{5cm}                     \textprimstress
\_sig\textprimstress , num2str(sig\_pres),...

\hspace{5cm}                     \textprimstress
\_SNR\textprimstress , num2str(SimParam.SNR),...

\hspace{5cm}                     \textprimstress
\_SIR\textprimstress , num2str(SimParam.SIR),...

\hspace{5cm}                     \textprimstress \_tr\textprimstress
, num2str(num\_of\_trials),...

\hspace{5cm}                     \textprimstress .mat\textprimstress
);

\hspace{1cm}    saveas(h, strcat(\textprimstress fig/\textprimstress
, newdir, \textprimstress /\textprimstress , filename,...

\hspace{5cm}                     \textprimstress .eps\textprimstress
));

\hspace{1cm}    delete(gcf);

end;

\setlength{\parindent}{0.9cm}

\noappendix
\nchapter{Список сокращений}

\hspace{-1cm}
АКП --- автокорреляционная последовательность; \\
АР --- авторегрессия;\\
АРСС --- авторегрессия скользящего среднего;\\
АСДЦ --- адаптивная система селекции движущихся целей; \\
АЦП --- аналого--цифровой преобразователь; \\
БМ --- блок масштабирования;\\
БПФ --- быстрое преобразование Фурье; \\
ГШ --- процедура ортогонализации Грама--Шмидта; \\
КГ --- когерентный гетеродин; \\
ЛСП --- линейный случайный поиск; \\
ММЭ --- метод максимальной энтропии; \\
МСКО --- минимум среднеквадратической ошибки; \\
ОКМ --- алгоритм обращения корреляционной матрицы помехи; \\
ОМП --- оценка максимального правдоподобия; \\
ОСПШ --- отношение сигнал помеха шум; \\
ОСШ --- отношение сигнал шум; \\
ПДО --- пространственно--доплеровская обработка; \\
ПВАО --- пространственно--временная адаптивная обработка; \\
ПРВ --- плотность распределения вероятностей; \\
ПУЛТ --- постоянный уровень ложных тревог; \\
РЛС --- радиолокационная станция; \\
РОКМ --- регуляризованный алгоритм обращения корреляционной
матрицы помехи; \\
СД --- селектор дальности; \\
СДЦ --- селекция движущихся целей; \\
СКО --- среднеквадратическая ошибка; \\
СПМ --- спектральная плотность мощности;\\
УПЧ --- усилитель промежуточной частоты; \\
УСП --- ускоренный случайный поиск; \\
УУСП --- управляемый ускоренный случайный поиск; \\
ФАР --- фазированная антенная решетка; \\
ФД --- фазовый детектор; \\
ЦОС --- цифровая обработка сигналов; \\
ЦРГФ --- цифровой режекторный гребенчатый фильтр; \\
ЧПВ --- схема черезпериодного вычетания; \\
ЭМ --- эквалайзер мощности;\\
CDMA --- code division multiple access (множественный доступ с
кодовым разделением каналов);\\
LMS --- least mean squares (метод наименьших средних квадратов); \\
NLMS --- normalized least mean squares (нормализованный метод наименьших средних квадратов); \\
RLS --- recursive least squares (рекурсивный метод наименьших квадратов); \\
STAP --- space--time adaptive processing (пространственно--временная обработка); \\
MEM --- maximum entropy method (метод максимальной энтропии);

\printbibliography
\end{document}